\documentclass{aa}  

\usepackage{graphicx}
\usepackage{txfonts}
\usepackage[utf8]{inputenc}
\usepackage{dcolumn}
\usepackage{natbib}
\usepackage{enumerate}
\usepackage{color}
\usepackage{grffile}
\usepackage{array,float}
\usepackage{multirow}
\usepackage{pdflscape}

\usepackage{afterpage}
\def\kms{\,km\,s$^{-1}$}      
\newcommand{\GP}{Galactic plane}
\newcommand{\mic}{$\,\mu$m}
\newcommand{\dd}{$^{\circ}$}
 
\newcolumntype{d}[1]{D{.}{\cdot}{#1}}
\newcolumntype{.}{D{.}{.}{-1}}
\newcolumntype{;}{D{.}{.}{0}}

\newcommand{\hi}{H{\sc i}}
\newcommand{\hisa}{H{\sc i}SA}
\newcommand{\hii}{H{\sc ii}}

\newcommand{\vlsr}{\textit{v}$_{\rm lsr}$}

\newcommand{\scimes}{\texttt{SCIMES}}
\newcommand{\Disperse}{\texttt{DisPerSE}}
\newcommand{\rms}{r.m.s.}
\newcommand{\submm}{submillimetre}

\newcommand{\mum}{$\mu$m}

\newcommand{\msun}{M$_\odot$}
\newcommand{\testfield}{science demonstration field}

\newcommand{\cco}{$^{13}$CO(2\,--\,1)}
\newcommand{\coo}{C$^{18}$O(2\,--\,1)}

\begin{document}

   \title{SEDIGISM: Structure, excitation, and dynamics of the inner
     Galactic interstellar medium\thanks{This publication is based on
     data acquired with the Atacama Pathfinder EXperiment (APEX) under
     programmes 092.F-9315(A) and 193.C-0584(A). APEX is a collaboration
     between the Max-Planck-Institut f\"ur Radioastronomie, the European
     Southern Observatory, and the Onsala Space Observatory.}}
	\titlerunning{The SEDIGISM Survey}
   
   \author{F.\,Schuller \inst{1,2} \and T.\,Csengeri \inst{1} \and J.\,S.\,Urquhart \inst{1,3}
   \and A.\,Duarte-Cabral \inst{4} \and P.\,J.\,Barnes \inst{5,6} \and A.\,Giannetti \inst{1}
   \and A.\,K.\,Hernandez \inst{7} \and S.\,Leurini \inst{1} \and M.\,Mattern \inst{1}
   \and S.-N.\,X.\,Medina \inst{1}
   \and C.\,Agurto \inst{2} \and F.\,Azagra \inst{2} \and L.\,D.\,Anderson \inst{8} 
   \and M.\,T.\,Beltr\'an \inst{9} \and H.\,Beuther \inst{10} \and S.\,Bontemps \inst{11} \and
   L.\,Bronfman \inst{12} \and C.\,L.\,Dobbs \inst{4} \and M.\,Dumke \inst{2} \and
   R.\,Finger \inst{12} \and A.\,Ginsburg \inst{13} 
   \and E.\,Gonzalez \inst{2} \and T.\,Henning \inst{10} \and J.\,Kauffmann \inst{1}
   \and F.\,Mac-Auliffe \inst{2} \and K.\,M.\,Menten \inst{1} \and 
   F.\,M.\,Montenegro-Montes \inst{2} \and  T.\,J.\,T.\,Moore \inst{14} \and E.\,Muller \inst{15}
   \and R.\,Parra \inst{2} \and J.-P.\,Perez-Beaupuits \inst{2} \and A.\,Pettitt \inst{16} \and
   D.\,Russeil \inst{17}  \and \'A.\,S\'anchez-Monge \inst{18} \and P.\,Schilke \inst{18} \and
   E.\,Schisano \inst{19} \and S.\,Suri \inst{18} \and L.\,Testi \inst{13} \and
   K.\,Torstensson \inst{2} \and P.\,Venegas \inst{2} \and K.\,Wang \inst{13} 
   \and M.\,Wienen \inst{1} \and F.\,Wyrowski \inst{1} \and A.\,Zavagno \inst{17}
 }
\authorrunning{F.\,Schuller et al.}
   \institute{Max-Planck-Institut f\"ur Radioastronomie,
   Auf dem H\"ugel 69, D-53121 Bonn, Germany
   \and
   European Southern Observatory, Alonso de Cordova 3107, Casilla 19001,
   Santiago 19, Chile 
  	\and
   	Centre for Astrophysics and Planetary Science, University of Kent,
    Canterbury, CT2\,7NH, United Kingdom
    \and
    School of Physics, University of Exeter, Stocker Road, Exeter, EX4 4QL, United Kingdom 
    \and
    Astronomy Department, University of Florida, P.O. Box 112055, Gainesville, FL 32611, USA
    \and
	School of Science and Technology, University of New England, Armidale, NSW 2351, Australia
    \and
    Astronomy Department, University of Wisconsin, 475 North Charter St., Madison, WI 53706, USA
    \and
	West Virginia University, Department of Physics \& Astronomy,
	P.O. Box 6315, Morgantown, WV 26506, USA
	\and
    Osservatorio Astrofisico di Arcetri, Largo Enrico Fermi 5, I-50125 Firenze, Italy
    \and
    Max-Planck-Institut f\"ur Astronomie, K\"onigstuhl 17, D-69117 Heidelberg, Germany
    \and
    Universit\'e Bordeaux, LAB, CNRS, UMR 5804, F-33270 Floirac, France
    \and
    Departamento de Astronom\'ia, Universidad de Chile, Casilla 36-D, Santiago, Chile
    \and
    European Southern Observatory, Karl-Schwarzschild-Str. 2,
    D-85748 Garching bei M\"unchen, Germany
    \and
    Astrophysics Research Institute, Liverpool John Moores University,
    146 Brownlow Hill, Liverpool, L3 5RF, United Kingdom
	\and
    National Astronomical Observatory of Japan, Chile Observatory,
    2-21-1 Osawa, Mitaka, Tokyo 181-8588, Japan
    \and
	Department of Physics, Faculty of Science, Hokkaido University, Sapporo 060-0810, Japan 
    \and 
   	Laboratoire d'Astrophysique de Marseille, Aix Marseille Universit\'e,
    CNRS, UMR 7326, F-13388 Marseille, France
    \and
	I. Physikalisches Institut, Universit\"at zu K\"oln, Z\"ulpicher Str. 77,
    D-50937 K\"oln, Germany
    \and
    Istituto di Astrofisica e Planetologia Spaziali, INAF,
    via Fosso del Cavaliere 100, I-00133 Roma, Italy}


 
  \abstract
   {The origin and life-cycle of molecular clouds are still poorly constrained, despite their importance for understanding the evolution of the interstellar medium. Many large-scale surveys of the \GP\ have been conducted recently, allowing for rapid progress in this field. Nevertheless, a sub-arcminute resolution global view of the large-scale distribution of molecular gas, from the diffuse medium to dense clouds and clumps, and of their relationship to the spiral structure, is still missing.}
   {We have carried out a systematic, homogeneous, spectroscopic survey of the inner \GP,
   in order to complement the many continuum Galactic surveys available with
   crucial distance and gas-kinematic information. Our aim is to combine this
   data set with recent infrared to sub-millimetre surveys at similar angular resolutions.}
   {The SEDIGISM survey covers 78\,deg$^2$ of the inner Galaxy ($-$60\dd $\leq$ $\ell$ $\leq$ 18\dd,
   $\vert$ $b$ $\vert$ $\leq$ 0.5\dd) in the J=2--1 rotational transition of $^{13}$CO.
   This isotopologue of CO is less abundant than $^{12}$CO by factors up to 100. Therefore,
   its emission has low to moderate optical depths, and higher critical density, making it
   an ideal tracer of the cold, dense interstellar medium.
   The data have been observed with the SHFI single-pixel instrument at APEX.
   The observational setup covers the \cco\ and \coo\ lines, plus several transitions
   from other molecules.}
   {The observations have been completed. Data reduction is in progress, and the final
   data products will be made available in the near future. Here we give a detailed
   description of the survey and the dedicated data reduction pipeline.
   To illustrate the scientific potential of this survey, preliminary results based
   on a \testfield\ covering $-$20\dd $\leq$ $\ell$ $\leq$ -18.5\dd\ are presented.
   Analysis of the \cco\ data in this field reveals compact clumps, diffuse clouds,
   and filamentary structures at a range of heliocentric distances. 
   By combining our data with data in the (1--0) transition of CO isotopologues from
   the ThrUMMS survey, we are able to compute a 3D realization of the excitation
   temperature and optical depth in the interstellar medium.
   Ultimately, this survey will provide a detailed, global view of the inner Galactic
   interstellar medium at an unprecedented angular resolution of $\sim$30\arcsec.
	}
   {}


   \maketitle
%

\section{Introduction}


Despite being our home, the global structure of the Milky Way is still poorly constrained.
There is substantial ongoing effort to simulate our Galaxy's spiral arms and bar potentials \citep[][among others]{mulder-liem, Khoperskov2013, Pettitt2014, Pettitt2015}.
However, only observations of the entire Galaxy can provide constraints for the gas distribution required for such models. 

Continuum surveys from the infrared to the millimetre, such as GLIMPSE \citep{benjamin2003_ori}, MIPSGAL \citep{carey2009}, WISE \citep{Wright2010}, Hi-GAL \citep{Molinari2010}, ATLASGAL \citep{schuller2009}, and the BGPS \citep{ref-bgps}, are sensitive to thermal emission from dust grains, associated with dense interstellar gas.
Radio-continuum surveys give complementary views of the heated, ionised gas (e.g.\ CORNISH, \citealt{hoare2012}, and THOR, \citealt{ref-bihr}).
These surveys provide catalogues with several thousand compact objects (e.g.\ ATLASGAL; \citealt{csengeri2014,urquhart2014c}),
up to over a million sources for infrared surveys \citep[e.g.\ MIPSGAL,][]{mipsgal-cat},
revealing the recent and ongoing star-formation activity throughout the inner Galaxy.
In addition, mid-infrared surveys reveal thousands of infrared dark clouds \citep[IRDC, e.g.][]{peretto-fuller}, which are seen in extinction against the bright Galactic background emission.

Spectroscopic follow-up observations are essential to measure radial velocities (\vlsr) of
the molecular clouds and clumps detected in far-IR and (sub-)mm surveys,
and to determine their kinematic distances; to constrain their virial mass and gravitational
state (bound or unbound); and to constrain the gas excitation, chemical abundances, and turbulence,
all of which are key parameters in theoretical models of star formation.
Currently available molecular surveys of the inner Galaxy include \citep[for a more complete list, see][]{heyer_dame}: the Galactic Ring Survey (GRS) in $^{13}$CO(1--0) \citep{jackson2006},
which covers a Galactic longitude range of  $\sim$17-55$^\circ$; the HOPS NH$_3$ and H$_2$O maser survey \citep{walsh2011}, which traces dense gas components over $-$70\dd $<$ $\ell$ $<$ 30\dd; the CHaMP survey that covers the Carina tangent (280$^\circ< \ell < 300^\circ$; 
\citealt{barnes2011,champ2016}) in the J $=$ 1--0 transitions of $^{12}$CO, $^{13}$CO, C$^{18}$O, and HCO$^+$; COHRS \citep{dempsey2013}, covering 10$^\circ< \ell < 55^\circ$ in $^{12}$CO(3--2); and CHIMPS \citep{rigby2015}, covering 28$^\circ< \ell < 46^\circ$ in $^{13}$CO(3--2) and C$^{18}$O(3--2). Most of these surveys cover relatively small ($\sim$20--40\dd) ranges in Galactic longitude.

Although they provide a global view of the Galaxy, the relatively low resolution ($\sim$1/8\,degree) and high optical depth of the $^{12}$CO(1--0) surveys by \citet{bronfman1988} and \citet{dame2001} limit our understanding of the dynamical structure of the inner Galaxy.
Several ongoing surveys will ultimately cover large pieces of the Galaxy in CO(1--0) and isotopologues at high angular resolution:
the complete and public ThrUMMS \citep[Three-mm Ultimate Mopra Milky Way Survey,][]{ref-thrumms} covers $-60$\dd $\, < \, \ell \, < \,$0\dd, $\vert b \vert \, < \,$1\dd at 72\arcsec\ resolution;
the ongoing Mopra Southern Galactic Plane CO survey (\citealt{burton2013,braiding2015}) has so far covered 10$^\circ\times$1$^\circ$, but is envisaged to cover 90$^\circ\times$1$^\circ$, with an angular resolution of 36\arcsec;
and the Milky Way Imaging Scroll Painting project
(MWISP\footnote{http://www.radioast.nsdc.cn/mwisp.php}), ongoing since 2011, plans to cover -10.25\dd $\leq \ell \leq$ 250.25\dd\ over $\vert b \vert \leq$ 5.25\dd, with a 52\arcsec\ resolution.

\begin{figure}
\begin{center}
\includegraphics[width=0.49\textwidth, trim= 0 0 0 0]{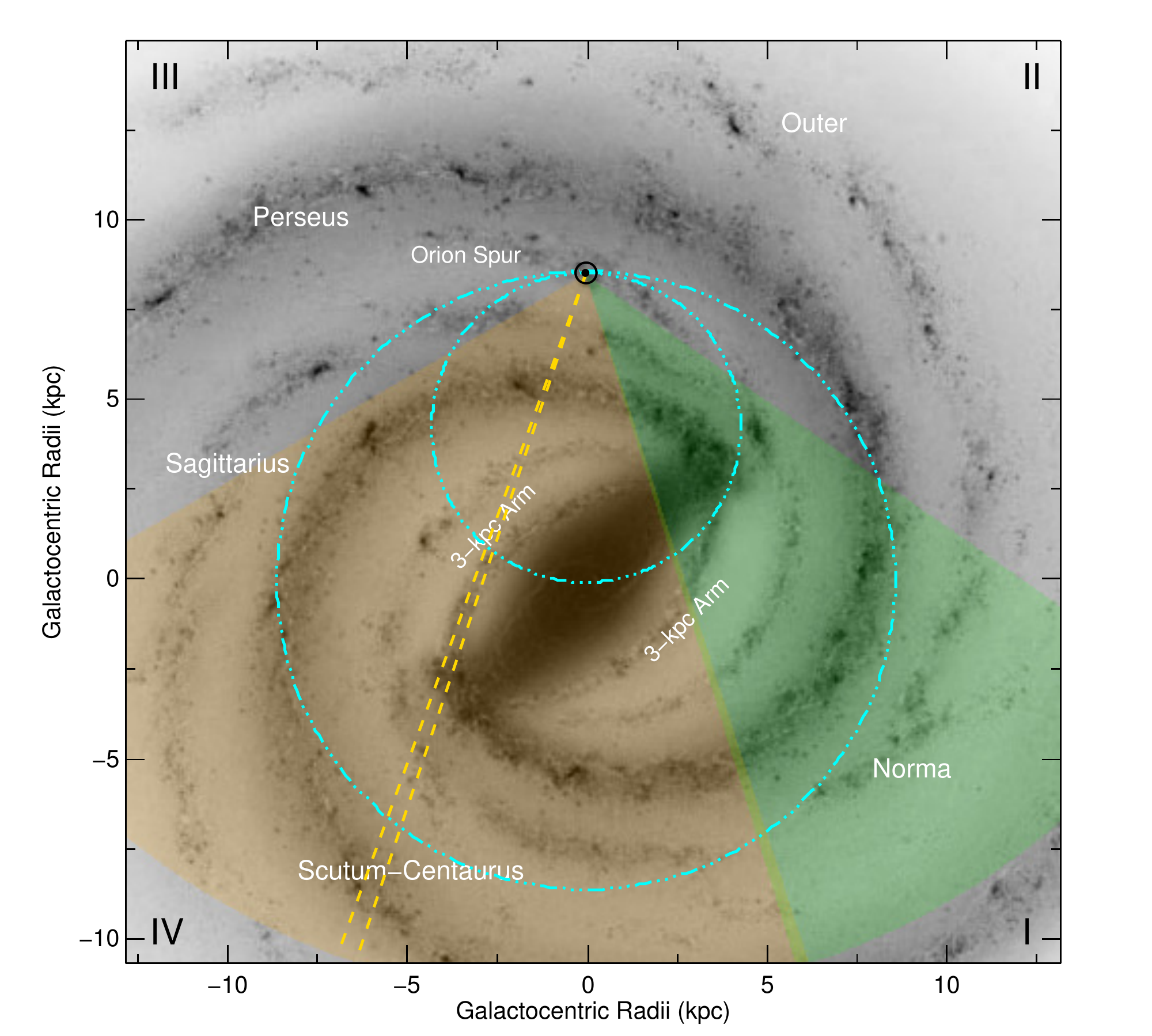}
\caption{Coverage of the SEDIGISM and GRS surveys, shown respectively
in orange and green shading, overlaid on the top down image of the Milky Way produced by Robert Hurt of the Spitzer Science Center in consultation with Robert Benjamin (see \citealt{churchwell2009} for more details). The large and small cyan circles indicate the Solar Circle and the position of the tangent points (maximum radial velocity),
respectively, while the dashed yellow lines demarcate the region selected as the \testfield\ (see Sect.\,\ref{sec:testfield}).
The position of the Sun is indicated by the $\odot$ symbol.}
\label{fig:coverage}
\end{center}
\end{figure}

To provide an even higher angular-resolution view of the inner Galaxy in mostly optically thin gas tracers, we have undertaken the SEDIGISM survey (Structure, Excitation, and Dynamics of the Inner Galactic Interstellar Medium). This survey covers 78\,deg$^2$ of the southern \GP: $-$60\dd $\leq$ $\ell$ $\leq$ 18\dd, with $\vert b \vert$ $\leq$ 0.5\dd, at 30$''$ resolution.
This longitude range has been selected to provide complementary coverage to the GRS survey
(see Fig.\,\ref{fig:coverage} for coverage map).
The prime target of the SEDIGISM survey are the \cco\ and \coo\ transitions, which are usually
optically thin in the Galactic ISM. Therefore, they are well suited to trace the dense molecular gas.
Thus, this survey provides fundamental information to constrain the Galactic structure,
in particular, the number and position of spiral arms.
It also provides crucial \vlsr\ measurements, allowing kinematic distances to be estimated, thus serving as an excellent resource for continuum surveys at comparable resolutions, such as ATLASGAL and Hi-GAL.

The ubiquitous presence of filaments in the Galactic ISM has been recognised with {\it Herschel} \citep[e.g.][]{Molinari2010,Andre2010} in both star-forming and quiescent clouds.
Filaments play a pivotal role in the formation of stars, as it is within them that instabilities develop, leading to the formation of clumps and cores \cite[e.g.][]{Federrath2016,Smith2016}.
In some more extreme cases, filamentary networks are thought to serve as channels for feeding mass onto protostellar cores in the early stages of star formation, providing the material needed for the formation of high-mass stars \citep[e.g.][]{Schneider2010,Peretto2013}.
Despite their importance, the exact formation mechanism of these structures is not yet fully understood.
While some theories suggest that shocks in the ISM are responsible for forming filaments, either through the classical turbulent motions of gas \citep{McKee2007,Federrath2016}, or through more extreme shocks from larger-scale converging flows (e.g. \citealt{heitsch2008}),
other theories suggest that magnetic fields play an important role in both forming and shaping these filaments, with the gas being guided through magnetic field lines \citep[e.g.][]{Nagai1998,Nakamura2008}. 

On much larger scales, long molecular filaments stretching up to hundreds of parsecs have been discovered in our Galaxy \citep{jackson2010,goodman2014,Ragan2014,Wang2015,Zucker2015,Abreu2016}.
There is no consensus yet on whether these are typically associated with the spiral arms, or instead found in inter-arm regions.
It is also unclear if these two types of filaments (the hundreds of parsec-scale filaments and the small pc-scale filaments associated with star formation), or their formation mechanism are related.
Galactic shear probably dominates the shaping of the large-scale filamentary clouds \citep[e.g.][]{Dobbs2015, DC2016}, whereas gravity, turbulence and magnetic fields are more likely to be relevant on the smallest scales; and most likely a mixture of all these processes in between \citep{Federrath2016}.

With a wide coverage of the \GP, and high spatial resolution, the SEDIGISM spectral-line survey
will be sensitive to filamentary structures on all scales, down to $\sim$1~pc at the distance
to the Galactic centre.
Thus, it will be key to providing the much needed kinematical information that will not only allow placing filaments within their Galactic context, but also provide important constraints on the physical properties and initial conditions leading to their formation.

The structure of the paper is as follows: we describe the observations in Sect.\,\ref{sect:obs}, and the data reduction pipeline in Sect.\,\ref{sect:pipeline}. We then present some results derived on a \testfield, including: extraction and characterisation of molecular clouds (Sect.\,\ref{sect:gmcs}); a study of dense gas clumps (Sect.\,\ref{sect:atlasgal}); a preliminary analysis of excitation and physical conditions, based on the combination of the SEDIGISM data with the ThrUMMS survey (Sect.\,\ref{sect:excitation}); and a study of filamentary structures (Sect.\,\ref{sec:filaments}). Finally, we summarise our conclsusions and highlight the perspectives of exploiting the full survey data in Sect.\,\ref{sect:conclusion}.

%
\begin{figure}[tp]
\centering \includegraphics[width=\columnwidth]{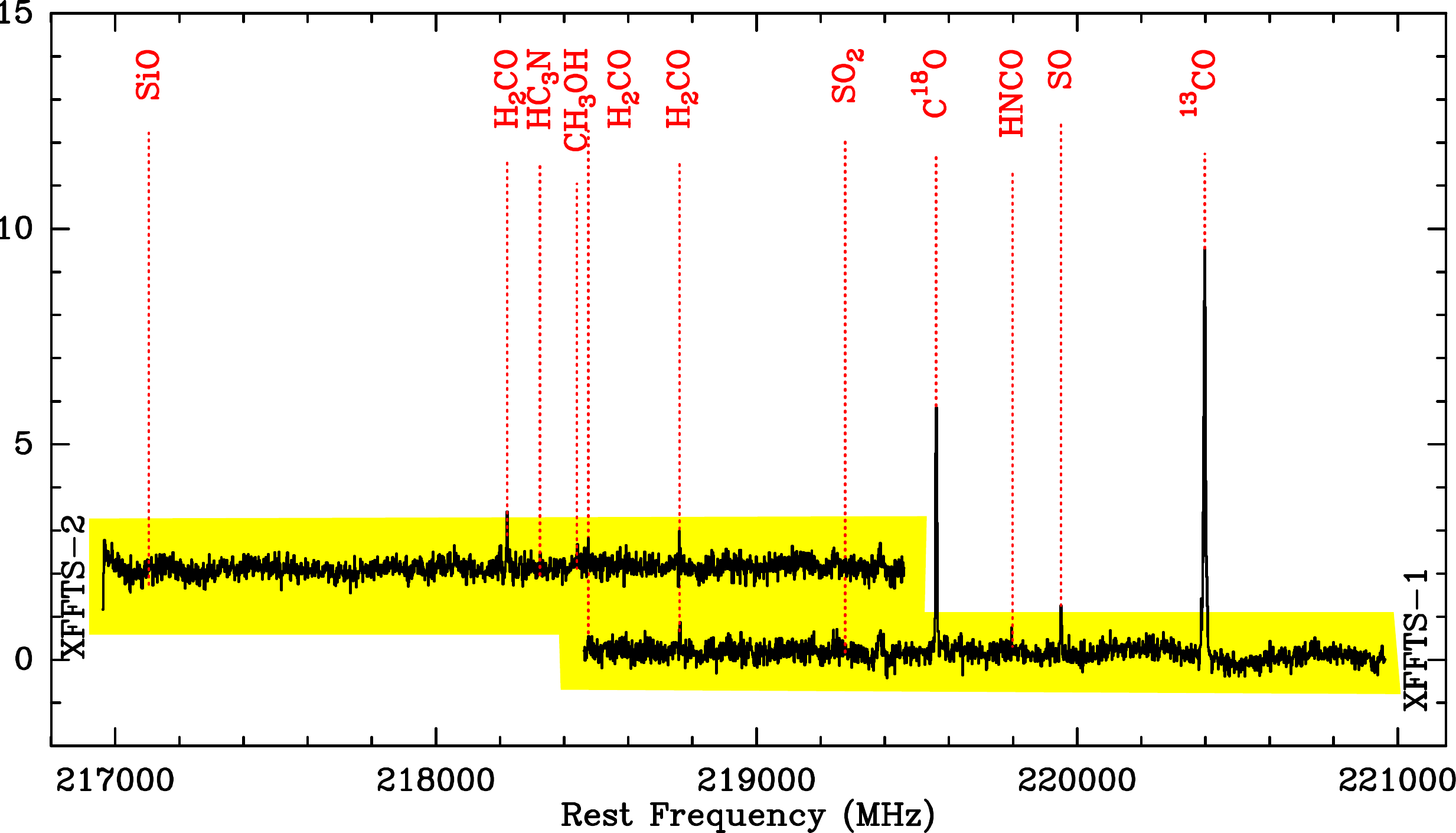}
  \caption{The 4\,GHz frequency coverage with the two backend units, XFFTS-1, and -2. The spectra have been extracted and averaged within the beam towards the brightest ATLASGAL source
in the \testfield, AGAL340.054$-$00.244 \citep[][see also Fig.~\ref{fig:other_lines}]{urquhart2014c}. Red labels mark the brightest spectral lines expected in the observed spectral coverage.
The spectrum recorded by XFFTS-2 has been offset for clarity.}\label{fig:setup2}
\end{figure}

%
\section{Observations}\label{sect:obs}

The data has been collected between 2013 and 2015 with the 12~m diameter Atacama Pathfinder Experiment telescope (APEX, \citealt{Guesten2006}), located at 5100\,m altitude on Llano de Chajnantor, in Chile.
The observations employed the lowest frequency module of the Swedish Heterodyne Facility Instrument (SHFI, \citealt{vassilev2008}).
This is a single-pixel heterodyne receiver with a sideband-separating mixer operating in a single
sideband mode. The back-ends consist of two wide-band Fast Fourier Transform Spectrometers (XFFTS; \citealt{Klein2012}). 
Each spectrometer covers 2.5\,GHz instantaneous bandwidth, with 32768 spectral channels.
At the frequency of our observations (219\,GHz), this translates to a velocity resolution of $\sim$0.1\,\kms.
The two spectrometers cover the 4~GHz IF bandwidth of the receiver with an overlap of 500~MHz, as shown
in Fig.\,\ref{fig:setup2}.
In addition to the CO-isotopologue lines, the setup also covers transitions from
several other molecules, including shock enriched molecules (SiO, SO) and dense gas tracers
(H$_2$CO, CH$_3$OH, CH$_3$CN) - see Table\,\ref{tab:linelist} for rest frequencies and energies.
These lines provide diagnostic tools of star formation activity
towards the densest regions, tracing for example molecular outflows, shocks, and infall motions.

\begin{table}[t]
  \caption{Transitions covered by the instrumental setup. Col.~1 lists the molecules; the
  transitions are described by the main quantum numbers in Col.~2; Cols.~3 and 4 give the rest frequencies
  and the lower state energies, respectively.}
  \label{tab:linelist}
  \begin{tabular}{ll..}
    \hline \hline
    Species & Transition & \multicolumn{1}{c}{Frequency} & \multicolumn{1}{c}{E$_{\rm L}$} \\
    & & \multicolumn{1}{c}{(GHz)} & \multicolumn{1}{c}{(K)} \\
    \hline
    CH$_3$CN & J=12--11, K=0 & 220.7473 & 58.3 \\
    $^{13}$CO & J=2--1 & 220.3987 & 5.3 \\
    SO & J=5--4 & 219.9494 & 24.4 \\
    HNCO & J=10--9, K=0 & 219.7983 & 47.5 \\
    C$^{18}$O & J=2--1 & 219.5604 & 5.3 \\
    SO$_2$ & J=22--23, K=7--6 & 219.2760 & 342.2 \\
    H$_2$CO	& (3$_{2,1}$ -- 2$_{2,0}$) & 218.7601 & 57.6 \\
    H$_2$CO	& (3$_{2,2}$ -- 2$_{2,1}$) & 218.4756 & 57.6 \\
    CH$_3$OH & J=4--3, K=2--1 E & 218.4401 & 35.0 \\
    HC$_3$N & J=24--23 & 218.3247 & 120.5 \\
    H$_2$CO & (3$_{0,3}$ -- 2$_{0,2}$) & 218.2222 & 10.5 \\
    SiO & J=5--4 & 217.1050 & 20.8 \\
    \hline
  \end{tabular}
\end{table}

The survey covers a total of 78\,deg$^2$ of the southern \GP\ ($-$60\dd $\leq$ $\ell$ $\leq$ 18\dd,
$\vert$ $b$ $\vert$ $\leq$ 0.5\dd), with a 28$''$ beam.
Our observing strategy consisted in dividing the full area to be surveyed into $0.5 \times 0.5$\,deg$^2$ fields. 
Each field was covered twice with on-the-fly mapping, scanning in two orthogonal directions:
along Galactic longitude and latitude. Using a scanning speed of 2$'$/s, and 15$''$ steps between lines
(i.e. almost half-beam sampling), the integration time per beam amounts to 0.34\,sec.
With these scanning parameters, we typically reach in the final, combined data a
1-$\sigma$ \rms\ noise of 0.8\,K (in main beam brightness temperature scale, $T_{\rm{mb}}$;
see \citealt{Guesten2006}) at 0.25\,\kms\ spectral resolution in average weather conditions,
meaning with a total amount of precipitable water vapour (PWV) up to 3\,mm.

\begin{figure*}[!htp]
\centering 
\includegraphics[width=0.85\textwidth]{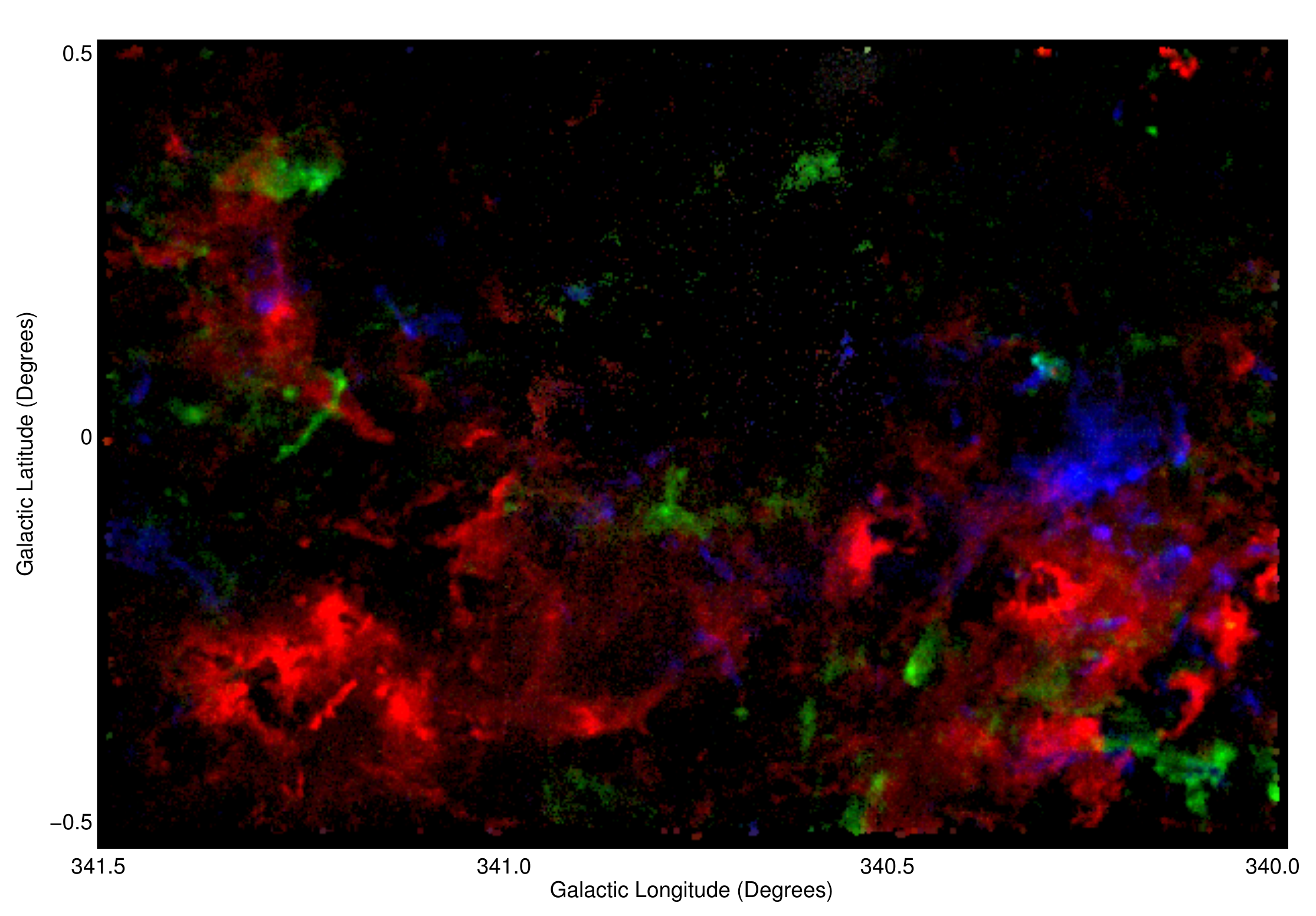}
\caption{Three-colour peak $^{13}$CO emission map of the \testfield. The velocity ranges used to produce the blue, green and red images are $-$130 to $-$110\,\kms, $-$110 to $-$60\,\kms\ and $-$60 to $+$5\,\kms, respectively.
The emission in these velocity ranges is dominated by the near-sides of the 3-kpc, Norma and Scutum-Centaurus arms, respectively.}
\label{fig:three_colour_co_map}
\end{figure*}

Using the RADEX code \citep{ref-radex}, assuming a kinetic temperature of 20\,K, representative of the wide
temperature range found in molecular clouds \cite[e.g.\ Sect.\ 5.1.2 in][]{heyer_dame}, a density of 10$^3$~cm$^{-3}$, and standard abundances [CO]/[H$_2$]$ \, = 10^{-4}$ and $^{12}$C/$^{13}$C$\, = 60$,
we estimate that the SEDIGISM survey can make a 3-$\sigma$ detection of gas with H$_2$ column densities above $\sim3\times 10^{21}$\,cm$^{-2}$ (or $\sim$60~\msun~pc$^{-2}$) in the \cco\ line,
and a 3-$\sigma$ detection of gas above 10$^{22}$\,cm$^{-2}$ (or $\sim$200~\msun~pc$^{-2}$) in \coo. 
The column density threshold for star formation being of the order of $5\times 10^{21}$\,cm$^{-2}$ \citep[e.g.][]{Lada2010}, this sensitivity is well suited to detect all the molecular structures associated with star formation and their surrounding medium.

\subsection{The \testfield}\label{sec:testfield}

To show a typical example of the data products, and to illustrate the potential of the survey, we selected a $1.5\degr \times 1\degr$ area between $\ell=340.0^\circ$ and $\ell=341.5^\circ$ (see the line of sight marked in Fig.\,\ref{fig:coverage}), hereafter referred to as the \testfield.
This is a representative sample of the full survey: the line of sight in this direction crosses several Galactic arms, so that structures are detected at various distances. 
The peak of the \cco\ emission in three representative ranges of \vlsr\ is shown in Fig.\,\ref{fig:three_colour_co_map}. 
The \cco\ and \coo\ cubes covering the \testfield\ are available for download from a dedicated server hosted by the MPIfR\footnote{http://sedigism.mpifr-bonn.mpg.de/}.

\begin{figure}[tp]
\centering 
\includegraphics[width=0.38\textwidth, angle=90]{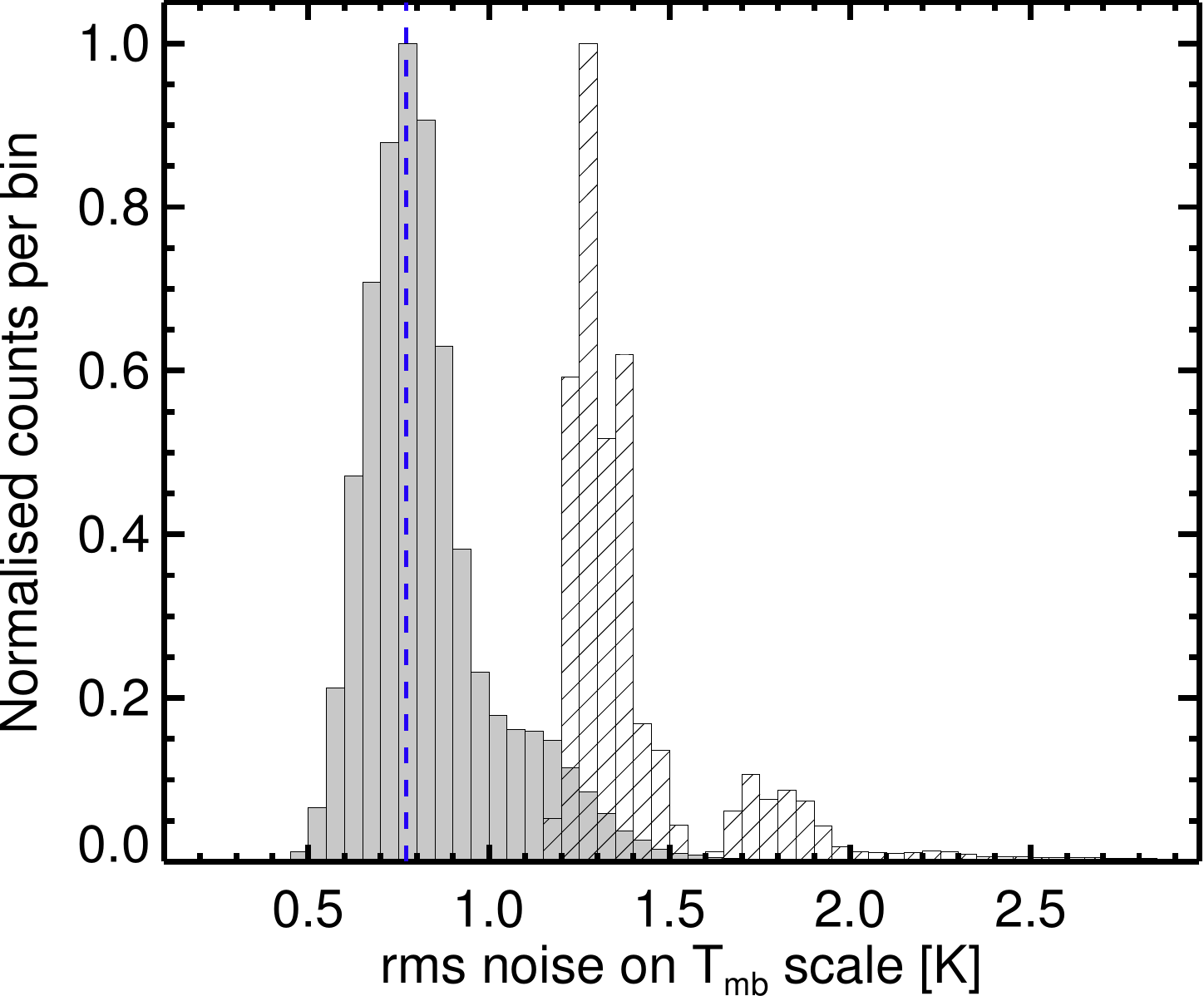}
\caption{Distribution of \rms\ noise values for the \testfield\ on $T_{\rm mb}$ scale.
The filled grey histogram shows the noise in the final, combined data;
the hatched histogram shows the noise measured in the individual scans
that were combined for the final data cube.
The distribution peaks at 0.78\,K, as indicated by the blue dashed line.
Both distributions have been normalised to a peak value of 1.}\label{fig:noise_distribution2}
\end{figure}

\begin{figure}[tp]
\centering 
\resizebox{\columnwidth}{!}{\includegraphics[angle=270]{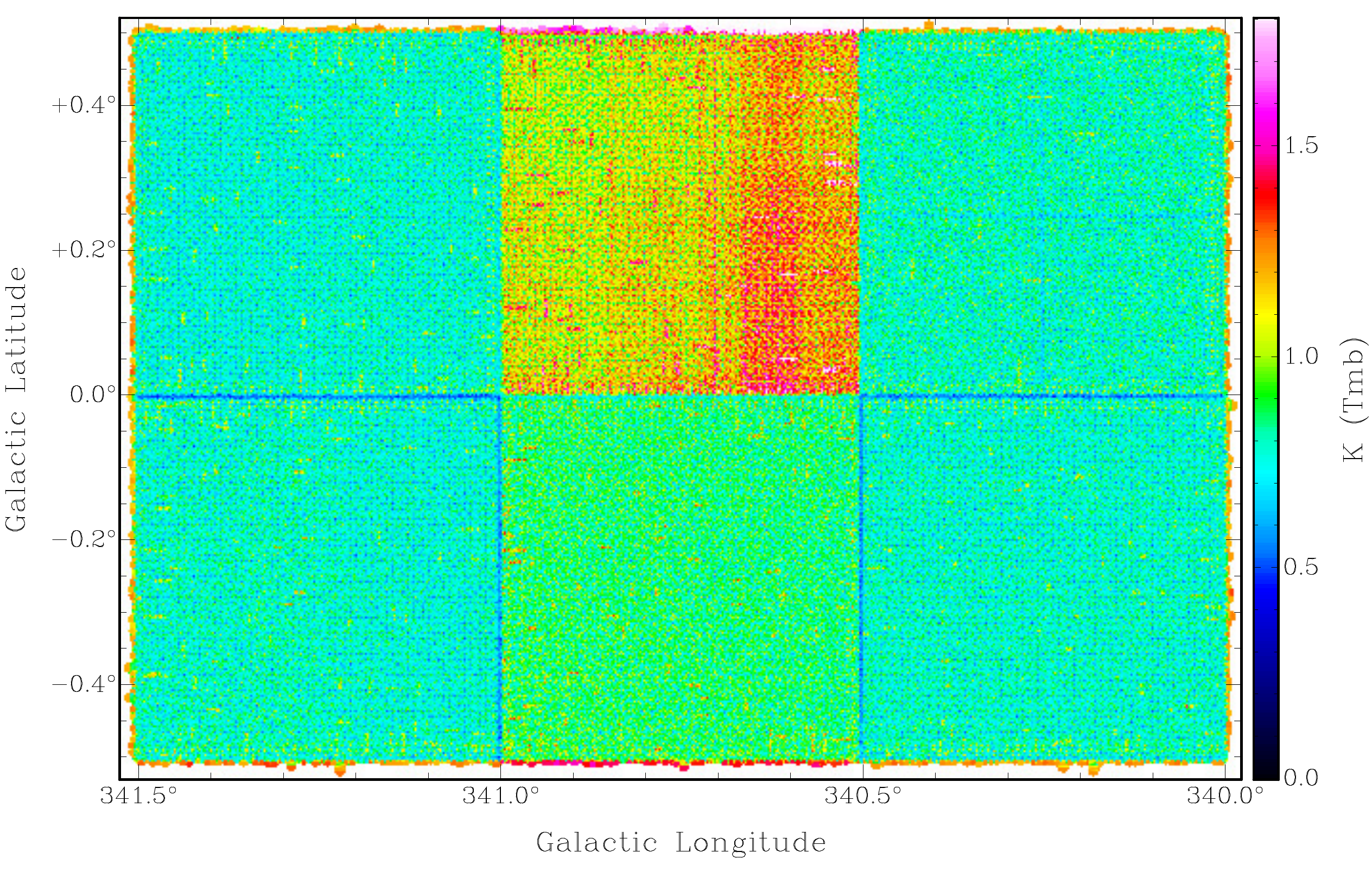}}
\caption{Spatial variations of the rms noise in the \testfield. }\label{fig_map_rms}
\end{figure}

The distribution of \rms\ noise values in this region is shown in
Fig.\,\ref{fig:noise_distribution2}, and the spatial variations of the rms can
be seen in Fig.\,\ref{fig_map_rms}.
Most observations of this field were done with PWV $<$ 2\,mm, resulting in \rms\ noise of order 1.3\,K in individual scans, and 0.7--0.8~K in the combined data. However, one sub-field, centred at ($\ell,b$) = (340.75, +0.25), was observed with PWV $\approx$ 3.4\,mm, which results in a slightly higher noise around 1.7\,K in individual scans, and $\sim$1.2\,K in the combined map.

\section{Data reduction}\label{sect:pipeline}

The data provided by the APEX telescope consist of spectra calibrated
in antenna temperature scale (T$_{\rm A}^{*}$), written in files readable
by the CLASS software from the GILDAS
package\footnote{http://www.iram.fr/IRAMFR/GILDAS/}.
We created a pipeline in CLASS to process each scan individually, starting
from the data expressed in T$_{\rm A}^{*}$ scale.
This pipeline automatically detects line emission in order to mask the corresponding channels before baseline subtraction. The main steps of the data reduction are detailed in the next section.

\subsection{Baseline subtraction and calibration}

The processing of each individual scan involves the following steps:
\begin{itemize}
\item First, a $\pm500$\,\kms\ velocity range centred on the \cco\ line rest frequency is extracted from the spectra; this range is expected to cover all the Galactic emission. 
Simultaneously, spectra centred on the other spectral lines are extracted from the data, but with a smaller velocity range of $\pm300$\,\kms.
\item The spectra are re-sampled to 0.25\,\kms\ velocity
resolution for the CO isotopologues, and 0.5\,\kms\ for the other lines, in order to improve the sensitivity for the weaker lines.
\item The calibration of observations done between March and June 2014 have been affected by technical issues with the receiver.
Appropriate scaling factors are thus applied to re-calibrate the
data\footnote{http://www.apex-telescope.org/heterodyne/shfi/calibration/calfactor/}.
These factors are 0.73 and 0.90 for the XFFTS-1 and XFFTS-2 backends, respectively,
for data taken between March 20 and April 23, 2014;
and 0.80 and 0.95 for data taken between April 23 and June 13, 2014.
\item For spectral lines covered by both backends, the data recorded with XFFTS-1 is used,
because this backend shows a more stable behaviour compared to XFFTS-2, even in relatively
poor weather conditions (i.e. PWV up to $\sim$4\,mm).
\item The \rms\ noise per channel, $\sigma_{\rm rms}$, is then determined by taking
the minimum of the median noise level in a sliding spectral window of 80 channels,
expected to be representative of line-free frequency ranges.
To increase the signal to noise ratio for the subsequent line finding, this is done on a
smoothed version of the data, where the spectra have been spatially averaged in a box of
500\arcsec\ around every given offset.
\item The channels that show a signal above a threshold of $3 \times \sigma_{\rm rms}$ 
are used to define the windows, which are subsequently excluded for baseline subtraction.
\item To increase the efficiency of the data reduction pipeline, the same windows are used
for three subsequent pointing offsets, since the data were dumped at every position 
spaced by $\sim$1/3 of the beam.
The same windows are applied to all extracted spectral lines.
\item A polynomial baseline of 3$^{rd}$ order is then subtracted from the
spectra extracted around all spectral lines.
\item Then, the spectra are converted to main beam temperature ($T_{\rm mb}$) scale using an 
efficiency value of 0.75\footnote{http://www.apex-telescope.org/telescope/efficiency/index.php}.
\item Finally, emission from the reference position is corrected for, when necessary (see below).
\end{itemize}

The observations for the SEDIGISM survey used absolute reference positions,
located at $\pm$1.5$^\circ$ in $b$, at the same $\ell$ as the centre of each
field. These reference positions are usually far enough from the \GP\
to be located towards emission-free regions of the sky, but not  all reference positions
are clear of emission in the \cco\ line.
Therefore, all the reference positions have been observed and reduced independently,
using a more distant reference position off the \GP.
Only towards those fields where the reference position shows emission in \cco,
this emission has been corrected for in the final maps.
We independently checked for emission in the \coo\ line toward
the reference positions as well, and corrected for it when necessary.
Still, we cannot exclude that negative artefacts due to imperfect correction for emission 
at the reference position, may still be present in some data.

Two line-rich sources (SgrB2 and IRC+10216) were regularly observed in 2014 and 2015,
using the same set-up as for the science observations. These sources can be used as 
spectral calibrators. We measure a dispersion of $\sim$7\% in the integrated area
below the \cco\ line between observations, for both sources. Therefore, we can safely
assume that the uncertainty on the temperature scale is better than 10\%, as is
typically the case with the SHFI instrument.

\subsection{Combining several scans}

\begin{figure}[tp]
\centering 
\resizebox{0.9\columnwidth}{!}{\includegraphics[angle=270]{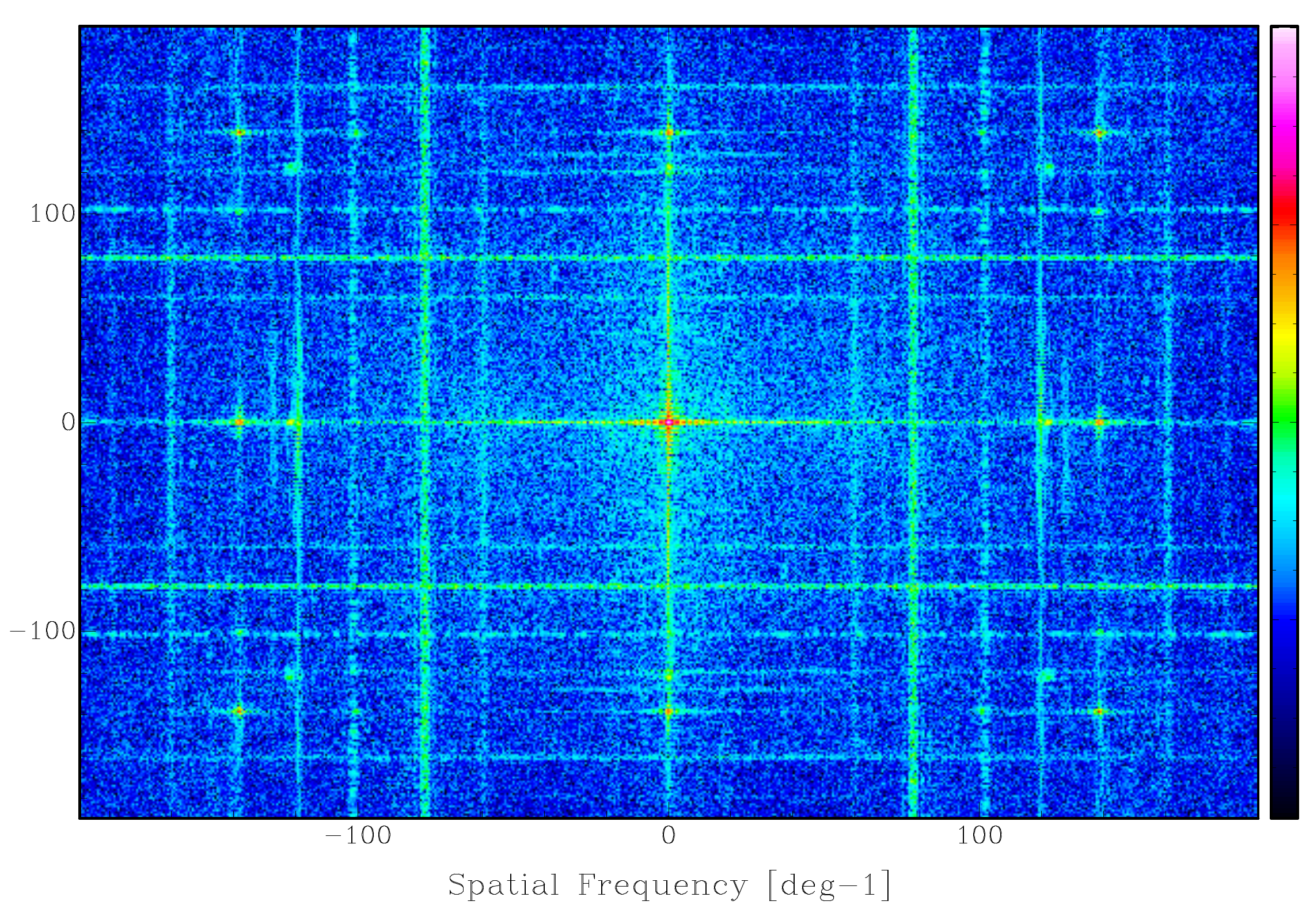}}
\caption{Two-dimensional power spectrum of the noise image shown in
Fig.~\ref{fig_map_rms}.\label{fig_rms_fft}}
\end{figure}

Two scans observed with orthogonal scanning directions cover each given field.
When the noise obtained by combining these two scans was significantly higher than average, we observed the field again at least in one scanning direction.
The reduced, calibrated data obtained from the different scans are finally combined and gridded using 9.5\arcsec\ cell size.
The gridding process includes a convolution with a Gaussian kernel of size one-third the telescope beam.
This provides data cubes with a final angular resolution of 30\arcsec.
Data cubes are generated for all the transitions listed in Table\,\ref{tab:linelist},
at 0.25\,\kms\ resolution for the CO isotopologues, and 0.5\,\kms\ resolution for the other species.

\begin{figure}[tp]
\centering \includegraphics[width=0.92\columnwidth]{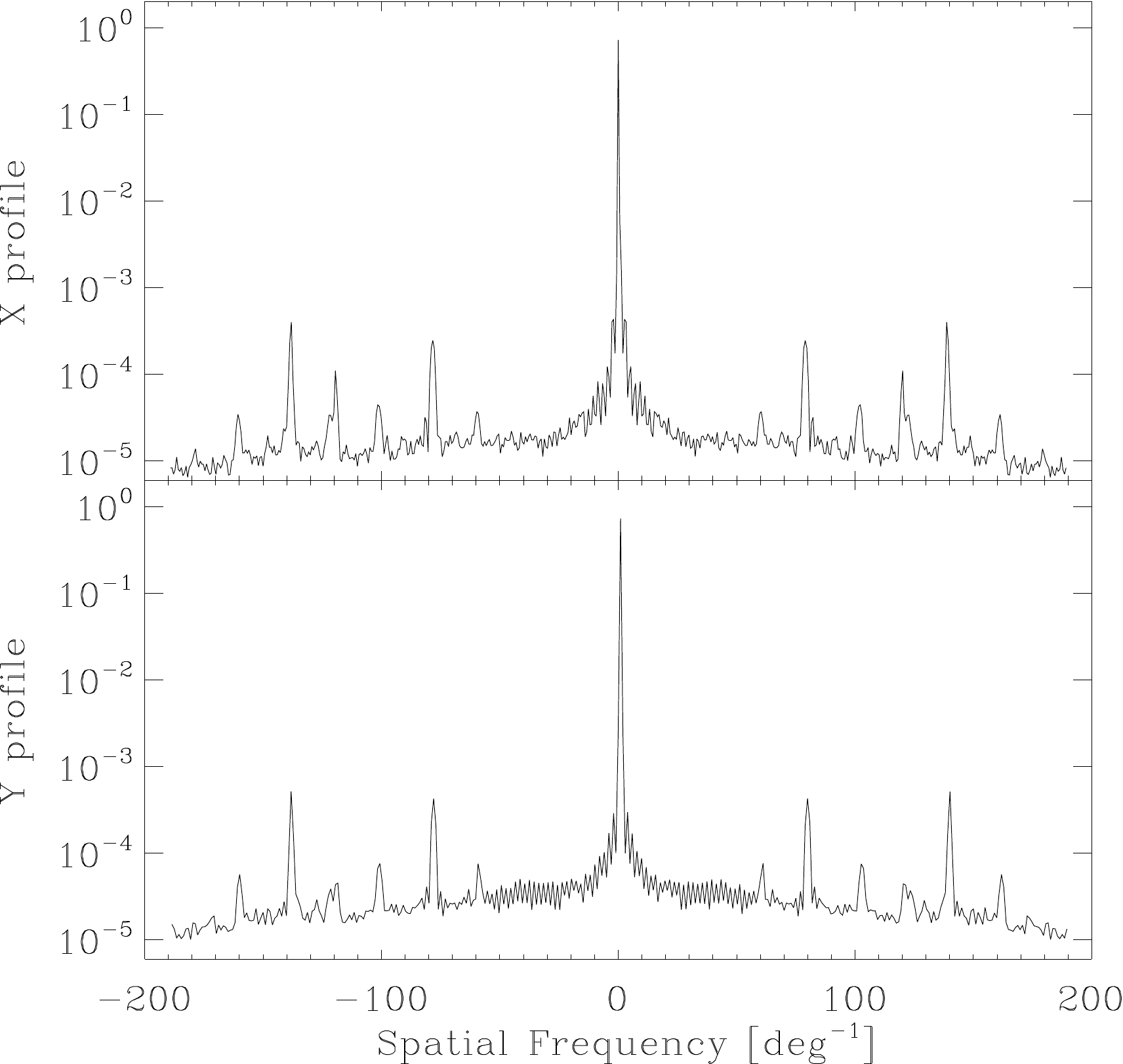}
  \caption{Power spectrum of the noise image, as shown in Fig.~\ref{fig_rms_fft},
  but averaged over Galactic latitudes (top) and longitudes (bottom).}
  \label{fig:fft_profile}
\end{figure}

Combining several scans observed with different scanning directions helps
in reducing striping artefacts in the data. However, the pixel-to-pixel noise is
not independent but shows high degree of correlation at some specific spatial
scales. This is illustrated in Fig.~\ref{fig_rms_fft}, which shows the
2D power spectrum (the square of the module of the 2D Fourier transform) of
the noise image shown in Fig.~\ref{fig_map_rms}. Some peaks are clearly visible
(see also Fig.~\ref{fig:fft_profile}) at frequencies corresponding to multiples of
the scanning parameters (steps between lines, and steps between dumps within
a line).

\subsection{Lines from other species}

\begin{figure}[ht]
\centering 
\includegraphics[width=0.9\columnwidth]{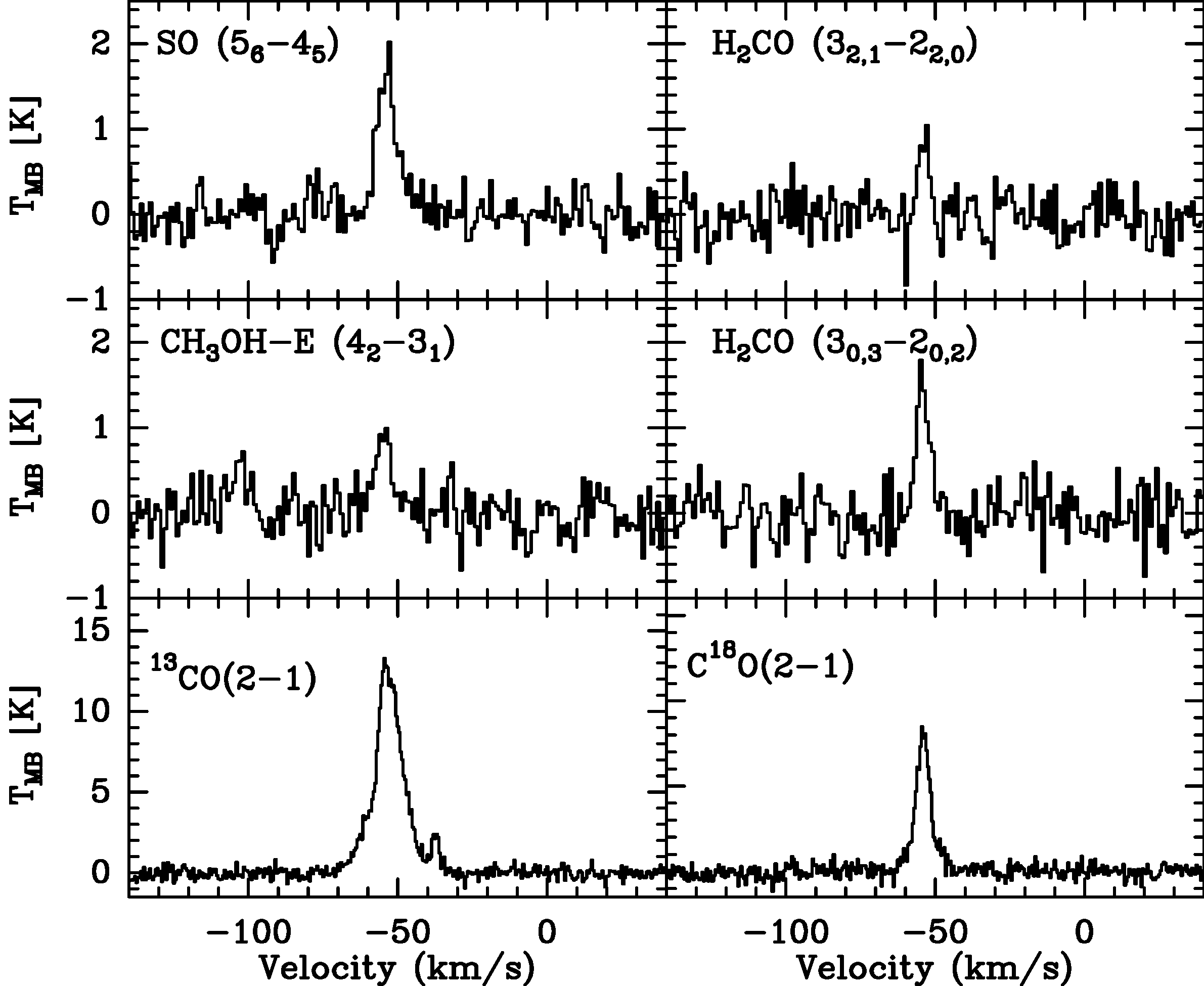}
\caption{Spectra extracted around the lines of SO, H$_2$CO, CH$_3$OH, $^{13}$CO
and C$^{18}$O towards AGAL340.054$-$00.244, which is the brightest ATLASGAL clump
in the field. The emission has been integrated within a radius of 15\arcsec\ around
the peak position of the dust clump.
}
\label{fig:other_lines}
\end{figure}

The velocity ranges where a signal was automatically detected in the \cco\ data have
been masked for subtracting baselines to the data covering all the other lines possibly
present in the data. Some lines from other species are indeed detected towards the
brightest clumps, including SO(5--4), H$_2$CO (three different lines), and one line of
CH$_3$OH. In the \testfield, these are seen only toward a handful of sources;
an example is shown in Fig.\,\ref{fig:other_lines}.
Due to low detection statistics, these lines will not be further discussed
in this overview paper.

%
%

\section{Molecular clouds and complexes}
\label{sect:gmcs}

\subsection{The extraction algorithm: \scimes\ }

\begin{figure*}[tbp]
\centering 
\includegraphics[width=\textwidth]{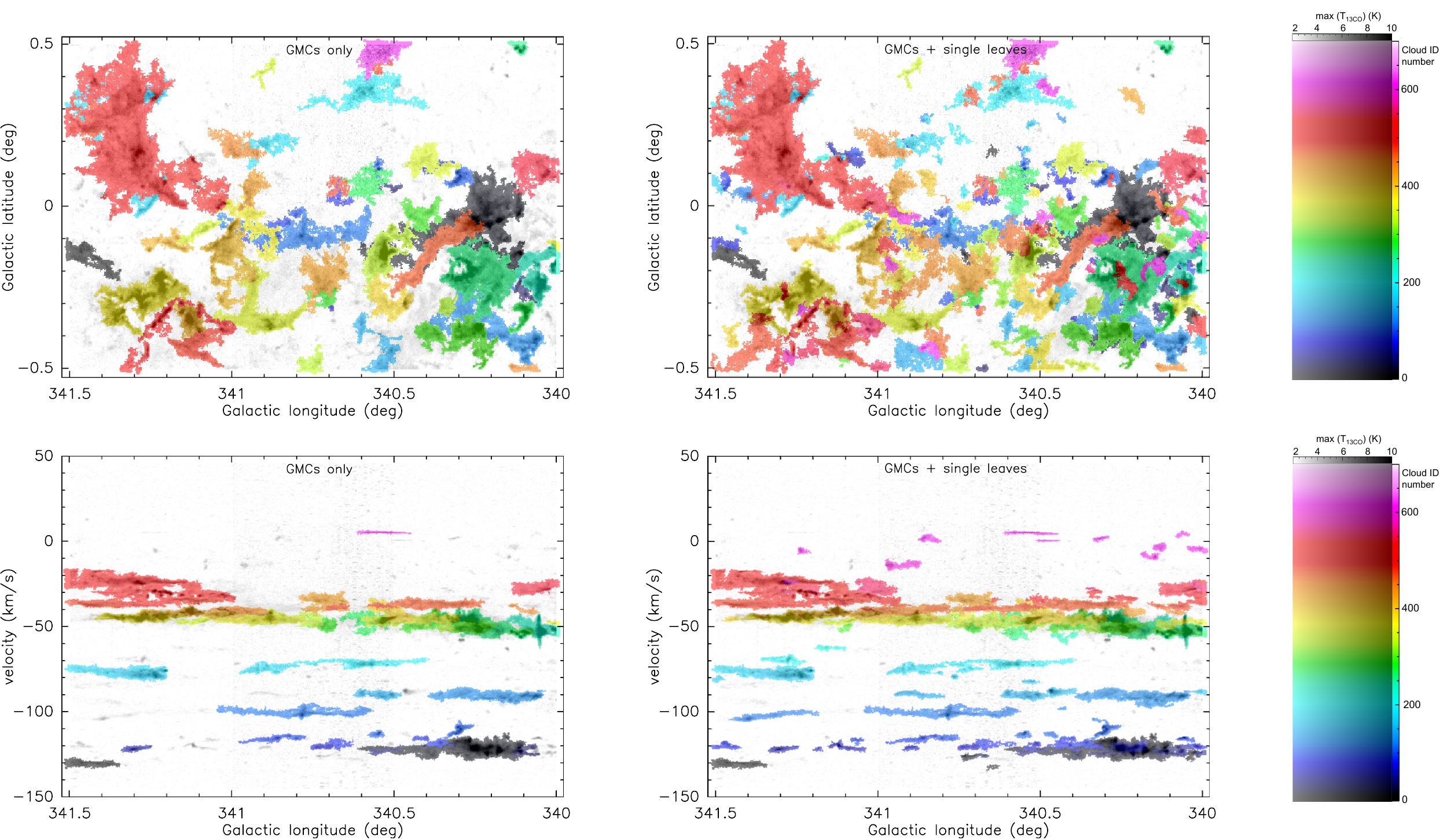}
\caption{Position and extent of the \scimes\ cloud complexes only (left), and all the clouds (right), i.e. the  molecular cloud complexes plus all the well resolved single-leaf clouds, in $\ell b$\ (top) and $\ell v$ space (bottom).
The colour-coding refers to the ID number of the clouds as per our catalogue, and relates to the third dimension of the cube (i.e. \vlsr). The coloured masks of the clouds are overplotted with transparency over the maps of the peak \cco\ emission, in grey scale (see the transparency-combined scale on the right).} 
\label{fig:gmcs}
\end{figure*}

To extract clouds from the SEDIGISM \testfield, we have used the \scimes\ algorithm \citep[Spectral Clustering for Interstellar Molecular Emission Segmentation; for details see][]{Colombo15}.
This tool is designed to identify molecular cloud complexes in 3D data cubes, based on cluster analysis.
While other available 3D cloud extraction algorithms tend to segment the emission into individual emission peaks/clumps inside molecular clouds,
the advantage of \scimes\ is that it is tailored to group different peaks together,
making it more suitable to extract large complexes of clouds.

In practice, this code considers the dendrogram tree of the 3D structures in the data cube \citep[as per the implementation of][to analyse astronomical data sets]{Rosolowsky2008}
in the broader framework of graph theory, and groups different leaves (in this case, the emission peaks) together into ``clusters'' of leaves, based on some criteria (e.g. intensity, luminosity, or volume).
For a more detailed description of the terminology and the algorithm, see \citet[][]{Colombo15}.

We ran the extraction of clouds on the \cco\ emission $\ell bv$ data cube of the SEDIGISM \testfield, after binning the velocities into 0.5\,\kms\ channels, resulting in a data cube with a noise level ranging between $\sim$0.6-0.8\,K.
For building the dendrogram tree of the data cube, we have therefore considered an \rms\ noise level ($\sigma$) of 0.7\,K.
We have used a value of 4-$\sigma$ as the minimum difference between two peaks for these to be considered as separate leaves,
and a lower threshold for detection of 2-$\sigma$, to maximise the connections between different structures at contiguous lower intensity levels.
We used both the intensity and volume as the \scimes\ clustering criteria.
This extraction provides an output data cube with a mask containing all the \scimes\ clusters found, a data cube with the mask of only the dendrogram leaves, and the entire catalogue of dendrogram structures.
As we were also interested in the larger clouds that may simply have little substructure within them (the level of sub-structuring can be a consequence of the resolution),
we have also included the single leaves in the dendrogram which had been excluded by the clustering algorithm,
but whose size was large enough to be well resolved (i.e. when 1-$\sigma$ of the semi-major axis of the cloud was larger than the beam size). 

\subsection{Catalogue of molecular clouds and measured properties}

We extracted a total of 182 molecular clouds, of which 58 are categorised as clusters by \scimes\ (i.e. complexes with at least 2 dendrogram leaves clustered together), and 124 are well-resolved single-leaf clouds.
The position and extent of these clouds can be seen as coloured masks in the $\ell b$ and $\ell v$ plots of Fig.\,\ref{fig:gmcs}, where the left panels show the \scimes\ molecular cloud complexes alone,
while the right panels also include the well resolved single-leaf clouds.

We measured the following properties for each cloud:
centroid position in Galactic coordinates, velocity, velocity dispersion, intensity-weighted semi-major and semi-minor axes, and respective position angle.
The aspect ratio is then computed as the ratio of the major/minor axis.
These properties are summarised in Table\,\ref{tab:scimes_gmcs} for a sub-sample of the largest GMCs (full catalogue in Table\,\ref{tab:full_catalogue}).
We also estimate the physical properties of our sample of clouds (see Sect.~\ref{sec:gmc_physical_properties}),
after determining their kinematic distances (Sect.\,\ref{sec:distances}).

In order to validate the cloud catalogue built using the \scimes\ algorithm, we compared the results of the \scimes\ extraction with independently, visually identified structures in the \cco\ $\ell bv$ cubes.
To do this, we started by identifying the brightest peaks of emission in the $\ell bv$ cube and determined their extent in velocity.
We then defined polygons around individual structures based on their morphology and calculated their corresponding average spectra, to which we fit single or multiple Gaussian profiles.
For the final identification we mapped the structures by integrating the emission over the FWHM of the fitted lines, and considering the emission above a threshold of $\sim$3-$\sigma$ noise level.
With this method, we identified a total of 25 bright structures.

By comparing the position and extent of these structures in $\ell bv$ to the \scimes\ clouds,
we find that ten of them have a one to one correspondence, while in five cases,
one visually identified structure corresponds to two \scimes\ clouds.
The remaining structures cover similar emission to the \scimes\ clouds, but the dissection
in position and velocity differs slightly, mostly with the \scimes\ algorithm tending to
segment the emission more in position space, and group it more efficiently in the velocity space.
Nevertheless, we find that in general the \scimes\ extraction provides a
satisfactory segmentation of the data.

\setlength{\tabcolsep}{2pt}

\begin{table*}
\caption{Properties of a sub-sample of clouds from the \scimes\ extraction, restricted to GMCs with at least 5 leaves.
The ID number shows the catalogue number associated with the cloud (same as the colour-scale in Fig.\,\ref{fig:gmcs}).
The GMC name is defined as SDG (for SEDIGISM) followed by the Galactic coordinates of the clouds' centroid.
Cols.\,3 and 4 show the intensity-weighted semi-major and semi-minor axes, $a$ and $b$, respectively;
Col.\,5 shows the position angle (P.\,A.), and Col.\,6 shows the aspect ratio (A.\,R.) defined as $a/b$.
In Cols.\,7-9 we show the centroid velocity, velocity dispersion, and average $^{13}$CO ($2-1$) integrated intensity across the area of the cloud.
Col.\,10 shows the number of dendrogram leaves, N$_{l}$, that make up each GMC.
Col.\,11 shows the adopted distances ($d$) and their uncertainties.
Cols.\,12-14 show the exact area defined by the clouds' masks, the equivalent radius ($R$, assuming circular geometry) and maximum length ($l_{\rm{max}}$).
Cols.\,15-17 show the total mass ($M$), the average surface density ($\Sigma$), and the virial parameter ($\alpha_{\rm{vir}}$). See Sect.~\ref{sec:gmc_physical_properties} for details.
}
\label{tab:scimes_gmcs_properties}
\label{tab:scimes_gmcs}
\begin{minipage}{\linewidth}
\small
\renewcommand{\footnoterule}{}  
\begin{tabular}{r l | ; ; ; c . c . c | c c c c c c c}
\hline 
\hline
& & \multicolumn{8}{c |}{Measured properties} & \multicolumn{7}{c}{Physical properties}\\
\hline
\multirow{2}{0.5cm}{ID} & \multirow{2}{1.0cm}{Name} &  \multicolumn{1}{c}{$a$} &  \multicolumn{1}{c}{$b$}  & \multicolumn{1}{c}{P.A.} & A.R. & \multicolumn{1}{c}{\vlsr}	& $\sigma_{v}$ & \multicolumn{1}{c}{$< W_{\rm{CO}} >$} & N$_{l}$ & $d$ & Area & $R$ & $l_{\rm{max}}$ & $M$ & $\Sigma$	& $\alpha_{\rm vir}$ \\ 
 & &  \multicolumn{1}{c}{(\arcsec)} &  \multicolumn{1}{c}{(\arcsec)} & \multicolumn{1}{c}{($^{\circ}$)} & & \multicolumn{1}{c}{(\kms)} & (\kms) & \multicolumn{1}{c}{(K\,\kms)} &  & (kpc) & (pc$^{2}$) & (pc) & (pc) & (10$^{3}$M$_{\odot}$) & (M$_{\odot}$pc$^{-2}$) &  \\ 
\hline
15 & SDG340.245$-$0.056 & 322 & 203 & -161 &  1.6 &  -122.0 &  2.3 &  9.7 & 14 & $  6.58\pm  0.28$ & 915 &   17.1 &   72.2 &    196.9 & 214 &   0.5\\
34 & SDG340.096$-$0.252 & 76 & 69 & 110 &  1.1 &  -123.2 &  1.8 &  4.0 & 5 & $  6.63\pm  0.29$ & 51 &    4.1 &   14.5 &      4.7 & 89 &   3.3\\
127 & SDG340.193$-$0.369 & 364 & 146 & 153 &  2.5 &   -90.7 &  1.6 &  5.1 & 9 & $  5.47\pm  0.26$ & 275 &    9.4 &   44.0 &     31.3 & 113 &   0.9\\
234 & SDG340.054$-$0.214 & 151 & 105 & 70 &  1.4 &   -51.9 &  2.6 & 15.8 & 5 & $  3.86\pm  0.36$ & 62 &    4.5 &   17.6 &     22.0 & 350 &   1.6\\
260 & SDG340.240$-$0.213 & 180 & 147 & 148 &  1.2 &   -48.9 &  3.5 & 17.4 & 9 & $  3.73\pm  0.37$ & 137 &    6.6 &   23.8 &     53.3 & 386 &   1.7\\
271 & SDG340.582+0.069 & 114 & 88 & -157 &  1.3 &   -47.1 &  3.6 &  3.3 & 6 & $  3.66\pm  0.38$ & 38 &    3.5 &   13.5 &      2.8 & 73 &  18.4\\
298 & SDG340.300$-$0.395 & 228 & 92 & 176 &  2.5 &   -49.1 &  2.9 & 14.3 & 7 & $  3.74\pm  0.37$ & 76 &    4.9 &   20.1 &     24.1 & 317 &   2.0\\
375 & SDG341.260$-$0.276 & 238 & 107 & -153 &  2.2 &   -44.4 &  1.2 & 12.5 & 8 & $  3.57\pm  0.40$ & 105 &    5.8 &   23.0 &     29.2 & 277 &   0.3\\
408 & SDG341.010$-$0.151 & 229 & 110 & 89 &  2.1 &   -42.3 &  1.1 &  6.1 & 6 & $  3.44\pm  0.41$ & 70 &    4.7 &   20.2 &      9.5 & 134 &   0.7\\
508 & SDG341.212$-$0.345 & 146 & 42 & -136 &  3.4 &   -30.4 &  1.6 &  5.5 & 5 & $  2.00\pm  0.49$ & 7 &    1.5 &    7.7 &      0.9 & 121 &   5.2\\
525 & SDG341.327+0.219 & 422 & 219 & 104 &  1.9 &   -23.6 &  2.2 &  6.2 & 13 & $  2.27\pm  0.56$ & 147 &    6.9 &   26.7 &     20.4 & 138 &   1.9\\
\hline
\end{tabular}
\end{minipage}
\end{table*}

\subsection{Distance determination}
\label{sec:distances}

\begin{figure}[ht]
\centering 
\includegraphics[width=0.45\textwidth]{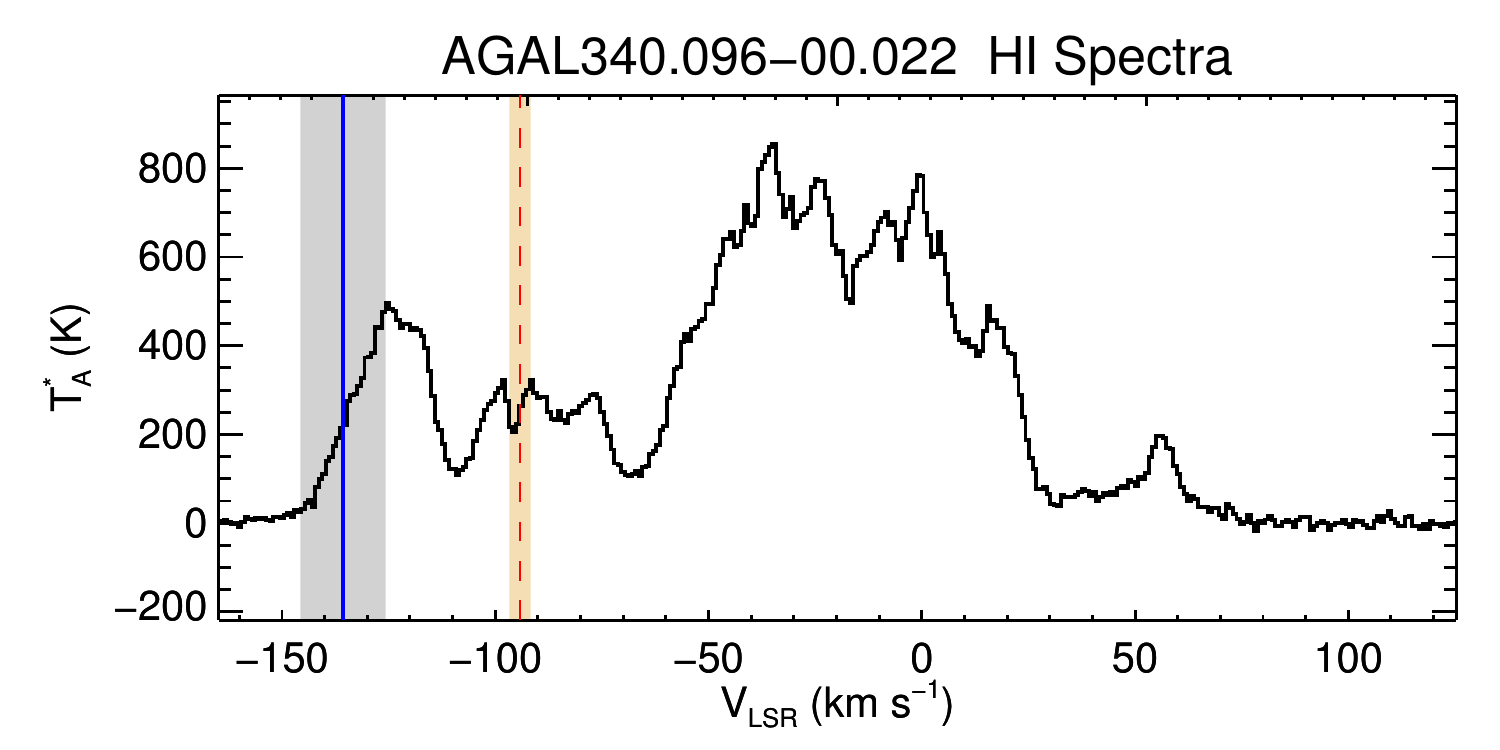}
\includegraphics[width=0.45\textwidth]{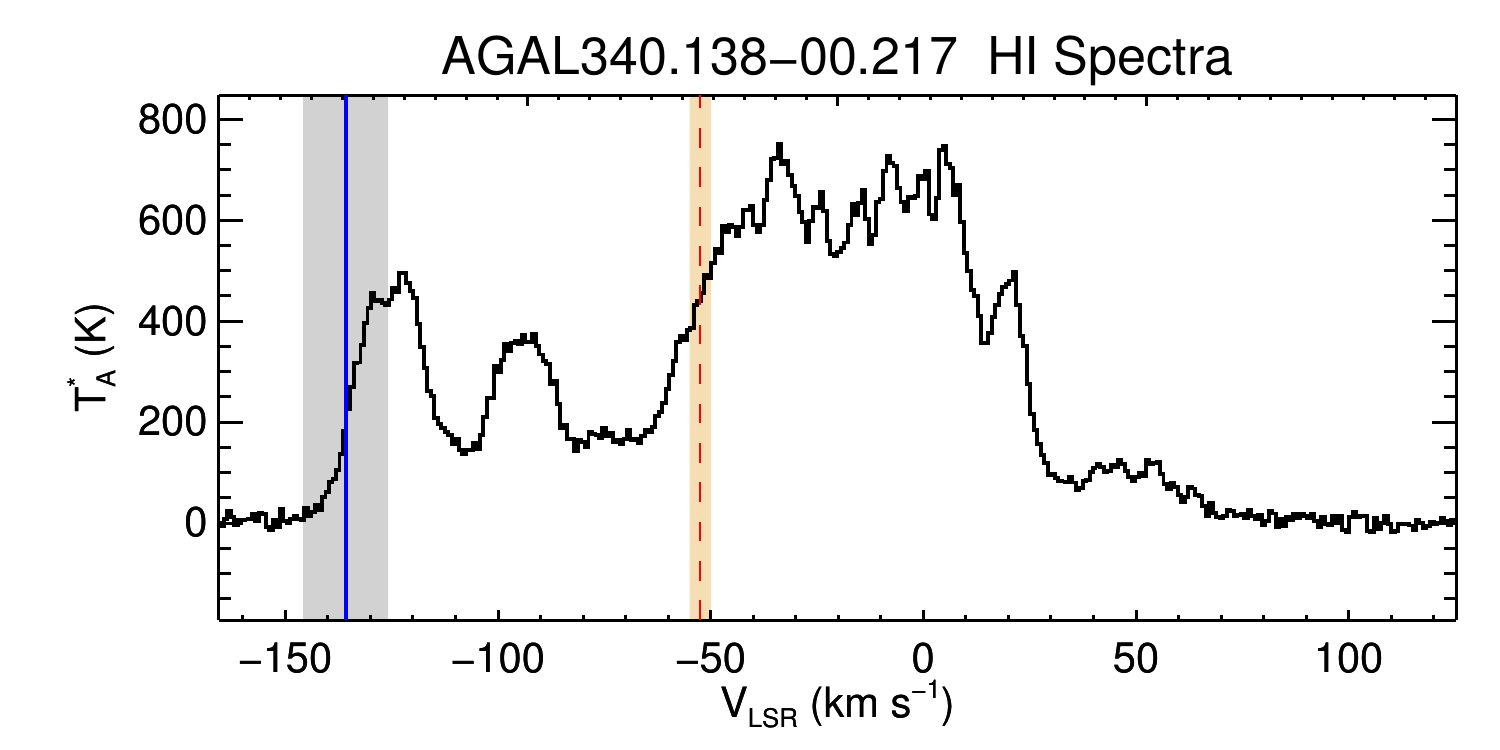}
\caption{Example \hi\ spectra extracted towards dense clumps identified from the ATLASGAL survey.
The velocity of the tangent position is determined from a fit to \hi\ data (e.g.\, \citealt{mcclure2007})
and is indicated by the blue line; the grey shaded region covers a velocity range of $\pm$10\,\kms.
Sources in this region are placed at the distance of the tangent point.
The dashed red vertical line shows the velocity of the source with the yellow shaded region showing the typical FWHM of the molecular lines.
The top and bottom panels show examples of a source located at the near and far distances, respectively.}
\label{fig:example_HI_spectra_scimes}
\end{figure}

We have derived the kinematic distances of all the molecular clouds in the \testfield\ using the Galactic rotation model of \citet{brand1993}. 
For sources located within the Solar Circle, there are two possible solutions equally spaced on either side of the tangent distance; these are known as the \emph{near} and \emph{far} distances.
To resolve these kinematic distance ambiguities, we have used the \hi\ self-absorption method (\hisa; e.g. \citealt{jackson2003,roman2009, wienen2015}).
This works on the premise that the cold \hi\ associated with a source at the near distance will produce a dip in the warmer \hi\ emission,
which arises from warm gas located throughout the Galactic mid-plane, at the same velocity as the source
(see Fig.\,\ref{fig:example_HI_spectra_scimes} for example profiles). 

For this purpose, we made use of H{\sc{i}} maps from the Southern Galactic
Plane Survey (SGPS; \citealt{mcclure2005}) to determine the existence of \hisa\ towards all
clouds, directly from the 3-dimensional data cube.
If a cloud showed a strong H{\sc{i}} absorption dip within the cloud mask when compared to
the immediate surroundings (from the H{\sc{i}} cube), the cloud was considered to have
strong \hisa, and was placed at the near distance.
Otherwise, the \hisa\ criterion was deemed ambiguous, in which case we made use of the distance
tool being developed by the Hi-GAL/VIALACTEA project.
The VIALACTEA automatic distance tool is still in development but
will be publicly available in the course of 2017
(Russeil et al. in preparation).
This tool combines pre-existing distance information gathered from catalogues found in the
literature (e.g. maser parallax distance, spectrophotometric distance of \hii\ regions,
\hisa\ solution for the kinematic distance ambiguity, IRDC associations, etc) and new 3D
extinction data cubes produced following \citet{marshall2006}.
If the distances provided by the Hi-GAL distance tool were robust for clouds with ambiguous \hisa\
(e.g. due to the match with an IRDC, which are considered to be predominately located in the foreground
with respect to the Galactic Centre), the distance was taken as the Hi-GAL distance tool solution.
If the distance ambiguity was not solved with either Hi-GAL or \hisa, but there was a match with
an ATLASGAL source for which the distance ambiguity had been solved independently 
(see Sect.~\ref{sect:atlasgal_distances}), we took the ATLASGAL distance.
In cases where none of these three methods could solve the distance ambiguity, we chose the
far distance as the final distance of the cloud, because clouds located at the far distance
are less likely to give rise to observable signatures (e.g. \hisa\ or IRDC). 

The distances that we have determined for each cloud are listed in Table\,\ref{tab:full_catalogue}.
Ten clouds in our sample are located on the Solar circle (i.e. $\vert$\vlsr$\vert \, < 10$\kms),
so that no reliable kinematic distance can be determined; these have been excluded from
our analysis.
For the remaining clouds, the quoted uncertainty simply comes from the rotation model and
an assumed uncertainty of 7\, \kms\ on the velocity, representative of typical
departure from circular motions induced by spiral density waves or local peculiar motions. However, there is also an intrinsic
uncertainty associated with the method of using velocities as proxy for the distance,
as well as uncertainties that depend on the origin of the distance determination
(e.g.\ maser or stellar parallax distances, or simply an HiSA/extinction distance).
For simplicity, we assume that the overall distance uncertainty is of the order of 30\%.

From the 35 clouds that contain at least one ATLASGAL clump, 24 clouds have distances that
agree with all three methods (\hisa\ method, Hi-GAL tool and ATLASGAL solution),
ten clouds have one method disagreeing (five Hi-GAL far distances were revised to a near
distance, and five far ATLASGAL distances were brought to the near distances), and
one cloud with a far distance assigned by both Hi-GAL and ATLASGAL was brought to
the near distance by our \hisa\ method.
For the remaining 147 clouds in our catalogue, 131 have distances that agree between the
\hisa\ method and the Hi-GAL distance tool determinations (although this number includes
26 clouds for which neither method had solved the distance ambiguity), and 16 distances have
been revised from the Hi-GAL $far$ distances to a $near$ distance with the presence of strong \hisa.
A tag indicating how the distance ambiguity was solved is included in the final
catalogue (Table\,\ref{tab:full_catalogue}).

\subsection{Physical properties of the clouds}
\label{sec:gmc_physical_properties}

\begin{figure*}[ht]
\centering 
\includegraphics[width=\textwidth]{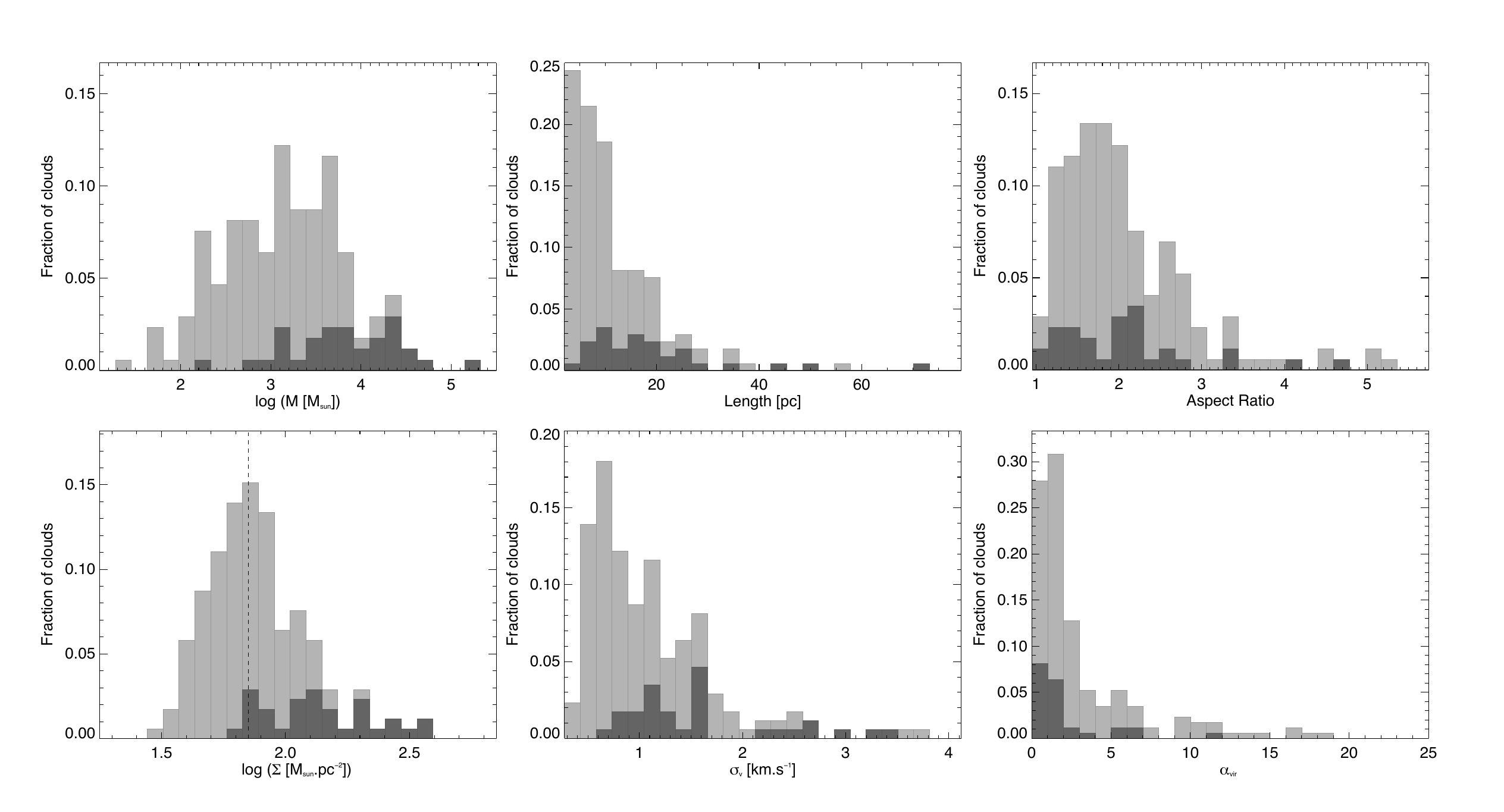}
\caption{Properties of the GMCs within the \testfield:
distributions of masses $M$ (top-left), maximal lengths (top-centre), aspect ratios (top-right),
average mass surface densities $\Sigma$ (bottom-left), velocity dispersions $\sigma_{v}$ (bottom-centre),
and virial parameters $\alpha_{\rm{vir}}$ (bottom-right).
The light-grey histograms show the distributions of properties for all clouds, excluding ten clouds 
with $\vert$\vlsr$\vert<10$\,\kms, as those have large distance uncertainties.
The dark-grey histograms show the sub-sample of clouds that have an ATLASGAL match.
The dashed vertical line on the average surface density plot shows our estimated
completeness limit, $\sim$70~\msun\,pc$^{-2}$, corresponding to
$\sim$3$\times10^{21}$\,cm$^{-2}$.}
\label{fig:hist_gmc_properties}
\end{figure*}

\begin{figure}[!ht]
\centering 
\includegraphics[width=0.45\textwidth]{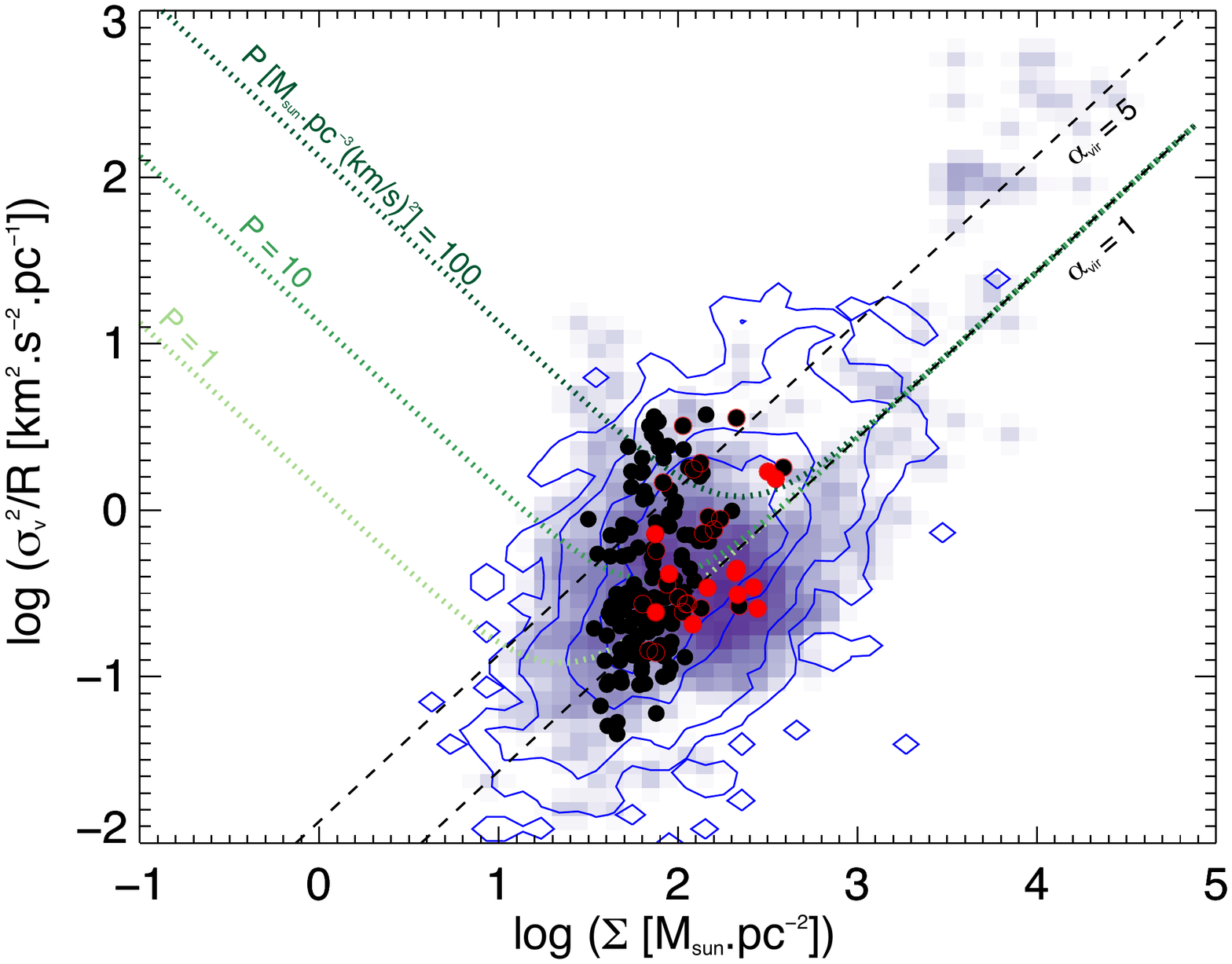}
\caption{Characteristic size-linewidth coefficient ($\sigma_v^{2}/R$) as a function of average gas surface density $\Sigma$ for all resolved GMCs extracted from the \testfield\
(black circles; those outlined in red highlight the clouds with an ATLASGAL counterpart,
and the filled red circles correspond to clouds showing signs of high-mass star formation
- see Sect.~\ref{sec:dense_gas} for details).
For comparison, the density of points in this plot from a compilation of extragalactic cloud samples
\citep[from][]{Rosolowsky2003,Ros07,Bolatto2008,Santangelo2010,Wong2011,Wei2012,Colombo2014}
and a compilation of Galactic cloud samples \citep[from][]{Heyer2009,Rathborne2009,wang2009,Ginsburg2012,Giannetti2013,Battersby2014,Walker2015}
are shown with blue contours, and with blue colour shading, respectively.
The green dotted curves show the expected force balance between kinetic, gravitational and external pressure,
for different values of external pressure, from $P = 1$ to $100$\,M$_{\odot}$\,pc$^{-3}$\,km$^{2}$\,s$^{-2}$ 
(which corresponds to $P/k \sim 5\times10^3 - 5\times10^5$ K\,cm$^{-3}$).
The black-dashed lines correspond to $\alpha_{\rm vir} = 1$ and $\alpha_{\rm vir} = 5$.}
\label{fig:heyer_plot}
\end{figure}

Having determined the distances to all the clouds, we estimate their physical properties,
which we list in Table\,\ref{tab:scimes_gmcs_properties} for our sub-sample (full catalogue
in Table\,\ref{tab:full_catalogue}).
In particular, Table\,\ref{tab:scimes_gmcs_properties} lists the exact area as covered
by the mask of each cloud, the equivalent radius (if taking a circular geometry), and the
projected length of the cloud, measured between the two most distant points within the cloud.

To estimate the cloud masses, we have converted the integrated intensities of \cco\ into
H$_{2}$ column densities using an $X_{\rm{^{13}CO(2-1)}}$ conversion factor, which we
have determined using an ancillary H$_{2}$ column density map derived from the Hi-GAL
survey data \citep{Molinari2010}.
This map was built by fitting a pixel-by-pixel grey body curve to the spectral energy
distribution from 160 to 500\,$\mu$m \citep{Elia2013}, assuming an opacity law  with a
fixed spectral index $\beta = 2,$ and $\kappa_{0} = 0.1$\,cm$^{2}$\,g$^{-1}$ at
$\nu_{0}$ = 1200\,GHz \citep{hildebrand1983}.
We then estimated the $X_{\rm{^{13}CO(2-1)}}$ factor for regions where there was an extracted cloud,
and obtained $X_{\rm{^{13}CO(2-1)}} \approx 1^{+1}_{-0.5} \times10^{21}$\,cm$^{-2}$\,(K\,\kms)$^{-1}$.
We note that this factor is in remarkable agreement with the one derived by solving the radiative
transfer equations for the J=2--1 and J=1--0 lines of $^{13}$CO simultaneously (see Sect.~\ref{sec:Xfactor}).
Interestingly, this is only a factor of 5 higher than the recommended value for the classical
$X_{\rm CO}$ \citep{Bolatto2013}, although the $^{12}$C/$^{13}$C isotopic ratio is typically of order 60; 
this is a direct consequence of the large difference in line opacity between $^{12}$CO and $^{13}$CO.

With this $X_{\rm{^{13}CO(2-1)}}$, and assuming a molecular weight $\mu_{mH}$ of 2.8, we derived
the cloud masses, the average gas surface density, $\Sigma$, across each cloud's area, and the
virial parameter, defined as \citep{bertoldi1992}:
\begin{equation}
\alpha_{\rm{vir}} = 5\sigma_{v}^{2}R / GM,
\end{equation}
where $G$ is the gravitational constant, $\sigma_{v}$ the measured velocity dispersion,
and $R$ is the equivalent radius.
Given the uncertainties on the distance estimates and on the X$_{\rm CO}$ factor, all
these quantities have an uncertainty of at least a factor two.

The distributions of these physical properties for our cloud sample are shown in
Fig.\,\ref{fig:hist_gmc_properties}, and their statistics are summarised in
Table\,\ref{tab:stats}.
From these, we can see that the mass distribution of the sample is relatively flat,
with masses spanning over four orders of magnitude, from a few tens up to 10$^{5}$\,\msun.
The other distributions are more strongly peaked, with aspect ratios $\la$2, and velocity dispersions around 1\,\kms.
The majority of the clouds have typical lengths of 10--20~pc, although the low-end distribution of lengths is limited by the resolution of the data.
Roughly $\sim$60\% of the clouds have $\alpha_{\rm vir} < 2$,  half of which with $\alpha_{\rm vir} < 1$.
Only a small fraction of clouds ($\sim$20\%) have $\alpha_{\rm vir} > 5$.
Considering a factor of two uncertainty on the mass estimates,
most of the clouds that we extract may well be self-gravitating structures.

\setlength{\tabcolsep}{6pt}
\begin{table}[t]
\caption{\label{tab:stats} Summary of the statistical properties of clouds within the \testfield.
Cols. 2 and 3 show the mean value of the property listed in Col. 1 and the associated standard deviation. Col. 4 shows the median value and the respective mean absolute deviation of the first and third quartiles.}
\begin{minipage}{0.5\linewidth}
\small
\renewcommand{\footnoterule}{}  
\begin{tabular}{l . . c }
\hline 
\hline
Property & \multicolumn{1}{c}{Mean} & \multicolumn{1}{c}{$\sigma$} & Median \\ 
\hline
$log [M$ (M$_{\odot}$)$]$	&	3.14 & 0.70 & $3.18 \pm 0.49 $ \\ 
Length (pc)					& 	11.8 & 10.1	&  $9.1 \pm 4.9$    \\ 
Aspect ratio				& 	2.1  & 0.80	&  $1.87 \pm 0.48$  \\ 
$\sigma_{v}$ (\kms)			&	1.12 & 0.64 &  $0.97 \pm 0.37$  \\
$log [\Sigma$ (M$_{\odot}$\,pc$^{-2}$)$]$ & 1.89 & 0.20 & $1.87 \pm 0.12$ \\ 
$\alpha_{\rm vir}$				&	3.1 & 3.6   &  $1.8 \pm 1.4$ \\ 
\hline
\end{tabular}
\end{minipage}
\end{table}

This can be better seen in Fig.\,\ref{fig:heyer_plot}, where we show the \citet{Heyer2009}
correlation between gas surface densities $\Sigma$ and the size-linewidth coefficient
(represented here as $\sigma_v^{2}/R$) for a compilation of Galactic and extragalactic GMCs,
as well as the clouds presented here.
On the top-left side of this plot, at low surface densities, clouds have large virial parameters,
and they are in a so-called pressure-confined regime. At higher gas surface densities, the gravity
becomes dominant and clouds cross over to a self-gravitating regime, with lower values of the 
virial parameter.
We can see that our sample of clouds from the SEDIGISM \testfield\ lies in a similar regime
as other Galactic and extragalactic clouds, mostly scattered around the $\alpha_{\rm vir} = 1$ line.

One should be aware, however, that the molecular clouds identified from the \cco\ emission
are likely tracing the high-density regions of larger molecular cloud complexes
\citep[e.g.][]{Langer2014,DC2016}.
In fact, within the \testfield, approximately 33\% of the total \cco\ emission above 3$\sigma$
was not assigned to any particular cloud, and is mostly part of a smoother, more diffuse
background connecting several clouds.
The extent of this background may be much larger than what we can detect here, not only because
some of the underlying molecular gas may be in fact CO-dark, but also because \cco\ cannot trace
the very low column densities.

To address this, we have investigated the emission towards the SEDIGISM \testfield\ with lower excitation energy transitions ($J$=1--0), and also higher abundance species ($^{12}$CO), to be more sensitive to the lower-column-density envelope of the clouds.
Using the $^{12}$CO(1\,--\,0) data of the \citet{dame2001} survey, \citet{garcia2014} found a total of only 11 GMCs within the \testfield, which comprise 172 of the \scimes\ clouds. However, the spatial resolution of the \citet{dame2001} survey is much coarser than that of the SEDIGISM survey (530$''$ compared to 30$''$).
Therefore, some of the GMCs identified by \citet{garcia2014} could be simply the result of the blending of several clouds that may not be physically connected.

We have also performed a comparison of our cloud catalogue with the clouds detected by the ThrUMMS survey
\citep{ref-thrumms} in the \testfield\ (with a much better resolution relative to the Dame survey although
still $\sim$2.5\,times lower than SEDIGISM), using the same cloud extraction algorithm as used here (\scimes).
We have done so using both the $^{12}$CO(1\,--\,0) and $^{13}$CO(1\,--\,0) data and found that the clouds
from \citet{garcia2014} are now sub-divided into smaller clouds, but they still tend to group
several SEDIGISM clouds together.
The grouping of numerous SEDIGISM clouds into larger GMCs as seen with $^{12}$CO or lower transitions
of $^{13}$CO seems to confirm the existence of a large scale diffuse molecular gas,
which connects the different peaks extracted from SEDIGISM.

\subsection{Galactic distribution of molecular clouds}
\label{sect:gal_distribution}


\begin{figure*}[t]
\centering 
\includegraphics[width=0.9\textwidth]{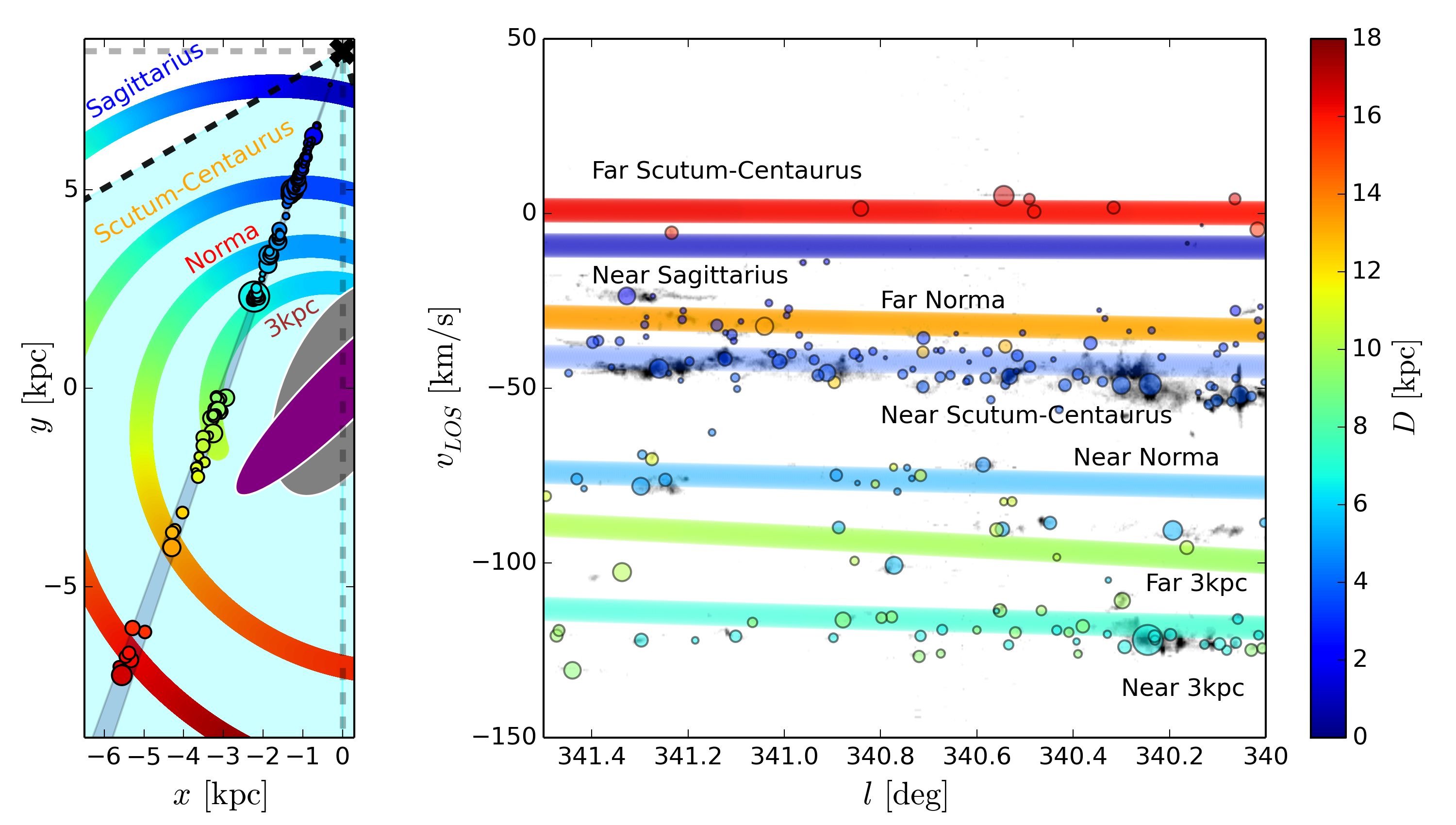}
\caption{Left: Top-down view of the Galaxy showing the SEDIGISM coverage (full survey in turquoise, and the \testfield\ as the small shaded line of sight). The different spiral arms from our model (see text for details) are shown and labelled with different colours. The positions of all the clouds in the \testfield\ are overplotted as coloured circles, whose colours indicate their assigned distances (see the colour bar to the right), and the size is proportional to the mass of the cloud.
Right: $lv$ plot of the peak intensity of $^{13}$CO in the \testfield\ (greyscale) overlaid with the positions of all the molecular clouds with assigned velocities (colours and sizes as in the left panel). The positions of the spiral arms are overplotted and labelled, also colour-coded with their distance.}
\label{fig:lv_diagram}
\end{figure*}

The positions of all the molecular clouds extracted from the \testfield\ are shown in
Fig.\,\ref{fig:lv_diagram}, overlaid on a customised model of the spiral arms.
If we were to take a typical 4-armed symmetric logarithmic spiral potential with a pitch angle of 12.5 degrees,
as found to be the best fit of the emission of the entire Galaxy by \citet{Pettitt2014}, based on fitting
hydrodynamical models to the low-resolution maps from \citet{dame2001},
the emission in the \testfield\ would not be fitted particularly well,
especially for the near Norma and near Scutum-Centaurus arms.
There are a number of possible reasons for this:
{\it i)} A global fit of log-normal spirals is unlikely to fit small fields accurately;
{\it ii)} The spiral arms shown are based on the stellar potential, and gas response is
much more complicated than simply tracing the bottom of the potential well;
and {\it iii)} These spiral arms are based on a simple circular rotation model, whereas observations
(and numerical simulations) suggest a more complex velocity field with many smaller scale undulations
and non-circular velocities.  

To improve the agreement with the data, we have altered the symmetric 4-armed model by slightly rotating
(by 15 degrees) the azimuthal locations of the Norma and Scutum-Centaurus arms; the resulting model is
shown in Fig.\,\ref{fig:lv_diagram}.
It is no surprise that a 4-fold symmetric, constant pitch angle spiral cannot reproduce all the features
simultaneously, as there are several examples in the literature of the Milky Way spiral arms deviating
from such an idealised form \citep{TC1993,russeil2003,Levine2006}.
Given the small longitude range covered by the \testfield, the loci of such modified spiral
arms are to be taken as illustrative only, showing the potential for the complete SEDIGISM
survey at high-resolution to provide strong constraints of the positions of the Galactic
arms towards the inner Galaxy.

Overall, the assignment of distances to our sample of clouds is broadly consistent with
the expected positions of spiral arms, with the exception of the far Norma arm, for which
we do not seem to associate many clouds, and the far Sagittarius arm (not shown in
Fig.\,\ref{fig:lv_diagram}), for which we do not find any cloud;
the run of this arm is only tentatively known at this location anyway.
The small number of clouds associated with the far Norma arm may be due to its close proximity
to the more prominent near Scutum-Centaurus arm in $lv$ space, and hence the preferred association
of clouds with the near distances; or simply because the Norma arm is relatively faint, even at the near distances, making it hardly detectable at the far side.
Interestingly the \testfield\ is relatively close to the bar end, and the 3-kpc Arm tangency
(see Fig.\,\ref{fig:coverage}); however it is difficult to say with certainty whether such
features are seen due to the small longitude extent.

\subsection{Turbulence within GMCs}
\label{sect:turbulence}

Here we show preliminary results of a statistical study of turbulence in the SEDIGISM data.
To describe the turbulence, we have applied a Velocity Channel Analysis (VCA) technique,
as described in \cite{ref-VCA}, on different sections of the \cco\ $\ell bv$
data cube. This technique consists of computing the spatial power spectra of the
two dimensional brightness distribution ($I_{2D}$) in velocity slices, and letting
the thickness $\Delta V$ of the slices vary. 
As the thickness of velocity slices increases, density fluctuations begin to dominate
the emissivity over velocity fluctuations.
It is expected that the power spectrum keeps steepening with increasing thickness $\Delta V$
up to a characteristic thickness, above which there is no significant change in the index.
When integrating over larger $\Delta V$, most of the velocity fluctuations average out,
so that the power spectrum traces only static density fluctuations.

\begin{figure*}[!ht]
\centering 
\includegraphics[width=0.95\textwidth]{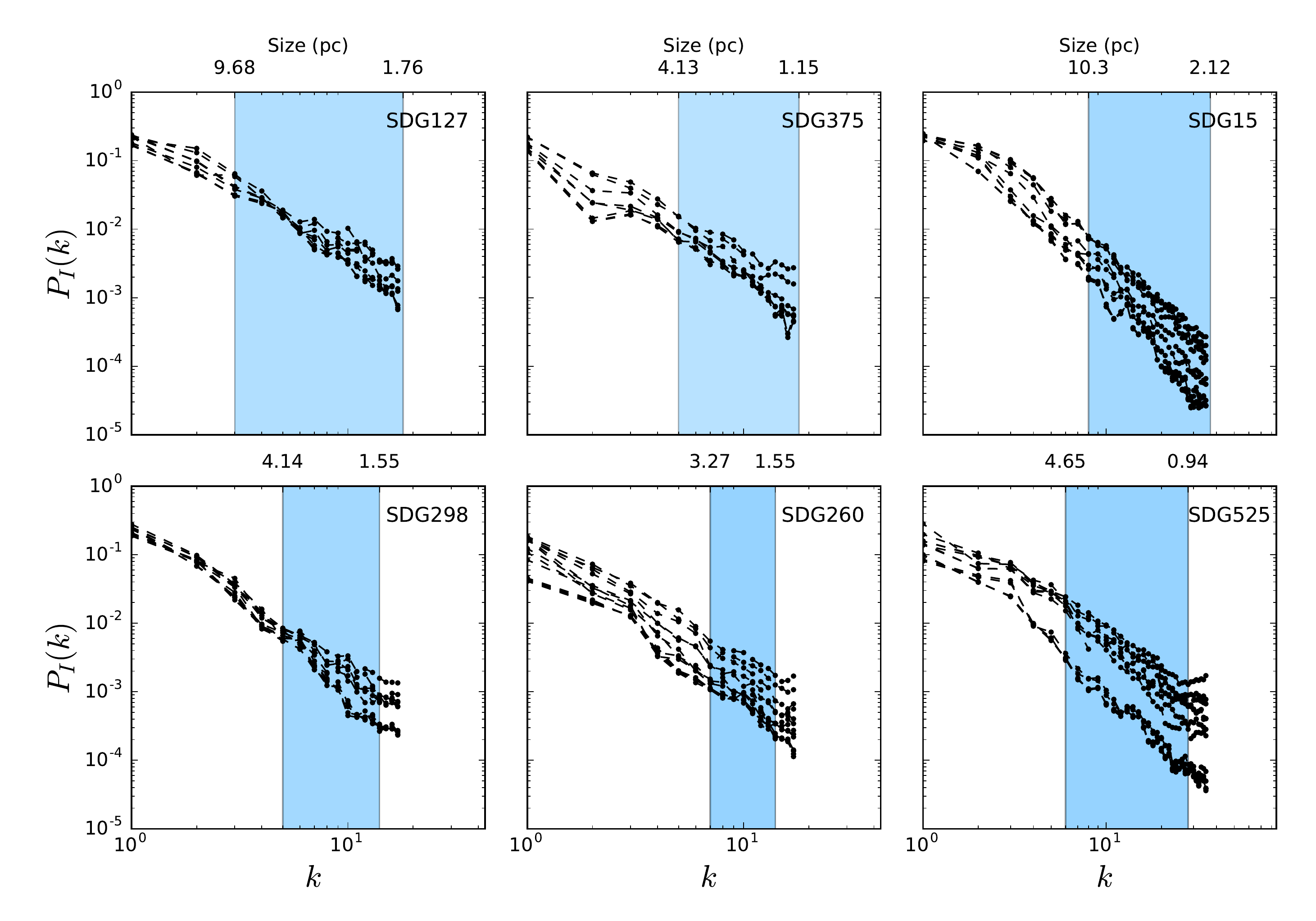}
\vspace{-3mm}
\caption{Normalised spatial power spectra for the six molecular clouds extracted
with \scimes\ with the highest numbers of leaves, for a range of velocity
thickness $\Delta V$: in each panel, the different curves correspond to increasing
thickness from top to bottom, with $\Delta V$ values in the range 0.5 to 18~\kms.
The blue area indicates the range of spatial scales over which a least square
fitting of the spatial power index was performed.
The scale on the lower X-axis gives the wave numbers ($k$), while that on the upper X-axis
indicates the corresponding scale in pc, using the distance to each GMC, as listed in Table\,\ref{tab:scimes_gmcs}.
Each panel is labelled with the ID of each GMC (as per Col. 1 in Table\,\ref{tab:scimes_gmcs}).}
\label{fig:turb1}
\end{figure*}

According to turbulence theory, the power spectrum is related
with the scale as $P_{I_{ND}}(k)\,\propto\, k^{\kappa_{ND}}$, where
$ND$ denotes the number of spatial dimensions, and $k$ is the wave number. Values
of $k$ go from 1, the whole length of the longer axis in the data,
to $Npix/2$, i.e. half the total number of pixels.
For incompressible, homogeneous, and isotropic turbulence, $\kappa_{3D} = -11/3$,
and $\kappa_{2D} = -8/3$ \citep{ref-kolmogorov}. In the limit of
shock-dominated, compressible turbulence, the spatial power index is $\kappa_{3D} = -4$,
$\kappa_{2D}=-3$ \citep{ref-burgers}. However, \cite{ref-VCA}
predict that the spectral index should saturate to $\kappa_{2D}=-3$
for an optically thick medium, and many observations support their predictions
(cf. \citealt{ref-burkhart}, who also provide a numerical confirmation for this value of -3).

We have applied the VCA method on the six molecular structures extracted with \scimes\
with the highest numbers of leaves (Sect.\,\ref{sect:gmcs} and Table\,\ref{tab:scimes_gmcs}),
corresponding to some of the most massive and highly sub-structured complexes.
The masks generated by \scimes\ were used to isolate the \cco\
brightness distribution of each GMC from the data: new cubes were
generated to cover each GMC, and the voxels outside of the masks were set to zero;
the size of each cube along each axis was adjusted to the nearest power of two. 
For our analysis, we let the thickness of the velocity slices vary from twice the velocity resolution
(0.5\kms) up to the thickest slice case, corresponding to integrated maps that include
the whole velocity range of the cloud in one channel.

We computed the power spectra as azimuthal median values of the 2D FFT of the data.
To properly consider the measured noise, we also calculated the power spectra for
emission-free channels. We corrected for the effects of measured noise and beam smearing
following the method described in \citet{BruntMacLow}.
The normalised power spectra for our sample of six  GMCs are shown in Fig.~\ref{fig:turb1}.
We only show the results for angular scales larger than two times the spatial resolution
because at smaller scales the effects of the beam size start to dominate.
Fig.~\ref{fig:turb1} also shows the range over which the least square fitting of the
spectral index was performed. Following \citet{ref-medina},
the power spectrum fits are made between a minimum scale corresponding
to 2.5 times the instrumental resolution, and a maximum scale corresponding
to the semi-major axis of each cloud.
Given the distances to the various GMCs, the range of scales over which we computed
the spectral indices corresponds to linear scales between 1 and 10 pc.
Therefore, this range is well suited to probe turbulence from the scale of
a complete cloud, where it may be externally driven (e.g.\ by supernovea),
down to scales where internal sources may contribute to turbulence (e.g.\ \hii\ regions,
stellar winds, proto-stellar outflows; \citealt{ElmegreenScalo,Dobbs2014}).

\begin{figure*}[ht]
\centering 
\includegraphics[width=0.9\textwidth]{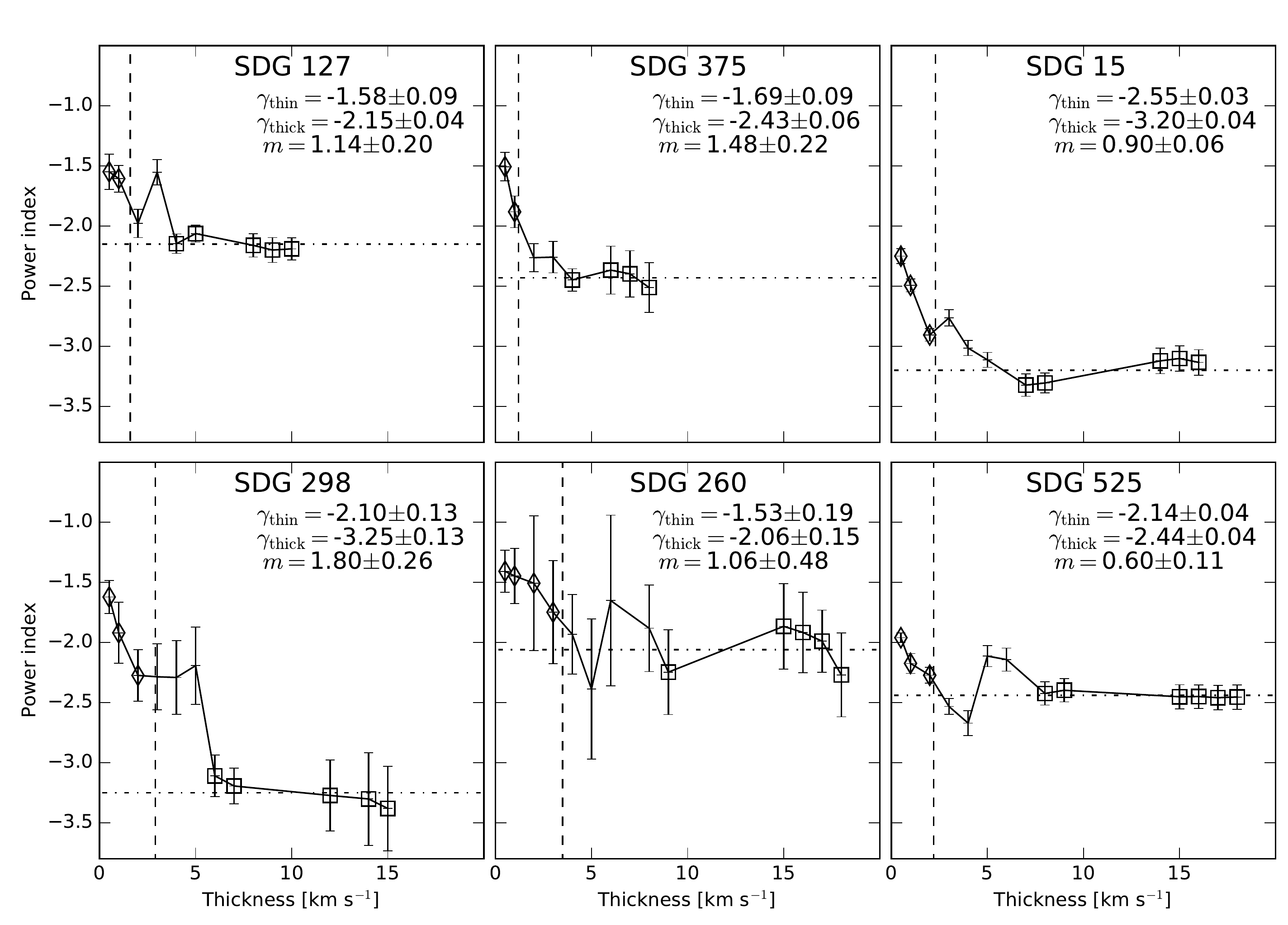}
\caption{Variations in the power spectrum indices with velocity thickness
for the different GMCs. The error bars correspond to the $1\sigma$ statistical
uncertainties on the fit. The dotted horizontal lines indicate the
saturation index for each GMC, corresponding to the thick regime where
density fluctuations dominate. The dashed vertical lines show the
respective velocity dispersion, $\sigma_V$, for each GMC.
The data points used to compute the thin and thick indices ($\gamma_{\rm thin}$
and $\gamma_{\rm thick}$), are indicated with diamonds and square symbols, respectively.
These values, as well as the index of the second order structure function
($m$), are indicated in each panel (see text for details).
}
\label{fig:turb2}
\end{figure*}

In Fig.~\ref{fig:turb2}, we show the variations of the spatial power
indices $\kappa_{2D}$ as a function of velocity slice thickness, $\Delta V$.
The spectra become steeper with increasing $\Delta V$, as predicted.
Following \citet{ref-VCA} (see also \citealt{ref-esquivel}), we computed the
spectral index for thin slices ($\gamma_{\rm thin}$) as the mean value of all indices
corresponding to a thickness $\Delta V < \sigma_V$. The thick index ($\gamma_{\rm thick}$)
is taken as the average of the indices where $\Delta V > \sigma_V$, 
but restricted to the regime where this index is almost constant.
Finally, the index of the second order structure function is computed as
$m \, = \, 2(\gamma_{\rm thin}-\gamma_{\rm thick})$ in the shallow cases (i.e.
where $\gamma_{\rm thick} > -3$), or $m \, = \, 2(\gamma_{\rm thin}+3)$ in
the steep cases (SDG~15 and SDG~298). We find values in the range 0.6--1.8,
in rough agreement with the index of the first order structure function
derived from principal component analysis ($\gamma_{\rm PCA}$, where $m = 2 \times \gamma_{\rm PCA}$)
and published in other studies (e.g. $\gamma_{\rm PCA} \approx \,$ 0.4--0.5,
\citealt{brunt-heyer2002b}, \citealt{RomanDuval2011}).\footnote{These studies
both derived a mean PCA index $\alpha_{\rm PCA}$ of 0.62. By rescaling this index
to compute the index of the true first order structure function
\citep[e.g.][]{brunt-heyer2002a,brunt-heyer2013,Brunt+2003},
we get values in the range 0.4--0.5.}
These results are based on a very limited sample of six GMCs, and should therefore
be regarded with caution.  We plan to perform a more systematic analysis on a
sample extracted from the full SEDIGISM survey in a forthcoming paper.

Interestingly, all six GMCs have similar characteristic scales of turbulence,
corresponding to velocity thickness typically between 4 and 8\kms,
above which velocity fluctuations average out and the spectral indices do not vary much.
This could indicate that similar processes are responsible for the turbulence
in all clouds in our (small) sample.
However, we do find significant variations in the spatial spectral indices between GMCs:
two of them (SDG~15 and SDG~298) have steep spectra, with indices that saturate around -3.5,
consistent with the picture of the Kolmogorov energy cascade, and with results reported elsewhere
(e.g.\ \citealt{Dickey2001,Muller2004}; see also \citealt{ElmegreenScalo}).
The other clouds show spectra that are significantly shallower, with 
$\kappa_{2D}$ between $-2.0$ and $-2.5$ for the thick case.
According to \cite{ref-VCA}, this can be 
explained by a significant contribution to the turbulence on small scales.

However, we cannot draw robust conclusions from these preliminary results.
For example, \citet{ref-burkhart}
have shown that the spectral indices also depend on the line excitations. In particular,
for a very optically thin, supersonic CO gas, with low density or low abundance,
the spectral index is shallower than the expectations for its column
density. Our results are consistent with this picture.
A more detailed analysis based on a Galaxy-wide sample of clouds within the full SEDIGISM survey,
capable of probing different environments in the Galaxy, will be presented in forthcoming papers.
In particular, it will be very interesting to look for possible differences between the Galactic
arms, and between the inter- and intra-arm regions.


\section{Dense gas and high-mass star formation} 
\label{sect:atlasgal}

In order to study the distribution of dense gas within the molecular clouds identified
in Sect.\,\ref{sect:gmcs}, and to identify potential sites of high-mass star formation,
we have investigated the distribution of compact ATLASGAL sources in the \testfield.
ATLASGAL has surveyed the inner \GP\ observing the dust continuum emission at 870\,$\mu$m
(\citealt{schuller2009}), with a peak flux 1-$\sigma$ sensitivity of $\sim$60\,mJy\,beam$^{-1}$,
which corresponds to a column density of $N$(H$_2$) = $1.5\times 10^{21}$\,cm$^{-2}$
(assuming a dust temperature of 20\,K and absorption coefficient $\kappa_\nu$ $=$ 1.85\,cm$^2$\,g$^{-1}$).
ATLASGAL is, therefore, an excellent tracer of the high-density gas within our sample of
molecular clouds, potentially pinpointing where high-mass stars are likely to form.

\subsection{Distribution of ATLASGAL clumps}
\label{sect:atlasgal_distances}

\begin{figure}
\centering 
\includegraphics[width=0.45\textwidth]{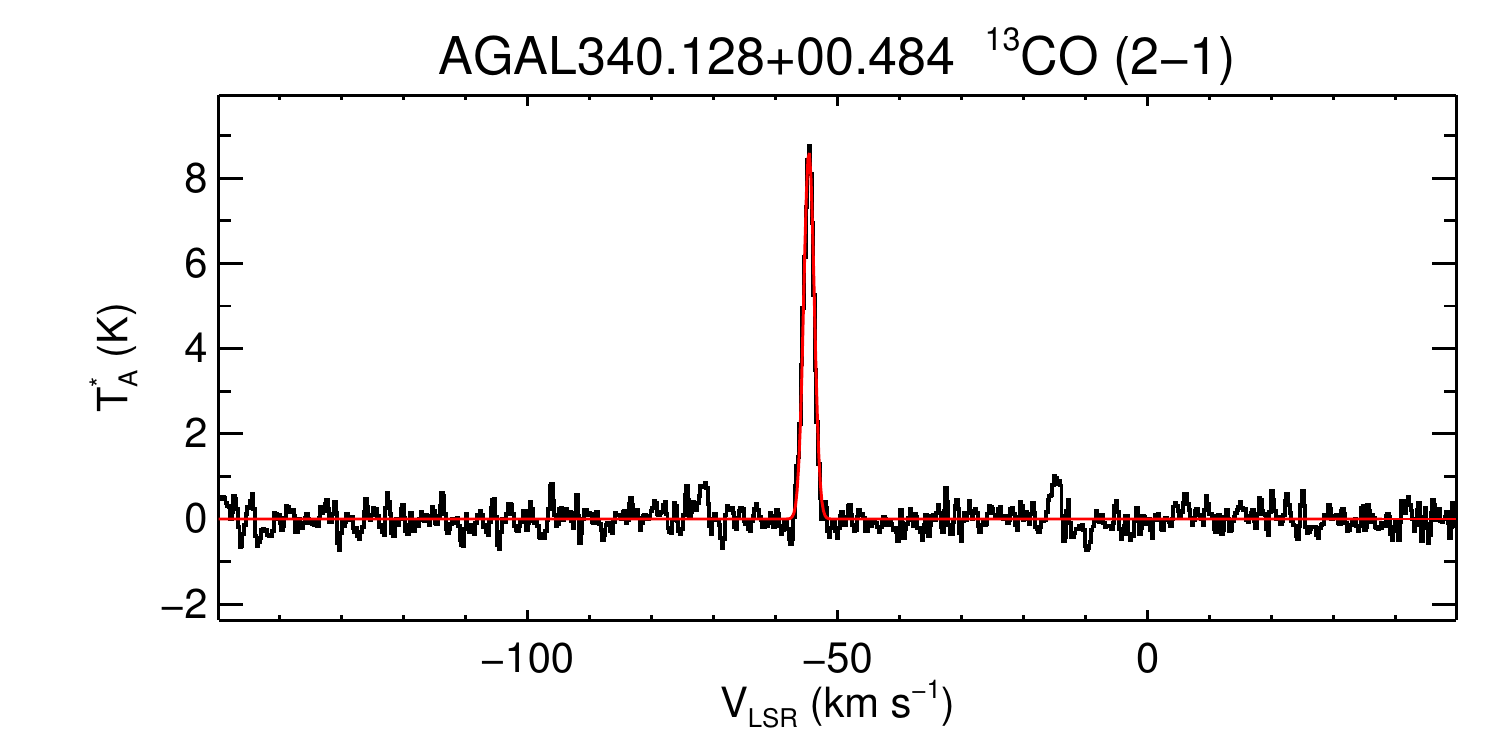}
\includegraphics[width=0.45\textwidth]{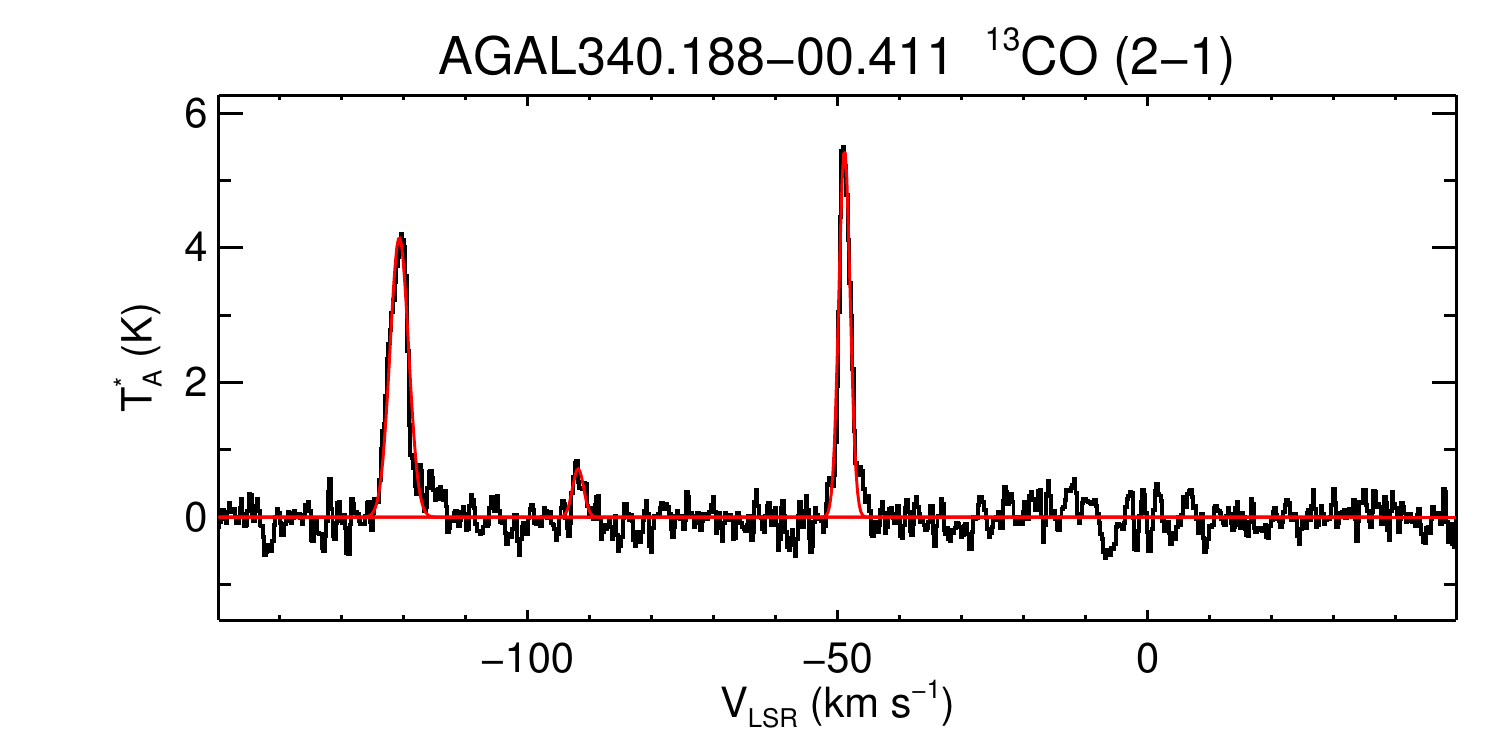}
\caption{Example $^{13}$CO spectra extracted towards two dense clumps identified from the ATLASGAL survey (black), one with a single velocity component, and another with multiple components along the line of sight. The results of the automatic Gaussian fitting are overlaid in red.}\label{fig:example_spectra}
\end{figure}

Within the \testfield, there are 140 ATLASGAL clumps from the Compact Source Catalogue
(CSC; \citealt{contreras2013,urquhart2014c}).
We have made use of the SEDIGISM data to estimate \vlsr\ for all the clumps,
which is essential to place them within their Galactic context, and to
associate them with the molecular clouds extracted in Sect.\,\ref{sect:gmcs}.
We did so by fitting the \cco\ spectra towards the peak of the \submm\ emission for all ATLASGAL clumps, using an iterative fitting programme that fits a Gaussian profile to the strongest emission feature, removes the fit from the spectrum, and repeats this process until there is no more emission above three times the \rms\ noise, measured from emission-free channels  (see examples presented in Fig.\,\ref{fig:example_spectra}).

\begin{figure}[ht]
\centering 
\includegraphics[width=\columnwidth]{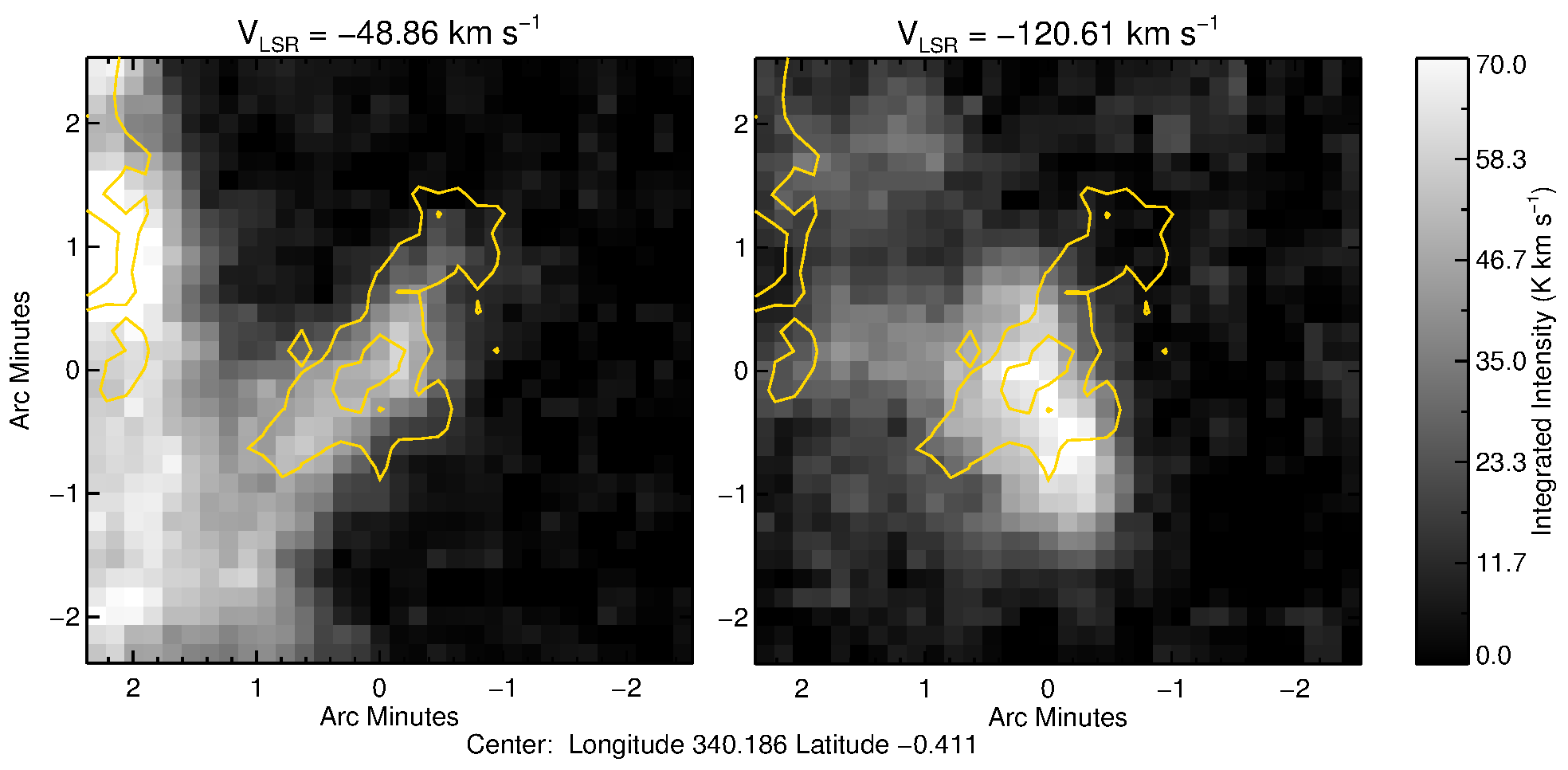}
\caption{Integrated $^{13}$CO maps of the two components seen towards AGAL340.188-00.411 (see lower panel of Fig.\,\ref{fig:example_spectra}). 
The emission has been integrated over a velocity range of twice the FWHM line width of each component.
The yellow contours show the distribution of the 870\,\mum\ emission mapped by ATLASGAL.
In this case, there is a better morphological correlation between the dust emission
and the CO emission found at $-48.9$\,\kms\ than with the CO emission at $-120.6$\,\kms.
Therefore, we assigned a velocity of $-48.9$\,\kms\ to this clump.
}
\label{fig:integrated_co_maps}
\end{figure}

Although we find multiple components towards 90\% of the sources, in the majority of cases,
the integrated intensity of the strongest component is at least twice that of the others
and is therefore considered to be the most likely to be associated with the clump,
as observed in other studies (e.g. \citealt{urquhart_13co_south}).
However, for 35 clumps, the multiple components have similar intensities;
for these we have either searched the literature for a velocity determined using other
high-density tracers, such as NH$_3$(1,1) or N$_2$H$^+$(1--0)
\citep{jackson2013,urquhart2014,wienen2015}, or compared the integrated \cco\ maps of
the different velocity components with the ATLASGAL dust emission maps,
choosing the velocity component that peaks at the position of the dust emission and
where the best correlation between spatial distribution of gas and dust is found
(see Fig.\,\ref{fig:integrated_co_maps} for an example of this method).
Using this combination of steps we are able to assign a velocity to 139 clumps in the \testfield.
For one source (AGAL340.096$-$00.022), we were unable to identify which velocity component was best representative of the clump.

We have then used the \citet{brand1993} Galactic rotation model to determine kinematic distances to each of the clumps again solving the kinematic distance ambiguities using the \hisa\ technique (as described in Sect.\,\ref{sec:distances}), and taking account of their association with IRDCs, and literature information.
Three sources are found to be located on the Solar circle (i.e. $\vert$\vlsr$\vert \, < 10$\kms) and so no reliable distance estimate is possible.
In total, we have resolved the ambiguities for 97 out of 140 clumps. We have compared our results with those of \citet{wienen2015} who performed a similar analysis for a sample of 51 clumps located in the \testfield\ and find an overall agreement of $\sim$90\%.

Since the ATLASGAL clumps trace the higher density peaks within clouds, it is likely that
small groups of ATLASGAL clumps are part of the same larger GMC complex.
Identifying these groups can help us to assign distances to clumps for which we have
not been able to resolve the distance ambiguity, and to derive more reliable distances by
using the systemic velocity of the cloud rather than a number of velocity measurements
that may vary over the cloud (cf.\ \citealt{russeil2003}).
We have therefore used a friends-of-friends analysis to identify groups of ATLASGAL
clumps that are coherent in $\ell b v$ space, allowing a maximum angular offset between
adjacent points of 8\,arcmins and velocity difference of 8\,\kms.\footnote{Here we are
following the method outlined by \citet{wienen2015}, but, with a more complete set of velocities
obtained from the SEDIGISM data, we are able to impose tighter constraints on the position
and velocity offsets.}
This revealed ten coherent groups of sources accounting for 123 of the clumps found in the \testfield\ (see Table\,\ref{tab:csc_groups}); this includes 34 clumps that we were unable to resolve the kinematic distance ambiguity for.

\setlength{\tabcolsep}{1pt}
\begin{table}
\begin{center}\caption{Summary of groups identified in the ATLASGAL CSC. Col.\,1 is the
group name constructed from the mean positions of the associated clumps.
Col.\,2 is the number of associated clumps. Cols.\,3-4 are the mean velocity of the group,
and the respective standard deviation (of inter-clump velocities).
Col.\,5 is the adopted distance to the group, i.e. the kinematic distance using the
systemic velocity of the group and the respective distance solution.
Col.\,6 is the ID number of the matched molecular clouds identified in Sect.\,\ref{sect:gmcs}
(ID number as listed in Tables\,\ref{tab:scimes_gmcs} and \ref{tab:full_catalogue})
and the number of leaves associated with them (in parentheses).
Col.\,7 lists the matches to the $^{12}$CO(1-0) GMC catalogue presented by \citet{garcia2014}.}
\label{tab:csc_groups}
\begin{minipage}{\linewidth}
\small
\begin{tabular}{l;...cc}
\hline \hline
\multicolumn{1}{c}{Group}&  \multirow{2}{*}{N$_{\rm{cl}}$} & \multicolumn{1}{c}{\vlsr} &	\multicolumn{1}{c}{$\Delta$\vlsr} &\multicolumn{1}{c}{$d$}  &\multicolumn{1}{c}{SDG } &\multicolumn{1}{c}{$^{12}$CO}\\

\multicolumn{1}{c}{name}&    & \multicolumn{1}{l}{(\kms)} & \multicolumn{1}{l}{(\kms)}  &\multicolumn{1}{c}{(kpc)} &\multicolumn{1}{c}{clouds}&\multicolumn{1}{c}{GMCs} \\

\hline
G340.249$-$00.266	&	53	&	-49.1	&	3.36	&	3.8	&	Multi\footnote{SEDIGISM clouds: 234 (5), 260 (9), 273 (2), 298 (7), 323 (2), 324 (1), 392 (2)} & NCEN25\\
G340.256$-$00.059	&	16	&	-122.5	&	2.22	&	6.7	&	Multi\footnote{SEDIGISM clouds: 15 (14), 71 (1)} & 3KPC3\\
G340.529$-$00.147	&	3	&	-46.3	&	0.43	&	3.7 &	331	(4) & NCEN25\\
G340.556$-$00.402	&	2	&	-90.1	&	0.46	&	5.6	& 	147 (4) & NORM7.2\\
G340.749$-$00.182 	&	6	&	-36.2	&	2.89	&	3.1	&	430 (2) & NCEN25\\
G341.022$-$00.152	&	2	&	-15.2	&	0.54	&	1.5	&	593 (1) & -- \\
G341.034$-$00.053	&	3	&	-38.7	&	3.74	&	3.3	& 	408 (6)  & NCEN25\\
G341.117$-$00.293	&	29	&	-43.0	&	2.78	&	3.5	& 	Multi\footnote{SEDIGISM clouds: 354 (3), 375 (8), 420 (3), 466 (1)}  & NCEN25\\
G341.295+00.336	    &	4	&	-78.3	&	0.77	&	5.2	& 	164 (3)  & NORM8\\
G341.310+00.209	    &	6	&	-25.5	&	2.76	&	2.4	&   525 (13) & NCEN26.5\\
\hline
\end{tabular}\\
\end{minipage}
\end{center}
\end{table}
\setlength{\tabcolsep}{6pt}

Given that the \hisa\ method has a reliability of $\sim$80\% (\citealt{busfield2006,anderson2009a})
we would expect some disagreement in the kinematic distance solutions for the ATLASGAL clumps
within these larger groups.
Indeed, we find this to be the case for the two largest groups, G340.249$-$00.266 and G341.117$-$00.293,
with 53 and 29 clumps respectively, where 21\% and 14\% of clumps (respectively) had been assigned a
far distance, whilst the others had been assigned a near distance.
Since these are within the expected fraction of unreliable \hisa\ solutions, we have assigned the near
distance to both groups and have therefore revised the distances of the fraction of clumps in disagreement.

\begin{figure*}[tbp]
\centering 
\includegraphics[width=0.49\textwidth]{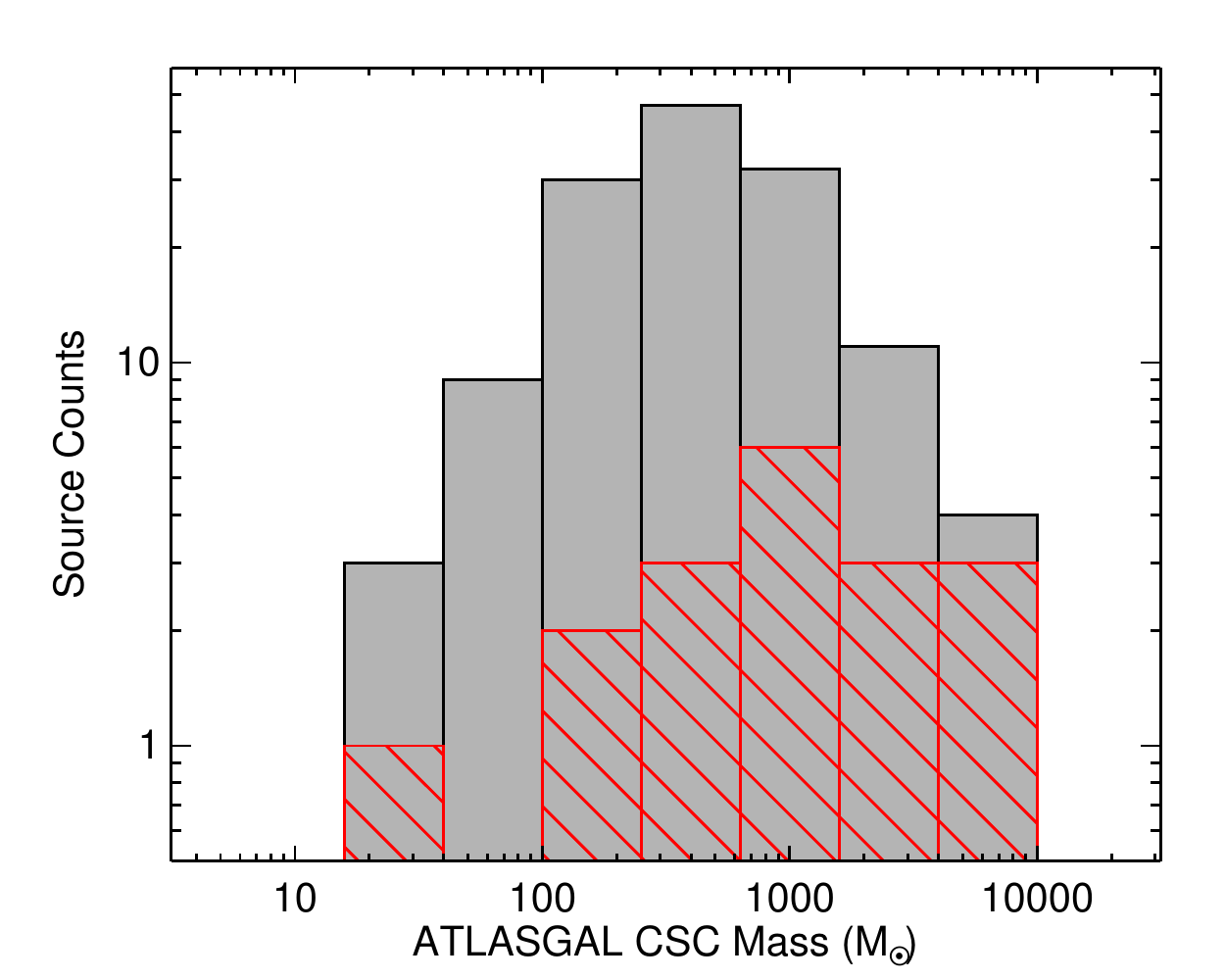}
\includegraphics[width=0.47\textwidth]{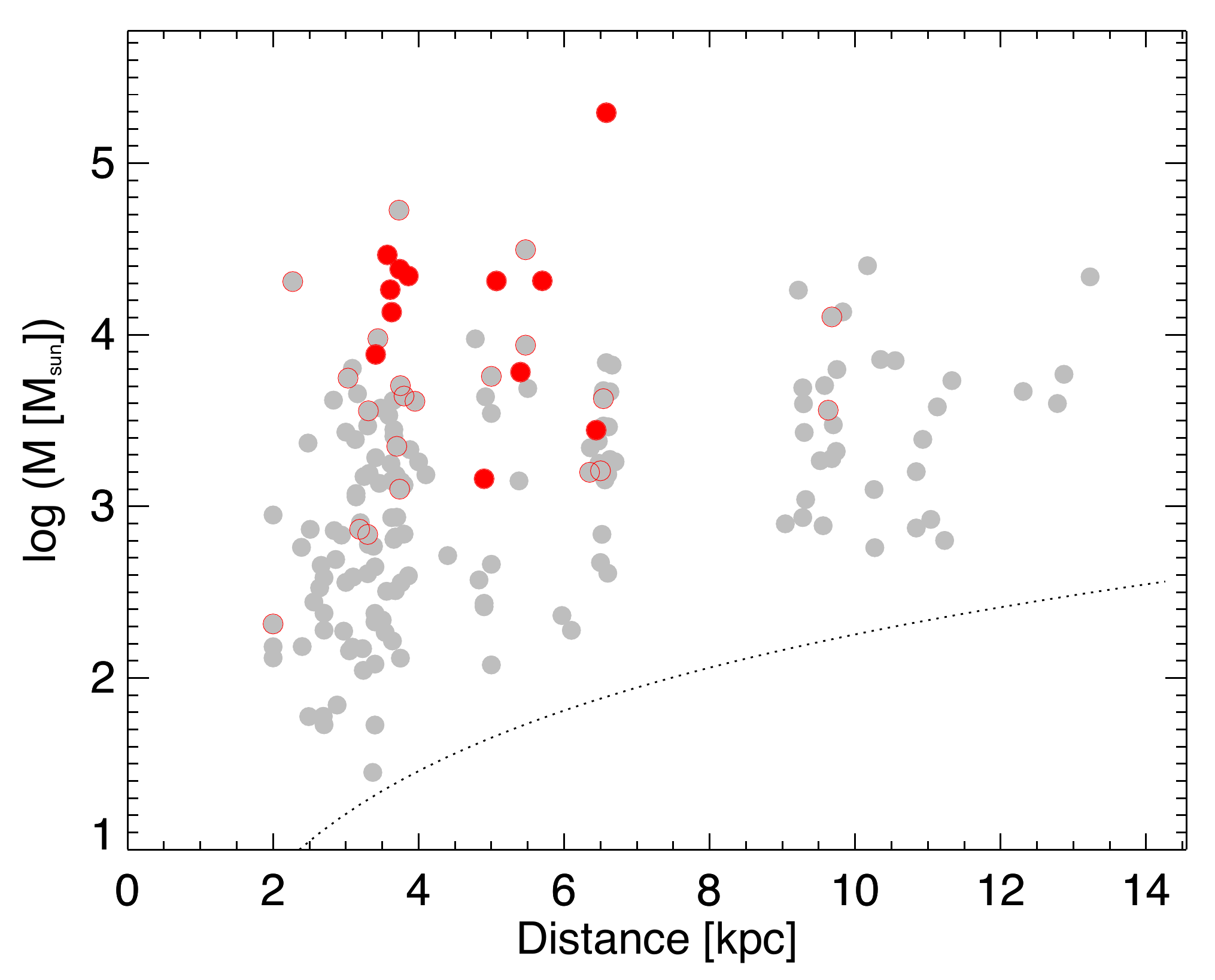}
\caption{{\em Left}: Mass distribution of all ATLASGAL clumps located in the \testfield\ with assigned distances.
The mass distribution of clumps associated with a high-mass star formation tracer is shown in red hatching.
The bin size is 0.4\,dex.
{\em Right}: Mass-distance distribution of all SEDIGISM clouds.
Clouds associated with ATLASGAL and with an HMSF tracer are shown in red-filled circles,
while the non-HMSF clouds are shown as grey circles (out of which those with an ATLASGAL match are red-outlined).
The dotted black line indicates the \cco\ mass sensitivity limit for an unresolved source (i.e. 28\arcsec\ radius).
}
\label{fig:mass_distribution_histogram}
\end{figure*}

\citet{garcia2014} have also determined distances to six of their lower-resolution $^{12}$CO GMCs within the \testfield; these are associated with 131 ATLASGAL clumps.
Of these, 126 clumps had assigned distances, and we find that 113 (i.e. $\sim$90\%) are in
agreement with those assigned by \citet{garcia2014}.
We had placed all nine remaining clumps at the far distance, consistent with the fact that
all of these are relatively isolated. 
Given that the five clumps for which we had not solved the distance ambiguity are positionally
correlated with GMCs identified by \citet{garcia2014} we have adopted their distance solution
for these sources.

Finally, after cross-matching with the SEDIGISM clouds (as described in Sect.~\ref{sec:atlasgal_gmc_match}), we further revised the distances of
seven ATLASGAL clumps, within six molecular clouds, to their $near$ distance solutions (as also mentioned in Sect.~\ref{sec:distances}).
By following these steps we have been able to determine a distance to 136 of the clumps located
within the \testfield. These adopted distances are listed in Table\,\ref{tbl:clump_properties},
along with the physical properties of the clumps.

\subsection{Correlation between ATLASGAL groups and GMCs}
\label{sec:atlasgal_gmc_match}

As discussed in Sect.\,\ref{sect:atlasgal_distances} we have found ten distinct groups of ATLASGAL clumps in the \testfield.
Comparing these groups with the catalogue of SEDIGISM molecular clouds, we find that the ten ATLASGAL groups
correspond to 20 SEDIGISM GMCs, although not all the ATLASGAL clumps in those groups fall within a SEDIGISM cloud.
The ID numbers of the matching GMCs are given in Table\,\ref{tab:csc_groups}.
We also find matches between 15 isolated ATLASGAL clumps and individual molecular clouds,
bringing the total number of SEDIGISM clouds with one or more associated ATLASGAL counterparts
to 35 (although this includes two clouds within the solar circle, i.e.
with $\vert$\vlsr$\vert \, < 10$\kms).
The total number of ATLASGAL clumps associated with SEDIGISM GMCs is 129.
The exact values of the distances as per the ATLASGAL clump catalogue and the final SEDIGISM
cloud catalogue differ typically by less than $\sim$10$\%$ simply because of the variation
of exact velocities used for the kinematical distance determination.
While this is well within the overall distance uncertainty, for the remainder of the
paper we will adopt the distances of the corresponding SEDIGISM clouds for these clumps,
for consistency.

In total, there are 11 ATLASGAL clumps not associated with a GMC.
One of these does not have a match because we could not assign a velocity.
The remaining ten ATLASGAL clumps that do not fall within any of the SEDIGISM clouds
are either small (and thus have not passed the criteria of the minimum size required
to be part of our cloud catalogue),
or, and most often, are in regions where the contrast is too low with respect to their
local background (i.e. not above the 4-$\sigma$ requirement to be considered as
independent peaks/leaves within the dendrogram); these regions, therefore, form part
of a smoother background that did not get assigned to any cloud.

The correlation of multiple SEDIGISM GMCs with a single ATLASGAL group, and the association
of a few of the ATLASGAL clumps to a background of more diffuse gas that connects the different
GMCs together, tends to confirm our suggestion (Sect.\,\ref{sec:gmc_physical_properties})
that the GMCs identified by \scimes\ from the $^{13}$CO data are tracing the high-density
regions of larger molecular cloud complexes.

\subsection{Dense gas within molecular clouds}
\label{sec:dense_gas}

\subsubsection{Mass distribution}

We estimate the isothermal masses of the ATLASGAL clumps using the \citet{hildebrand1983} method assuming that the total clump mass is proportional to the integrated flux density measured over the source:

\begin{equation}
\label{eqn:mass}
M_{\rm{clump}} \, = \, \frac{d^2 \, S_\nu \, R}{B_\nu(T_{\rm{dust}}) \, \kappa_\nu},
\end{equation}

\noindent where $S_\nu$ is the integrated 870~\mic\ flux density taken from the ATLASGAL CSC, $d$ is the distance to the source, $R$ is the gas-to-dust mass ratio, which we assume to be 100, $B_\nu$ is the Planck function for a dust temperature $T_{\rm{dust}}$, and $\kappa_\nu$ is the dust absorption coefficient taken as 1.85\,cm$^2$\,g$^{-1}$ (\citealt{schuller2009} and references therein). We use a dust temperature of 20\,K, consistent with previous studies in the literature (e.g. \citealt{motte2007, hill2005}).

In the left panel of Fig.\,\ref{fig:mass_distribution_histogram} we show the clump mass distribution for all 136 clumps with distances.
We also show (in red hatching) the mass distribution for the clumps associated with a high-mass star formation (HMSF) tracer from previous studies, such as a methanol maser, a massive young stellar object or a compact \hii\ region (\citealt{urquhart2013a,urquhart2014b}).
Only 18 clumps have been associated with HMSF, and these are mostly associated with methanol masers (14 in total, nine of which have only methanol maser associations, indicating a relatively early stage of their evolution).
Three clumps are solely associated with compact \hii\ regions (more evolved), four clumps are associated with two different tracers, and two clumps have all three tracers, hosting multiple evolutionary stages.
In the right panel of Fig.\,\ref{fig:mass_distribution_histogram} we show the SEDIGISM cloud mass distribution as a function of distance, where grey-filled circles show clouds without a known HMSF tracer,
grey circles with red outline show clouds with at least one ATLASGAL clump but no HMSF tracer, and red-filled circles show clouds with an ATLASGAL and an HMSF tracer.

\setlength{\tabcolsep}{3pt}
\begin{table*}
\begin{center}\caption{Properties of the ATLASGAL CSC clumps. Col.~1 gives the CSC name.
Cols.~2 and 3 give the adopted \vlsr\ and the \cco\ FWHM line-width.
Col.~4 gives the distance assigned to the clump.
Col.~5 is the clump mass, derived from the integrated 870~\mic\ flux density.
Col.~6 is the virial mass, computed as $5 \delta v^2 R / G$.
Cols.~7 and 8 list associations with \scimes\ clouds (Sect.~\ref{sect:gmcs} and Table~\ref{tab:full_catalogue}),
and with $^{12}$CO clouds from \cite{garcia2014}, respectively.
Col.~9 indicates matches with HMSF tracers (MMB: methanol maser;
HII: compact \hii\ region; YSO: massive young stellar object).
Col.~10 indicates the ATLASGAL CSC group (see
Table~\ref{tab:csc_groups}) to which each clump is associated, if any.
}
\label{tbl:clump_properties}
\begin{minipage}{\linewidth}
\small
\begin{tabular}{l.....cccc}
\hline \hline
\multicolumn{1}{c}{ATLASGAL }&  \multicolumn{1}{c}{\vlsr}& \multicolumn{1}{c}{$\delta v$} &\multicolumn{1}{c}{$d$} & \multicolumn{1}{c}{$M_{\rm{clump}}$} &	\multicolumn{1}{c}{$M_{\rm{vir}}$}    &\multicolumn{1}{c}{SDG}  &\multicolumn{1}{c}{$^{12}$CO} &\multicolumn{1}{c}{MSF} &\multicolumn{1}{c}{ATLASGAL}\\
\multicolumn{1}{c}{clump}&  \multicolumn{1}{c}{(\kms)}& \multicolumn{1}{c}{(\kms)} &\multicolumn{1}{c}{(kpc)} & \multicolumn{1}{c}{(10$^{2}$\msun)} &	\multicolumn{1}{c}{(10$^{2}$\msun)}    &\multicolumn{1}{c}{cloud}  &\multicolumn{1}{c}{GMC} &\multicolumn{1}{c}{assoc.} &\multicolumn{1}{c}{group}\\
\hline
AGAL340.119$-$00.022	&	-121.9	&	2.2	&	6.7	&	16.3	&	3.6	&	15	&	3KPC3	&	MMB	&	G340.256$-$00.059	\\
AGAL340.182$-$00.047	&	-123.7	&	5.3	&	6.7	&	13.8	&	21.4	&	15	&	3KPC3	&	MMB	&	G340.256$-$00.059	\\
AGAL340.249$-$00.046	&	-121.3	&	5.3	&	6.7	&	76.0	&	53.1	&	15	&	3KPC3	&	MMB/HII	&	G340.256$-$00.059	\\
AGAL340.359+00.129	&	-119.0	&	3.3	&	6.6	&	3.4	&	3.3	&	88	&	3KPC3	&	HII	&	--	\\
AGAL340.784$-$00.097	&	-101.4	&	3.9	&	6.0	&	20.0	&	11.7	&	116	&	NORM7.4	&	MMB	&	--	\\
AGAL340.508$-$00.442	&	-90.5	&	2.8	&	5.6	&	18.8	&	3.8	&	147	&	NORM7.2	&	--	&	G340.556$-$00.402	\\
AGAL340.466$-$00.299	&	-89.3	&	2.6	&	5.7	&	5.9	&	3.7	&	154	&	NORM7.2	&	HII	&	--	\\
AGAL340.054$-$00.244	&	-53.0	&	8.8	&	3.8	&	43.0	&	87.8	&	234	&	NCEN25	&	MMB/YSO/HII	&	G340.249$-$00.266	\\
AGAL340.248$-$00.374	&	-50.4	&	5.4	&	3.8	&	48.5	&	37.8	&	298	&	NCEN25	&	MMB/HII	&	G340.249$-$00.266	\\
AGAL340.536$-$00.152	&	-46.7	&	3.8	&	3.7	&	8.6	&	11.9	&	331	&	NCEN25	&	MMB	&	G340.529$-$00.147	\\
AGAL340.656$-$00.236	&	-21.7	&	4.2	&	2.1	&	0.4	&	1.7	&	-1	&	--	&	MMB	&	--	\\
\hline
\end{tabular}\\
Note: Only a small portion of the data is provided here, the full table is available in electronic form at the CDS via anonymous ftp to cdsarc.u-strasbg.fr (130.79.125.5) or via http://cdsweb.u-strasbg.fr/cgi-bin/qcat?J/A\&A/.
\end{minipage}
\end{center}
\end{table*}
\setlength{\tabcolsep}{6pt}

Although the statistics are low, there is a trend for the HMSF regions to be in the most massive clumps, within the most massive clouds.
We note that a larger proportion of the most massive clumps are associated with star formation (\citealt{urquhart2014}) and that these tend to be warmer than more quiescent clumps (e.g. \citealt{urquhart2011,wienen2012}).
This may lead to clump masses being over estimated; however, at the angular scales of the structures probed here (i.e.\ $\sim$30\arcsec) the temperature range from starless clumps to those hosting \hii\ regions or photo-dissociation regions is 15 to 25\,K
(\citealt{dunham2011b,urquhart2011,wienen2012,Deharveng2015,Guzman2015}).
A difference of $\pm$5\,K around the temperature used to estimate the clump masses
corresponds to a maximum variation of approximately 30\% in the dust masses and so
does not have a significant impact on the overall clump mass distribution.

\subsubsection{Dense gas fraction}

In order to investigate whether the presence of high-mass star formation was related to the amount of dense gas within each cloud, we determined the dense gas fraction (DGF) of the SEDIGISM GMCs using:
\begin{equation}
{\rm{DGF}} = M_{\rm{clump}}/M_{\rm{GMC}},
\end{equation}

\noindent where $M_{\rm{clump}}$ is the clump mass, as defined above.
When several clumps are associated with a single GMC, the masses of all the clumps are summed together.
We assume that the dust emission is tracing the dense gas at column densities above
$\sim$7.5$\times10^{21}$\,cm$^{-2}$, corresponding to the 5-$\sigma$ limiting sensitivity for ATLASGAL.

\begin{figure}[t]
\centering 
\includegraphics[width=0.45\textwidth]{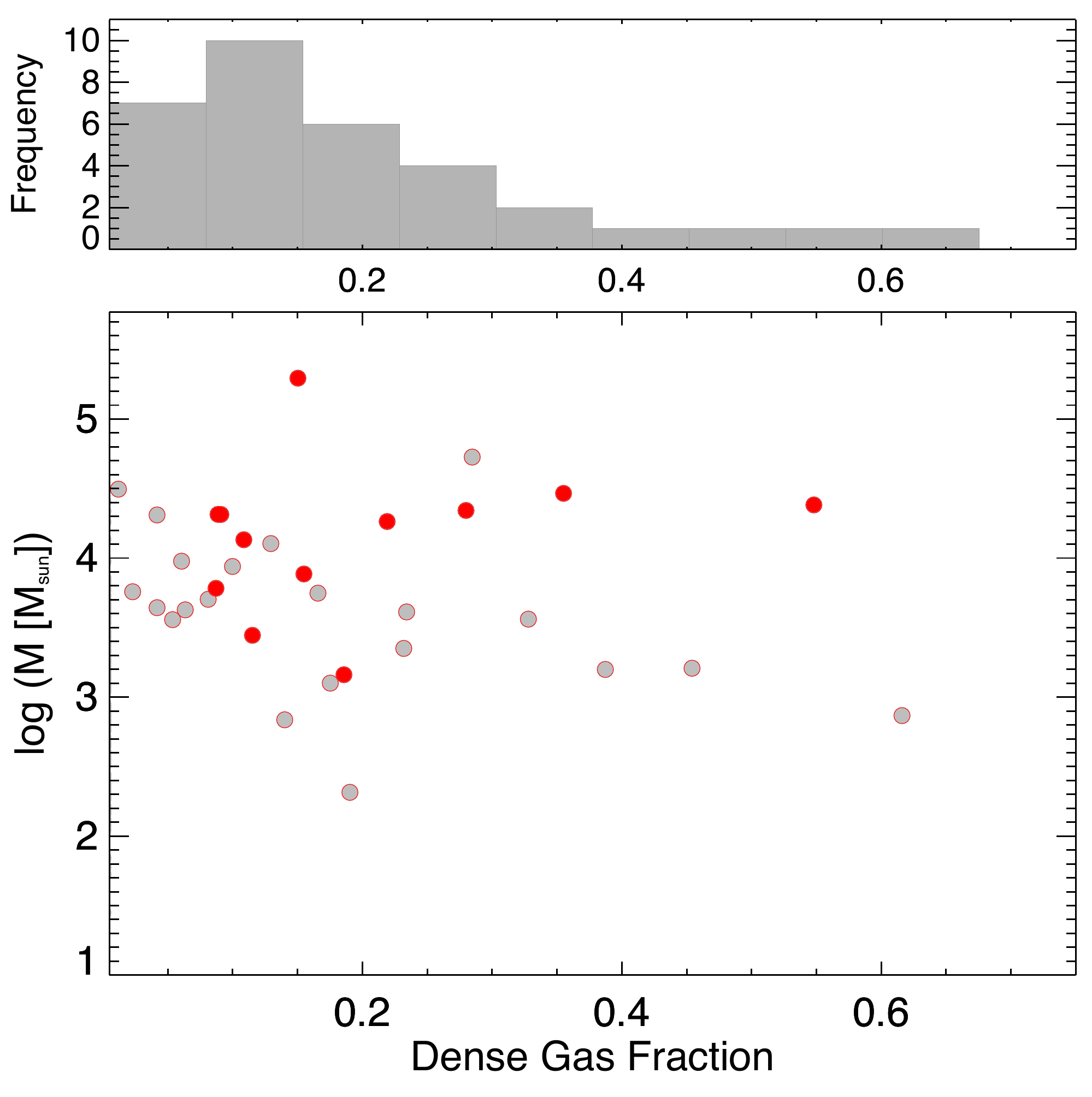}
\caption{{\em Top}: Histogram of the dense gas fraction (DGF) of the GMCs associated with ATLASGAL clumps.
{\em Bottom}: Distribution of the DGF as a function of cloud mass for all GMCs that have an ATLASGAL counterpart,
colour-coded according to the existence of an HMSF marker (red for clouds with an HMSF tracer, and grey for no HMSF).}
\label{fig:dgf_histogram}
\end{figure}

In Fig.\,\ref{fig:dgf_histogram} we show the histogram of the dense gas fraction of the GMCs,
and the distribution of DGF as a function of cloud mass.
The median value of DGF for the clouds with an ASTLASGAL counterpart is
$\sim$15\% (and a mean value of $\sim$21$\%$), which is 2-3 times higher than reported
in the literature \cite[i.e.\ 2-7\%, cf.][]{eden2013,Ragan2014,battisti2014,csengeri2015}.
However, this represents only a small fraction of the clouds extracted with \scimes\
(33 out of 182). The remaining ones would have a DGF close to zero, so that the mean DGF
value would be significantly lower when including all clouds ($\sim$4$\%$).
In addition, the \cco\ transition has a higher critical density than
$^{12}$CO(1--0), which was used in some of these previous studies, and is therefore not
necessarily capturing all of the material associated with the extended envelope. The SEDIGISM
\cco\ data have a column density sensitivity of a few 10$^{21}$\,cm$^{-2}$ and, as a
consequence, the DGFs estimated are likely to be upper limits to the true values.
Finally, one could expect that for clouds with angular sizes smaller or similar to
the maximum recoverable scale in the ATLASGAL data \citep[uniform emission on scales larger
than $\sim$2.5$'$ is filtered out during data reduction,][]{schuller2009}, the signal
detected by the bolometers could encompass most of the cloud rather than
simply the denser parts \citep[see e.g.\ the discussion in][]{battisti2014}.
This would also result in over estimating the true DGF.
We investigated this effect in our sample, and we found that clouds with small angular
sizes have a wide range of DGFs (between 0 and 60\%), but also that some
well resolved clouds, with angular sizes an order of magnitude larger than the maximum
recoverable scale of ATLASGAL, have a relatively large DGF (up to $\sim$60$\%$).
Therefore, we do not think that this effect is significant in our sample.

\begin{figure*}[t]
\centering 
\includegraphics[width=\textwidth]{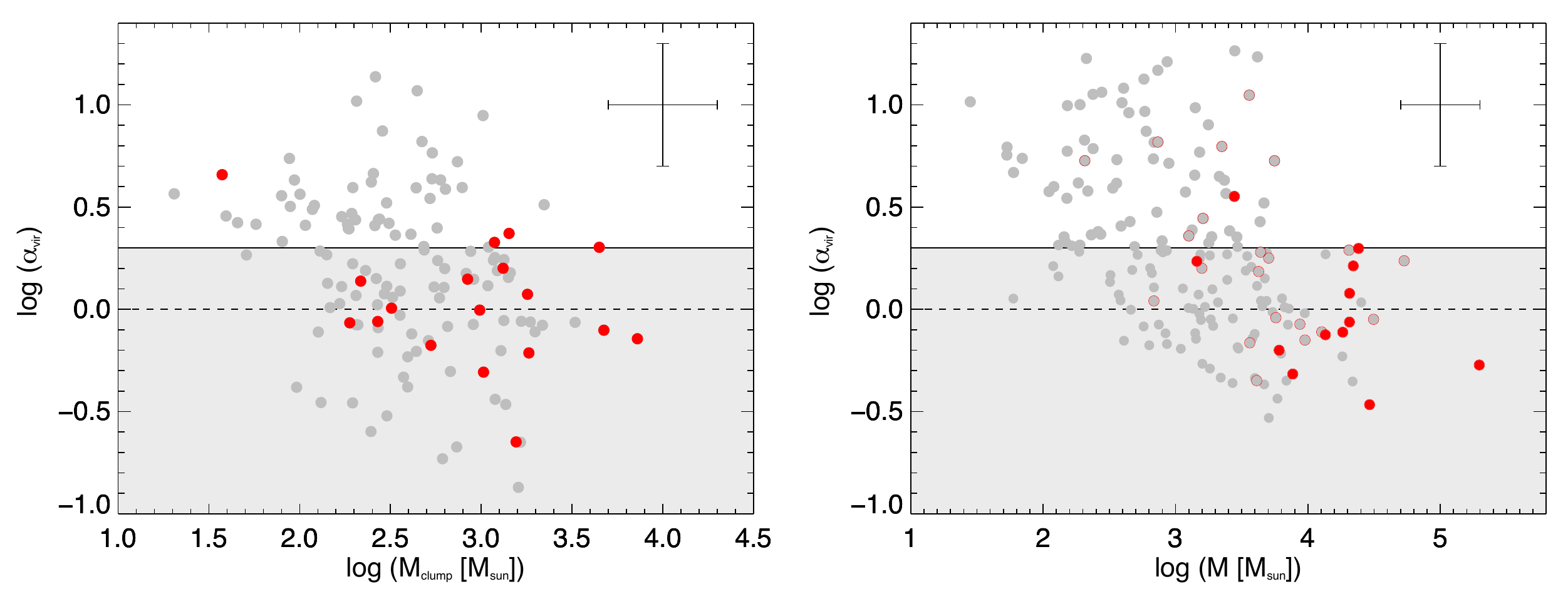}
\caption{Virial parameter ($\alpha_{vir}$) as a function of clump mass ($M_{\rm{clump}}$) in the left panel, and cloud mass ($M$) in the right panel.
This is shown for the HMSF and non-HMSF sub-samples; these are indicated as red and grey circles, respectively.
In the right panel, GMCs with an ATLASGAL match but no HMSF tracer are shown as grey circles with a red outline.
The solid horizontal line indicates the critical value of $\alpha_{vir} = 2$, for an isothermal sphere in hydrostatic equilibrium without magnetic support.
The dashed black line indicates an $\alpha_{vir} = 1$. The light grey shading indicates the region where clouds are unstable
and likely to be collapsing without additional support from a strong magnetic field.
Representative error bars, corresponding to a factor two uncertainty, are shown
in the upper-right corner of each panel.}
\label{fig:virial_mass_plot}
\end{figure*}

We do not find any particular trend of DGF with the cloud masses nor with the existence of high-mass star formation. Instead, the existence of HMSF seems to correlate better with the total mass of the cloud.
We caution that the statistics within the SEDIGISM \testfield\ are low for any firm conclusions, and therefore, this will be further investigated in a subsequent paper using the complete survey.

\subsubsection{Virial parameter}

Finally, we estimated the virial parameters for all clumps using the same formula described in Sect.\,\ref{sec:gmc_physical_properties} and used this to evaluate their stability. 
In the left panel of Fig.\,\ref{fig:virial_mass_plot} we show the virial parameters as a function of clump mass. There is a clear trend for decreasing virial parameters with increasing mass, indicating that the most massive clumps are also the most gravitationally unstable, and the more likely to be undergoing collapse unless they are supported by strong magnetic fields.
Similar trends have been reported in the literature \cite[e.g.][]{barnes2011, kauffmann2013, urquhart2015b}.
We note that all of the HMSF clumps are found in the unstable part of the parameter space, which would suggest that, even if magnetic fields can stabilise the clumps globally (e.g. \citealt{pillai2015}),
they are unlikely to be able to do this on smaller size scales, as star formation is clearly ongoing in some of these clumps.

Interestingly, if we plot the virial parameter of the SEDIGISM clouds as a function of cloud mass,
we find a similar trend as for the clumps (Fig.\,\ref{fig:virial_mass_plot}, right panel),
even though we have more clouds in the unbound regime, as a consequence of the lower
surface densities traced by \cco. 
Despite that, this figure shows that most clouds that are unstable on clump scales, 
and most particularly, the clouds that already show signs for HMSF, 
do seem to be gravitationally unstable even at the larger scales of the GMCs,
in line with the idea that HMSF is preferentially taking place in globally
collapsing clouds \citep[e.g.][]{barnes2010,Schneider2010,Peretto2013}. 
A similar trend for the most massive clouds to appear more likely to be bound is 
also seen in simulations \cite[e.g.][]{Dobbs2011}.



\section{Excitation, optical depth, and physical conditions}\label{sect:excitation}
\subsection{Excitation and column density}
\label{sect:coldens}

The two CO isotopologues' $J$=2--1 emission lines observed with SEDIGISM, in
particular when combined with the three iso-CO $J$=1--0 lines from
ThrUMMS \citep{ref-thrumms}, allow us to derive a detailed, spatially- and
velocity-resolved distribution for various physical and chemical
properties in all the observed molecular clouds, including: optical
depths, excitation temperatures, molecular abundances, and column
densities.  \citet{ref-thrumms} already demonstrated the diagnostic
power of such an approach with the ThrUMMS data alone, finding a new
value of the conversion factor between CO emissivity and mass column
density, which suggests that the total molecular mass of the Milky
Way may have previously been substantially underestimated.  They did
this assuming only a common LTE excitation between the three main
iso-CO species, and a fixed intrinsic abundance ratio $R_{13}$ $=$
[$^{12}$CO]/[$^{13}$CO].  While the latter may indeed also vary, their
results on the mass distribution are relatively insensitive to the
exact value assumed for $R_{13}$.

The validity of a common $T_{ex}$ between the very optically thick
$^{12}$CO lines and the more typically optically thin $^{13}$CO and
C$^{18}$O lines is a more relevant issue, but the SEDIGISM data now
allow a straightforward resolution to this issue as well.  Building
upon the method described by \citet{Kramer1999} and
\citet{Hernandez2011}, we have developed a root-finding algorithm
to compute $\tau$, $T_{ex}$ and the column density at each voxel
of the data cube.
We arrive at our solutions by matching the
column density calculated from each $^{13}$CO transition across a
range of $T_{ex}$.
The column density is given by the usual plane-parallel radiative
transfer equation:
\begin{equation} 
	N = \frac{3h}{8\pi^3\mu^2}~\frac{Q(T_{\rm ex})e^{E_l/kT_{\rm ex}}}{J_u(1-e^{-h\nu/kT_{\rm ex}})}~\int\tau_{ul}{\rm d}V,
    \label{eq:coldens}
\end{equation}
\noindent
where the total $N$ is calculated separately for each transition
line ($J_u=2$ or 1), $\mu$ is the dipole moment of the CO molecules,
$Q(T_{\rm ex})$ is the rotational partition function, and $E_l$ is the
energy of the lower state of transition $J_u \rightarrow J_u-1$.

The optical depth, $\tau_{ul}$, is derived through the plane-parallel
radiative transfer equation:
\begin{equation}
T_{\rm mb} = \frac{h\nu}{k}(f_{\rm T_{ex}}-f_{\rm T_{bg}})(1-e^{-\tau_{ul}}). 
\label{eq:tau}
\end{equation}
\noindent
Here $T_{\rm mb}$ is the main beam brightness temperature, $T_{bg}$ is
the background temperature of 2.73 K, and
$f_T=[\rm{exp}(\textit{h}\nu/(kT))-1]^{-1}$. Since $T_{\rm mb}$
for both transitions are observed with either the ThrUMMS or SEDIGISM
surveys, we can use Eqs.~(\ref{eq:coldens}) and (\ref{eq:tau}) to
express the ratio between the $J$=2--1 and $J$=1--0 column densities
as a function of $T_{ex}$:
\begin{equation}
\eta_{21}(T_{ex}) =\Big| \log{ \Big( \frac{N_{\rm tot,21}}{N_{\rm tot,10}}}\Big) \Big|,
\label{eq:eta21}
\end{equation}
\noindent
where $N_{\rm tot,21}$ and $N_{\rm tot,10}$ are the total column
densities calculated from each line transition using Eq.~(\ref{eq:coldens}).

\begin{figure}[tp]
\centering 
\includegraphics[width=0.4\textwidth,angle=90]{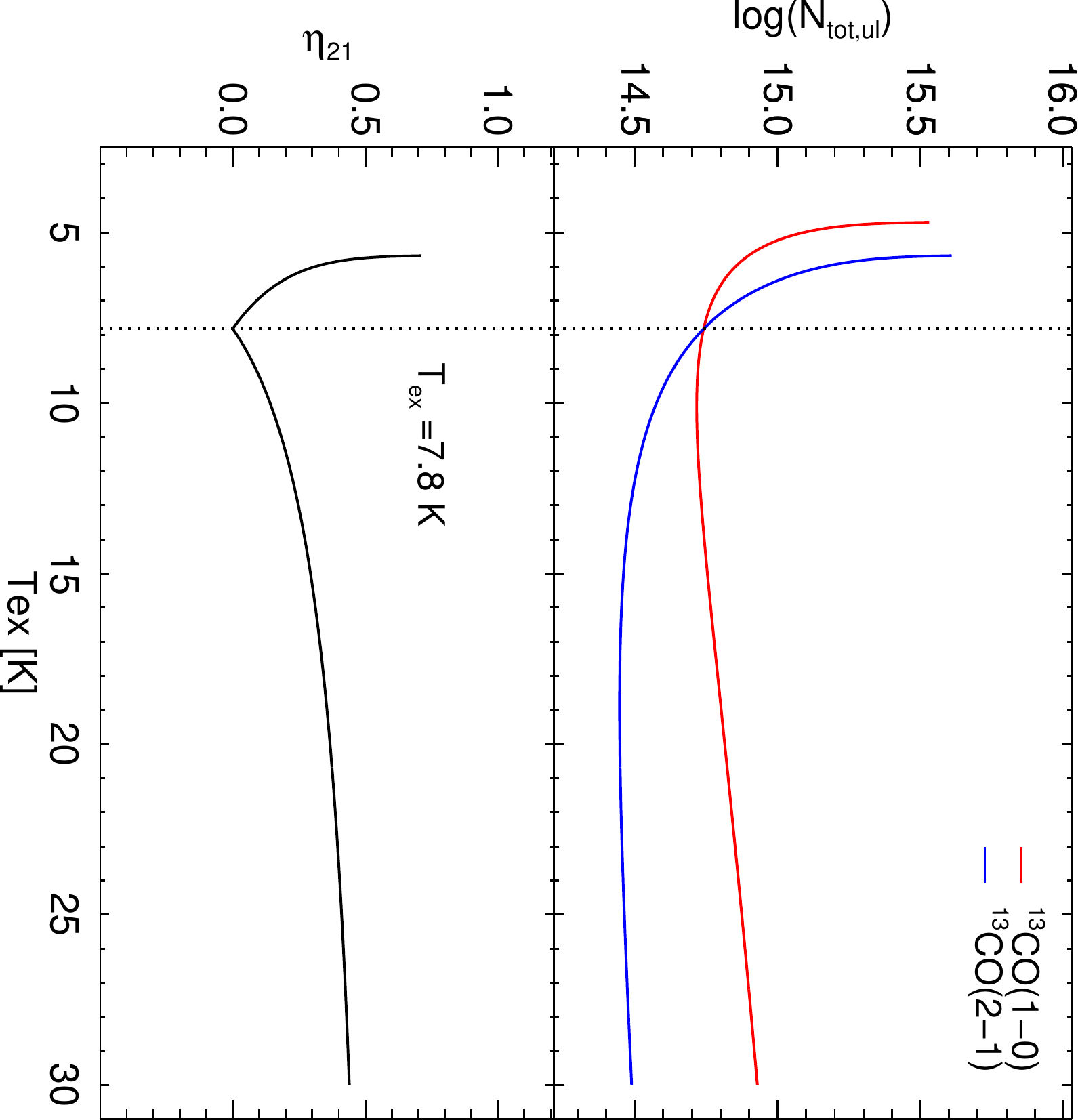}
\caption{\textit{Top:} The total $^{13}$CO column density was
  estimated over a range of $T_{ex}$ from each line transition
  separately using Eq.~(\ref{eq:coldens}). \textit{Bottom:} The
  distribution of $\eta_{21}$ (Eq.~(\ref{eq:eta21})) over the same range of $T_{ex}$.
  Since we have defined $\eta_{21}$ as the absolute difference between the
  two total column density estimates, the $T_{ex}$ in this voxel is determined
  by locating the global minimum, as shown by the vertical dotted black
  line. For this example voxel, we find $T_{ex} \, =$ 7.8~K.}
\label{fig:voxTex}
\end{figure}

\begin{figure*}[t]
\centering 
\includegraphics[width=0.99\textwidth]{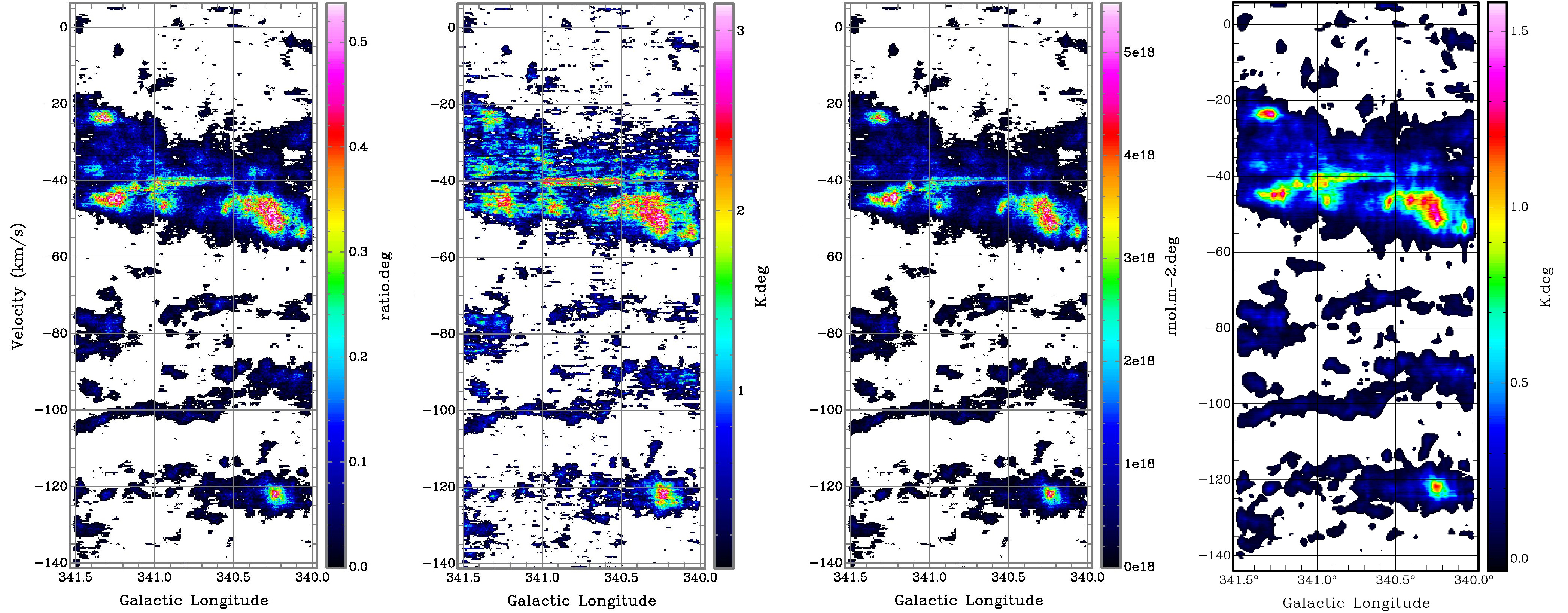}
\caption{Radiative transfer solutions, obtained as described in \S\ref{sect:coldens}, for $\tau$(2--1) (left panel) and $T_{ex}$ (second panel) of the $^{13}$CO emission data cubes.
The third panel shows the resulting column density $N$($^{13}$CO), derived from the optical depth and excitation temperature cubes. 
The data in each panel have been integrated over the 1-degree extent in latitude, including any zero-valued pixels, to form the respective longitude-velocity moment maps shown here.
The right panel shows a similar $lv$ moment map of the \cco\ emission as seen in the SEDIGISM data convolved to the ThrUMMS resolution.
}
\label{fig:thrumms_results}
\end{figure*}

\begin{figure*}[ht]
\centering 
\includegraphics[width=0.95\textwidth]{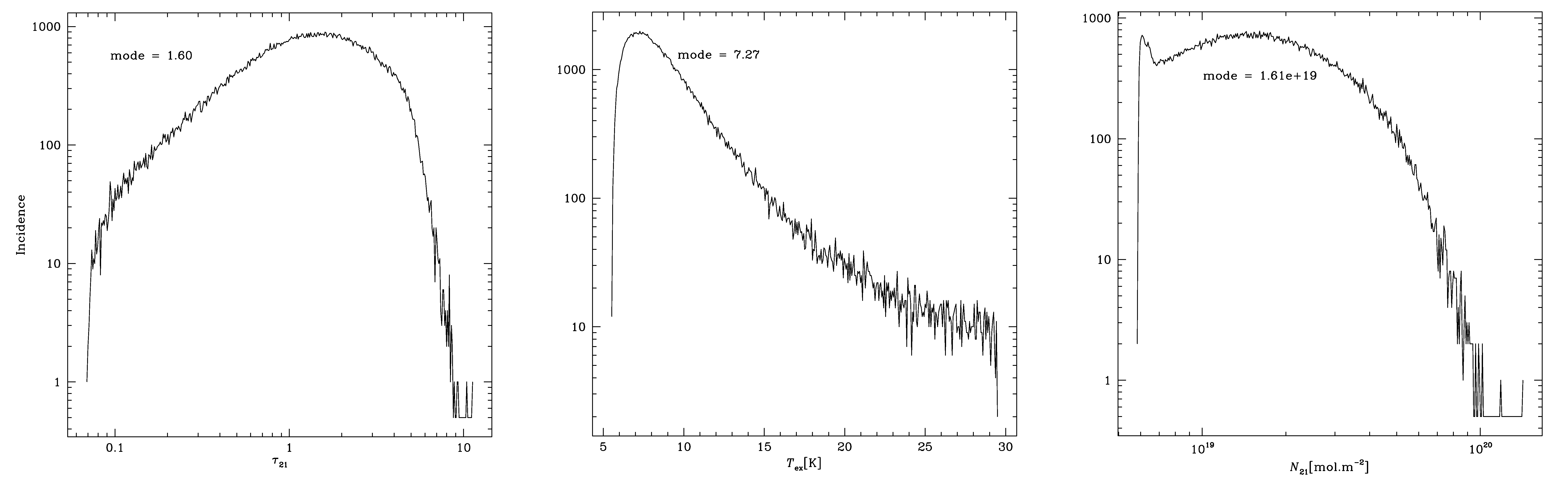}
\caption{Distributions of $\tau$(2--1) (left panel), $T_{ex}$ (middle panel) and $N$($^{13}$CO) (right panel) for the 140,000 usable voxels in the \testfield.}
\label{fig:TF-histog}
\end{figure*}

To improve computing time we simplified the iterative method of \citet{Hernandez2011},
who estimated the three-dimensional $T_{ex}$ distribution throughout a highly
filamentary  IRDC using C$^{18}$O $J$=2--1 and C$^{18}$O $J$=1--0.
This modification is possible since Eq.~(\ref{eq:eta21}) is a function
with a global minimum within a domain of $T_{ex}\ge2.73$~K, which
represents when $N_{\rm 21}$ and $N_{\rm 10}$ are equal (i.e. when
$\eta_{21}=0$). Thus, the voxel $T_{ex}$ can be estimated by simply
minimizing $\eta_{21}$ within $T_{ex}$ of range 2.73 to 30 K, the
typical excitation temperature range for GMCs
\citep[e.g.][]{ref-thrumms}.
By equating N$_{21}$ and N$_{10}$, we are assuming that their excitation
temperatures are equal.  It is possible that the excitation will differ between
the two transitions and produce unequal, possibly sub-thermal, excitation temperatures.
However, for lower density cloud regions, \citet{ref-jimenez} combined 
$^{13}$CO\, J=1--0 data with the J=3--2 and J=2--1 LVG analysis,
finding that the overall column  densities were within a factor of 2.
We find that most voxels with S/N $> \sim$4 have line ratios that allow
convergence to a single assumed $T_{ex}$. For some noisier voxels, the algorithm
fails to converge to a $T_{ex}$ solution due to "unphysical" line ratios assuming
a shared $T_{ex}$; these voxels are then omitted from our analysis.
While mathematically  this might suggest differential thermalisation, we
discount this possibility due to the low S/N at these locations.

For the present study, we use the $^{13}$CO data from both surveys. We first
convolve the 30$''$ resolution SEDIGISM cubes to the 72$''$ resolution
of the ThrUMMS data.  $T_{ex}$ was estimated for all voxels with
$T_{\rm mb}$ measurement above zero to avoid unphysical column density
estimates and improve computing time.  Fig.~\ref{fig:voxTex} presents
the $T_{ex}$ solution for one example voxel.  For each voxel with a
$T_{ex}$ solution, we are able to compute the opacities ($\tau_{21}$
and $\tau_{10}$) and total column density, N($^{13}$CO). Finally, by
performing this analysis for each voxel, we are able to derive the
three-dimensional spatially- and velocity-resolved distribution of the
physical conditions of the $^{13}$CO gas.

Fig.\,\ref{fig:thrumms_results} presents the results for $\tau_{2-1}$, $T_{\rm ex}$ and $N(^{13}$CO) in the \testfield, as longitude-velocity maps integrated along $b$.
The distributions of these three quantities on a voxel basis are shown in Fig.~\ref{fig:TF-histog}.
Interestingly, the $T_{\rm ex}$ and $\tau$ distributions, while each contributing to the column density,
are distinctly different in several places.
That is, some locations with high $N$ are mostly due to a high excitation
while other locations derive their high $N$ from a high opacity.
The latter is especially interesting since we see that the highest column density
clumps reach peak $^{13}$CO opacities of $\sim$8, which certainly shows
that common assumptions about optically thin emission can lead one's analysis
and physical interpretation astray.
Maps of single lines cannot by themselves give us this physical insight.

\subsection{The $^{13}$CO X-factor}
\label{sec:Xfactor}
From this radiative transfer solution, we can directly compute an important result which bears on much of the new science presented here, as well as confirming previously published results. As shown by \citet{ref-thrumms}, a voxel-by-voxel calculation of the ratio of $^{13}$CO column density to integrated intensity, N/I, can be used as a direct probe of the spatially-resolved X-factor, relating integrated intensity (here for $^{13}$CO, but usually for $^{12}$CO in the literature) to the total molecular hydrogen column density.  When this ratio is plotted as a function of I, one can also reveal the nature of the conversion law, that is whether it is "flat" (constant X, the standard method for many decades) or a more complex function of other parameters \cite[e.g.][]{narayanan,ref-thrumms}.

\begin{figure}[tp]
\centering 
\includegraphics[width=0.45\textwidth]{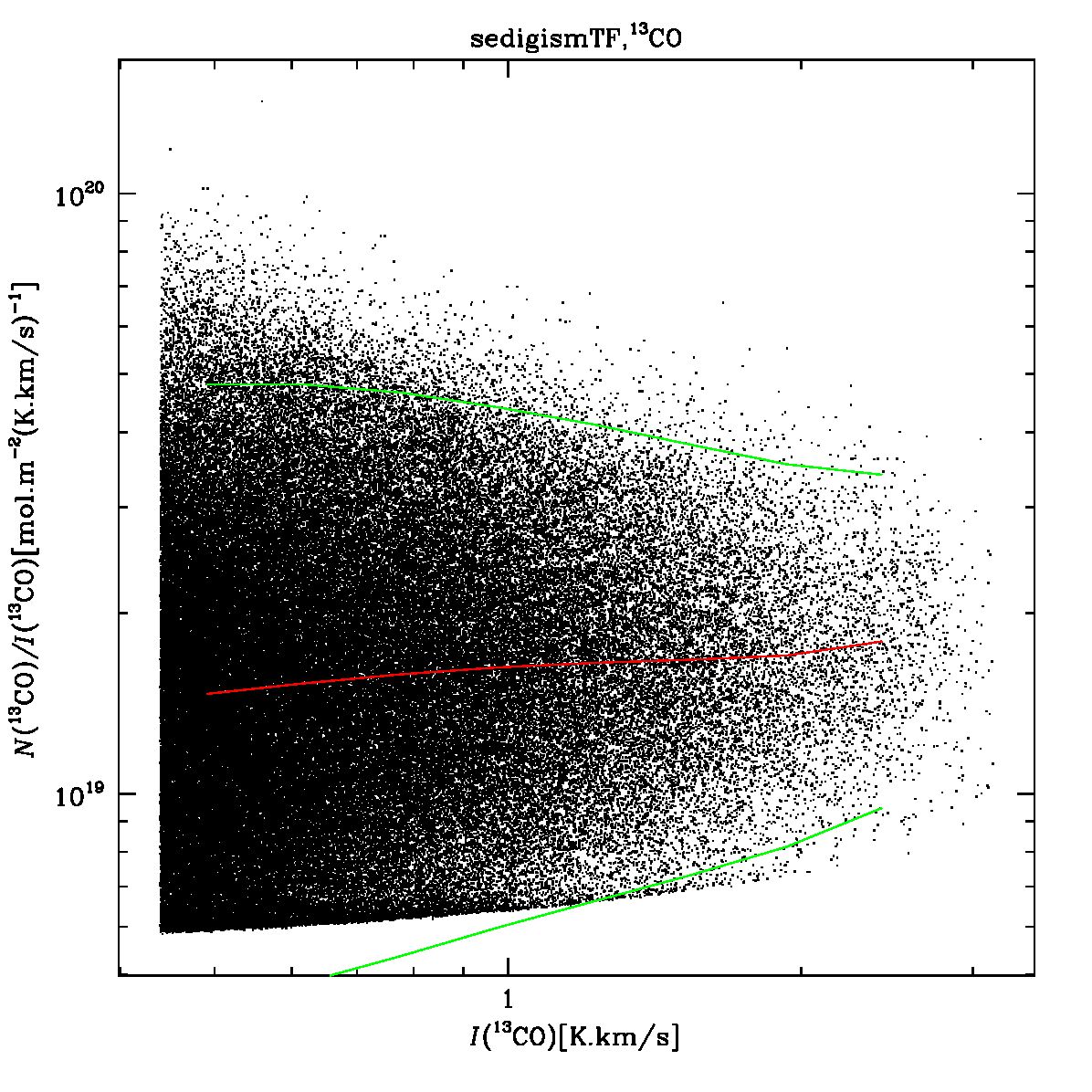}
\caption{Ratio of column density from radiative transfer computation to \cco\ integrated
intensity, N/I, as a function of the same integrated intensity, for the
$\sim$140,000 voxels in the \testfield\ with I above a 5-$\sigma$ level of 0.44~K~\kms.
The red line shows the trend of median values (which are very similar to the means) for
the data binned into ten equal log(I) intervals while the green lines show
the 2-$\sigma$ excursions from the median values.}
\label{fig:Xfactor}
\end{figure}

We present this comparison here in Fig.~\ref{fig:Xfactor}. Unlike the result for the $^{12}$CO(1--0) line \cite[$X$ $\propto$ $I^{0.4}$,][]{ref-thrumms}, we see that X is statistically flat for \cco\, with a mean value of 1.8$\, \times 10^{15}$~cm$^{-2}$~(K~\kms)$^{-1}$ across the 140,000 voxels in the \testfield\ with I($^{13}$CO) $>$ 5-$\sigma$ .

To convert this to a true X-factor for the \cco\ line, we need to multiply this mean by two gas-phase abundances, [$^{12}$CO]/[$^{13}$CO] and [H$_2$]/[$^{12}$CO].
For simplicity, we take the first ratio as 60 to conform with \citet{ref-thrumms}, and the second as 10$^4$ \cite[e.g.][]{dame2001,Bolatto2013}.
Then we obtain a mean X factor in the \testfield, based only on the $^{13}$CO data from the ThrUMMS and SEDIGISM surveys, 
of 1.08$ \, \pm$ 0.19 $\times 10^{21}$~cm$^{-2}$~(K~\kms)$^{-1}$.

This result agrees very well with the factor derived from the comparison between Hi-GAL and \cco\ data (see Sect.~\ref{sec:gmc_physical_properties}).
Although fortuitous to some extent, based on the inclusion of somewhat uncertain conversion parameters in both estimates, this close agreement is remarkable.  In addition, since we have used the same approach as in \citet{ref-thrumms} to compute the $^{13}$CO column density, the above agreement in X estimates gives strong confirmation to the ThrUMMS $^{12}$CO conversion law as well.  In particular, the clear difference in the behaviour of the conversion law for $^{13}$CO (flat) compared to $^{12}$CO (a function of I($^{12}$CO) at least) lends credence to Barnes et al's (2015) argument that their $^{12}$CO conversion law arises from the extremely high opacity in the $^{12}$CO(1--0) line ($\tau$ up to $\sim$400).

A direct implication of our approach is that we will be able to make reliable maps
of the total molecular column density, from a combination of SEDIGISM and ThrUMMS data,
across the entire 10$^9$ voxel data set of these surveys.
With distances to the various kinematic features, these column densities are then readily
converted into masses of individual clouds, as identified e.g.\ by the \scimes\ algorithm in Sect.~\ref{sect:gmcs}.
Also, combining these results with the C$^{18}$O SEDIGISM data and the J=1--0 $^{12}$CO data
from ThrUMMS will further allow us to construct spatially- and velocity-resolved maps of
the various molecular abundances, and relate any abundance variations we may see
to environmental or other factors.

\section{Filamentary structures}\label{sec:filaments}

In this section, we investigate the presence of filamentary structures in the \testfield\, and outline the potential of SEDIGISM to verify their coherence in velocity, and derive their
properties: velocity dispersion, length, column density, mass, and linear mass density.
In Sect.\,\ref{sec:fil-agal}, we discuss the filament candidates identified in the ATLASGAL and Hi-GAL continuum surveys \citep{schuller2009,Molinari2010}; in Sect.\,\ref{disperse}, we show the results of applying the \Disperse\ algorithm \citep{sousbie2011} directly on the $^{13}$CO data cube.

\subsection{Filament candidates in ATLASGAL and Hi-GAL}\label{sec:fil-agal}

Recently, the ATLASGAL and Hi-GAL surveys were used to identify filament candidates in the \GP\ through the analysis of their continuum emission at 870\,$\mu$m \citep{Li2016}, and at 70, 160, 250, 350 and 500\,$\mu$m \citep{Schisano2014,Wang2015}, respectively.
The two surveys deliver a unique dataset to compile an unbiased catalogue of filament candidates throughout the Galaxy. Indeed, ATLASGAL provides high angular resolution at a wavelength sensitive to the cold dust and is unaffected by background contamination and saturation, which complicate the analysis of Hi-GAL data.  On the other hand, Hi-GAL is more sensitive to emission from low-density
structures, down to values of $\sim$10$^{21}$\,cm$^{-2}$ at 16\,K, while ATLASGAL has
a 5-$\sigma$ column density sensitivity of $\sim$7.5$\times$10$^{21}$\,cm$^{-2}$
for a dust temperature of 20\,K.

\begin{figure}
   \centering
\includegraphics[width=0.49\textwidth]{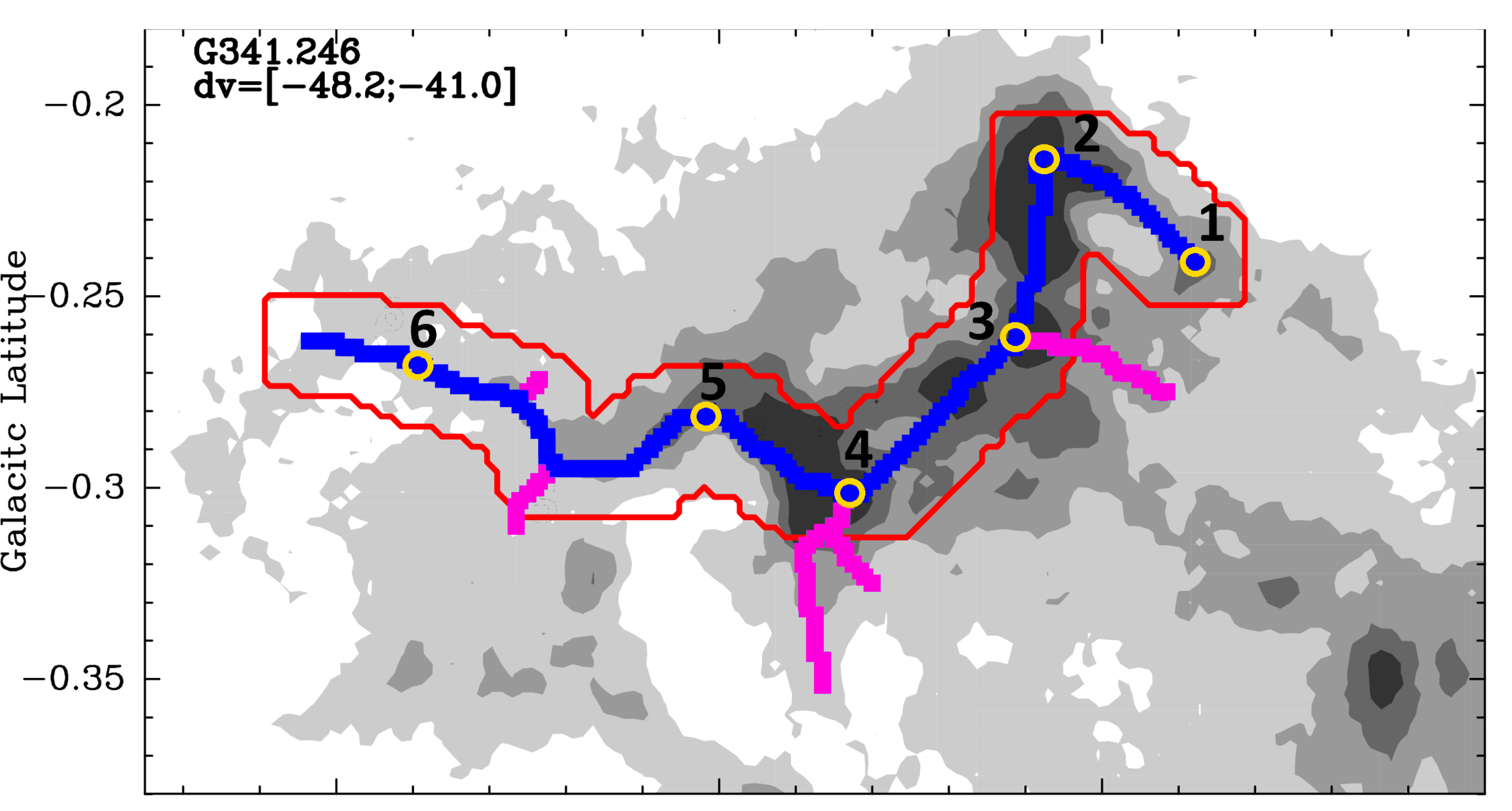}
\includegraphics[width=0.49\textwidth]{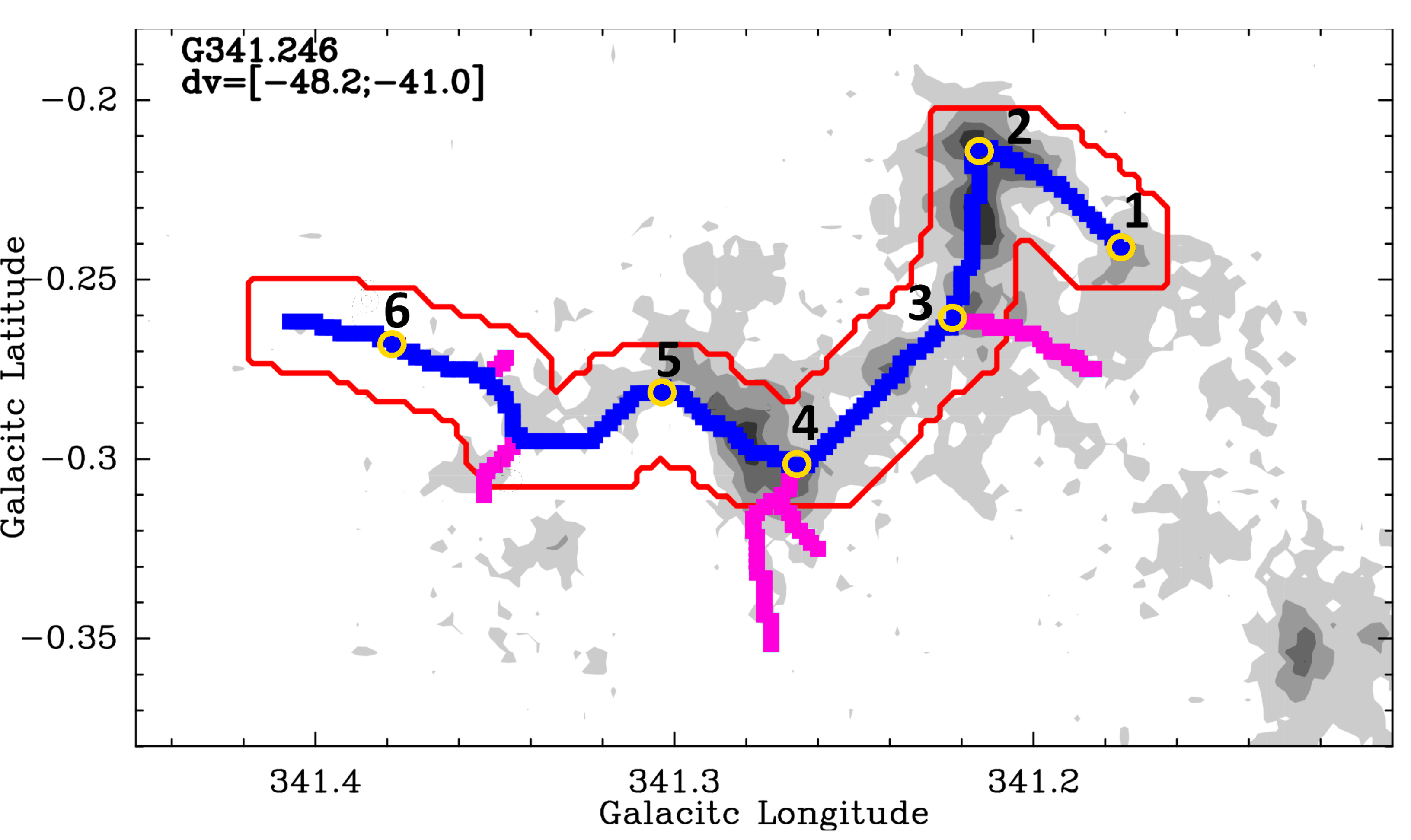}
\caption{Integrated intensity map of the \cco\ (top) and \coo\ (bottom) transitions towards G341.246--0.267 in the velocity range of [$-48.2,-41.0$] km\,s$^{-1}$. Grey levels are from 10-$\sigma$ (9.4\,K\,km\,s$^{-1}$) for $^{13}$CO in steps of 10-$\sigma$, and from 5-$\sigma$ (4.7\,K\,km\,s$^{-1}$) for C$^{18}$O in steps of 5-$\sigma$. The blue solid line marks the main spine of the filament identified; the magenta lines mark the other sub-branches identified on the ATLASGAL dust emission map by \citet{Li2016}.  The yellow empty circles mark the positions  where the six spectra shown in Fig.\,\ref{spectra_G341} are extracted. The red thin  line marks the dilation box used to compute the length and the mass of the filament.}\label{fig:atlasgal-example2}
\end{figure}

\begin{figure}
\includegraphics[width=0.49\textwidth]{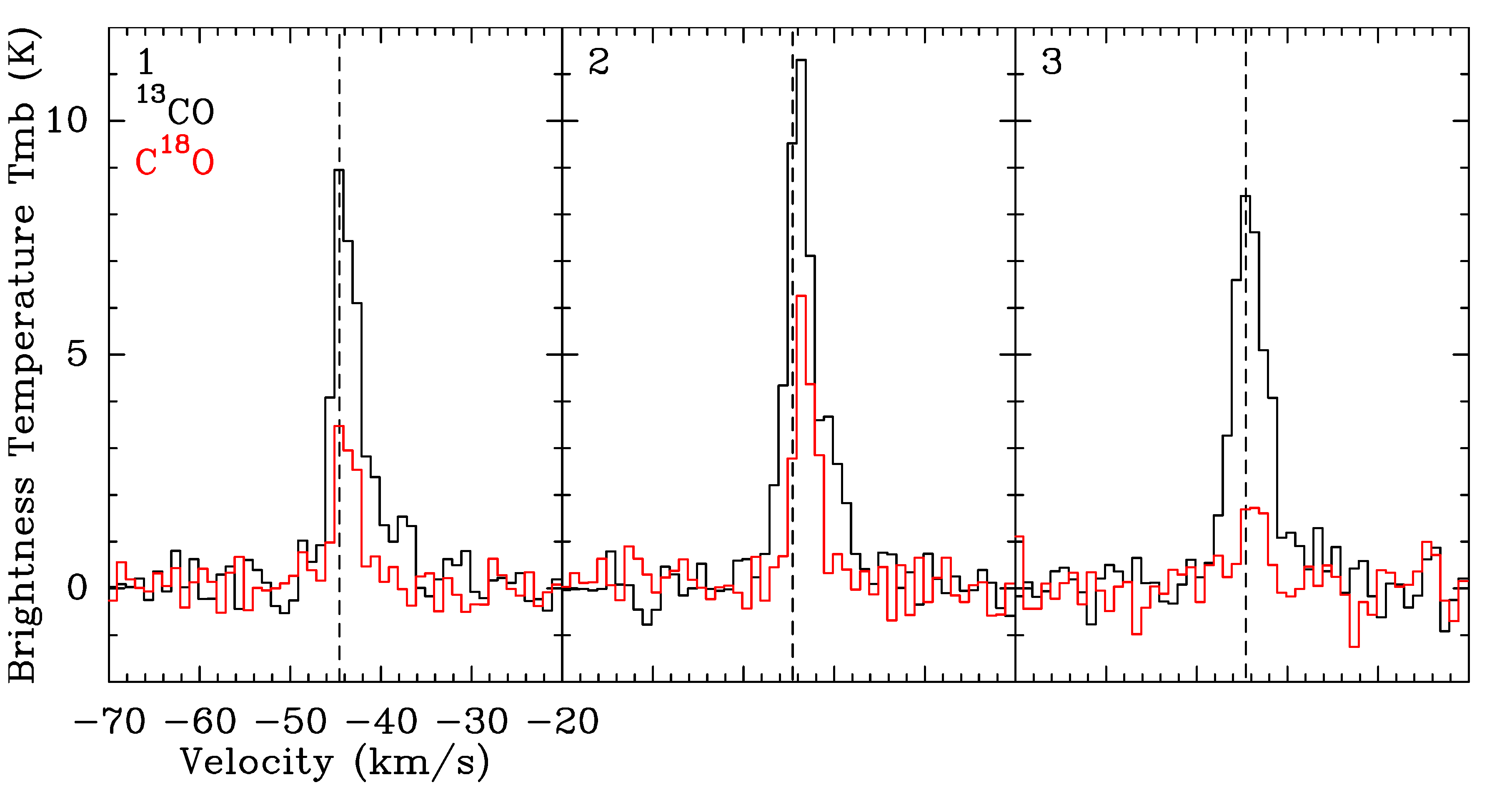}
\includegraphics[width=0.49\textwidth]{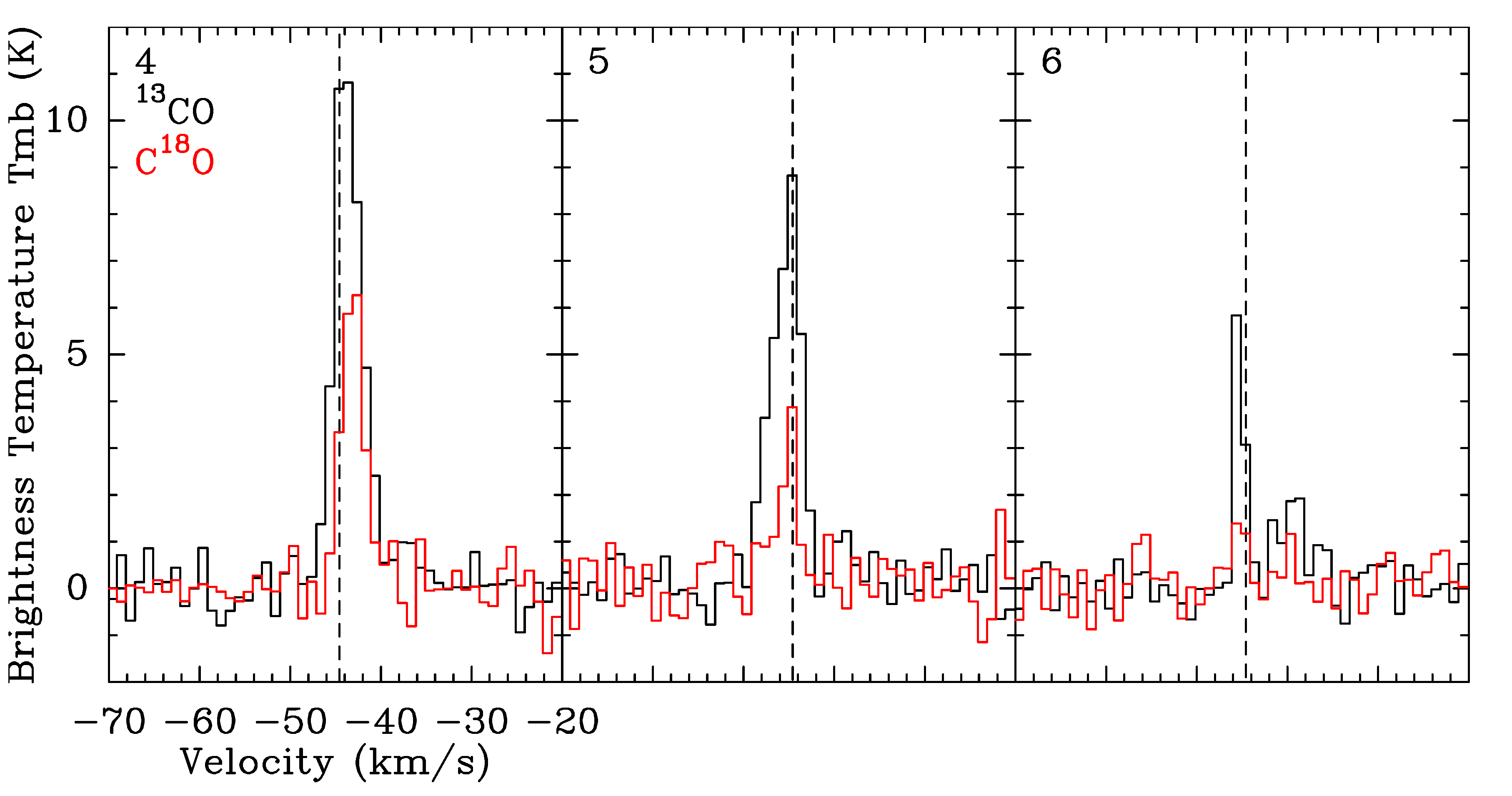}
\caption{\cco\ (black) and \coo\ (red) spectra extracted at six positions along the spine of the G341.246--0.267 filament (Fig.\,\ref{fig:atlasgal-example2}).}\label{spectra_G341}
\end{figure}

\cite{Li2016} identified twelve filamentary structure candidates in the SEDIGISM \testfield\ based on the ATLASGAL data, nine of which are single filament candidates (elongated linear structures with typical aspect ratios larger than three), and the other three being networks of filaments (several filaments that seem to be connected to each other).
A study of Hi-GAL column density map of the region reveals 88 filament candidates.
Details on the two catalogues and on the methods used to identify the structures and extract their dust properties are given by \citet{Li2016}, \citet{Schisano2014} and Schisano et al. (in prep.). All twelve structures detected in ATLASGAL are also found in the Hi-GAL sample, hence, in the following discussion, we will focus on this common candidate list.

\begin{table*}[ht]
\setlength{\tabcolsep}{4pt}
\begin{center}
\caption{Catalogue of ATLASGAL filaments (top) and networks (bottom) in the \testfield.
Col.~1 gives the filament name, as in \cite{Li2016}. 
Col.~2 is the central velocity of \cco, measured in the SEDIGISM data;
the velocity range used to compute column densities and masses is given in Col.~3, and the integrated intensity is in Col.~4.
Col.~5 is the projected area, computed for the assigned distance given in Col.~6. The total mass that we computed is given in Col.~7.
Col.~8 shows the length of each filament, and Col.~9 is the velocity dispersion.
Finally, the measured and the virial linear mass densities are given
in Cols.~10 and 11, respectively.}
\label{tab:atlasgal-catalog}
\begin{minipage}{\linewidth}
\begin{tabular}{lrrrcccrcrr}
\hline
\hline
Name &\multicolumn{1}{c}{ $\varv$ }&\multicolumn{1}{c}{$\Delta \varv$}&\multicolumn{1}{c}{ $\int{T}{d\varv}$ }&\multicolumn{1}{c}{ $A$ }&\multicolumn{1}{c}{ $d$ }&\multicolumn{1}{c}{ $M_{\rm ^{13}CO}$ }&\multicolumn{1}{c}{ $l$ }&\multicolumn{1}{c}{ $\sigma_v$ }&\multicolumn{1}{c}{ $(M/l)_{\rm obs}$ }&\multicolumn{1}{c}{ $(M/l)_{\rm vir}$}\\
 &\multicolumn{1}{c}{ (km\,s$^{-1}$)} &\multicolumn{1}{c}{ (km\,s$^{-1}$)} &\multicolumn{1}{c}{ (K\,km\,s$^{-1}$) }&\multicolumn{1}{c}{ (pc$^2$)} &\multicolumn{1}{c}{ (kpc)} &\multicolumn{1}{c}{ ($\rm 10^3~M_\odot$)} &\multicolumn{1}{c}{ (pc)} &\multicolumn{1}{c}{ (km\,s$^{-1}$)} &\multicolumn{1}{c}{ ($\rm M_\odot/pc$)} &\multicolumn{1}{c}{ ($\rm M_\odot/pc$)}\\
\hline
\multicolumn{10}{c}{Filaments}\\
 G340.301--00.387 &  -49.3 & --57.2,--41.4   & 22.18 &  45.82 &  3.8  & 28.0$^\dagger$ &  23.61 & 1.56 & 1187 & 1127\\
 G340.316+00.079  & -111.4 &--116.5/--106.3  &  5.59 &  39.63 &  6.2  &  4.7 &  11.15 & 0.51 &  424 &  121\\
 G340.482--00.306 &  -45.0 &--51.3/--38.7    &  9.54 &  15.16 &  3.8  &  4.5$^\dagger$ &   7.08 & 1.40 &  635 &  909\\
 G340.511--00.471 &  -43.5 &--50.2/--36.8    & 17.50 &  11.33 &  3.5  &  4.4 &   5.58 & 1.08 &  785 &  546\\
 G340.981--00.013 &  -46.9 &--50.3/--43.5    &  7.11 &  19.40 &  3.4  &  3.1 &  11.42 & 1.27 &  272 &  750\\
 G341.244--00.265 &  -44.6 &--48.2/--41.0    & 17.43 &  39.81 &  3.6  & 22.4 &  21.34 & 1.04 & 1049 &  507\\
 G341.415+00.244  &  -37.4 &--39.6/--35.2    &  8.27 &   7.17 &  3.2  &  1.3 &   4.26 & 0.83 &  312 &  320\\
 \hline
 \multicolumn{10}{c}{Networks}\\
G340.200--00.035   & -122.2 &--127.3/--117.1  & 22.70 & 116.52 &  6.6  &  59.2&    \\
G340.236--00.153  &  -50.9 &--58.3/--43.5    & 18.55 & 209.95 &  3.8  & 122.0&    \\
G340.941--00.319  &  -45.9 &--50.4/--51.4    & 10.37 &  74.70 &  3.6  &  30.1&    \\
G341.306+00.339   &  -78.3 &--82.3/--74.3    & 12.71 &  56.84 &  5.2  &  16.0&    \\
\hline
\end{tabular} \\
Note: $^\dagger$ For G340.301--00.387 and G340.482--00.306, the masses from the $^{13}$CO data are likely overestimated by up to 20\% and 34\%,
respectively, due to contamination from emission associated with other structures.
\end{minipage}
\end{center}
\setlength{\tabcolsep}{6pt}
\end{table*}

As a first step, we verified the coherence in velocity of the filament candidates making use of the SEDIGISM \cco\ and \coo\ data.
For this purpose, we extracted spectra along the skeleton of each filament (i.e. the centre positions of each filament as identified by \citealt{Li2016}), and we analysed the position-velocity diagrams.
We then averaged the spectra in a dilation box of width equal to three SEDIGISM beams (see Fig.\,\ref{fig:atlasgal-example2}) to  derive the central velocity and the width of the detected spectral features.
Finally, we computed integrated intensity maps of the \cco\ and \coo\ lines for each observed velocity component.
We then compared the morphology of the molecular line emission with that of the dust emission to verify their association.

Eleven of the twelve ATLASGAL structures in the \testfield\ were detected in the SEDIGISM data.
The undetected filament (G340.600+00.067) shows only weak dust emission and
is located in the slightly noisier area of the field. Ten identified
structures have a coherent velocity component along the spine (the main
part of the skeleton as identified in ATLASGAL - see Fig.\,\ref{fig:atlasgal-example2} for one example).
The eleventh candidate (G340.630–00.093) shows several velocity components at all positions along the spine and the association with the dust structure is not clear.
Therefore, we exclude this object from the current analysis. Another six structures
(G340.482–00.306, G340.511–00.471, G340.981–00.013, G341.415+00.244, G340.236–00.153, G341.306+00.339) show
additional velocity components which may contribute to the dust
emission.  One of them, G340.236--00.153,  is defined as a network of filaments
by \citet{Li2016} but splits into two networks of filaments at --51.3\,km\,s$^{-1}$ and $-122.1$\,km\,s$^{-1}$ in the $^{13}$CO data. These are labelled as G340.236--00.153  and G340.200--00.035 in Table\,\ref{tab:atlasgal-catalog}.

To estimate the mass of each filament, we computed the $^{13}$CO column density at
each pixel in the dilation box around the spine
by integrating in velocity over a range equal to twice the average FWHM of the $^{13}$CO
component associated with the filament.
Column densities were computed using the $X_{\rm{^{13}CO(2-1)}}$ factor derived from Hi-GAL data (Sect.~\ref{sec:gmc_physical_properties}).
%
We then computed the mass of each filament in its dilation box through the equation:
\begin{equation}
M({\rm H_2}) = \sum_{i} N_i(\rm H_2) A_i \, \mu m_{\rm p},
\end{equation}
where $N_i(\rm H_2)$ is the H$_2$ column density computed for pixel $i$, $A_i$ its area,
$\mu=2.8$ the mean molecular weight,  and $m_{\rm p}$ the proton mass.
We checked for contamination of other structures in the relevant velocity ranges and only in G340.301--00.387  and G340.482--00.306 there is contamination from structures at red-shifted velocities from the main velocity of the filament  up to 20\% and 34\%, respectively.

\begin{figure*}
  \includegraphics[width=0.48\textwidth]{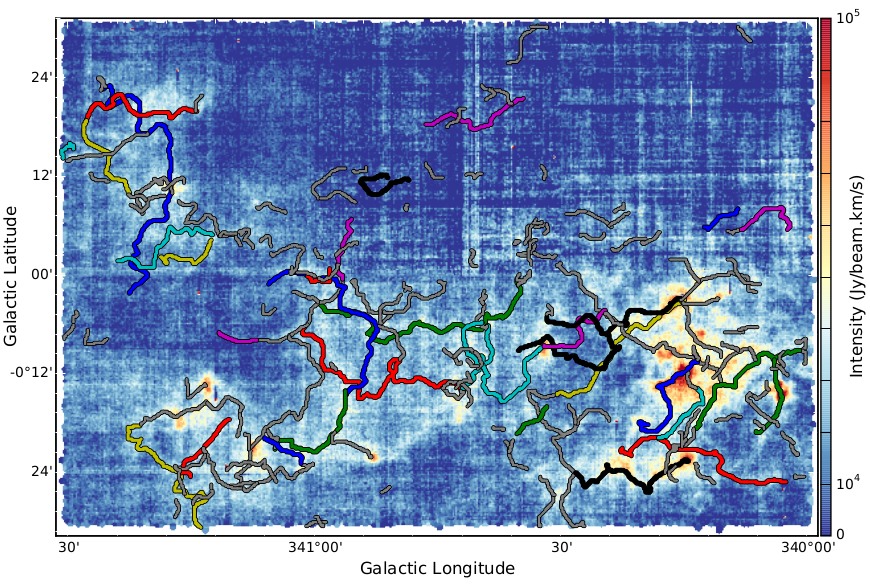}
  \includegraphics[width=0.52\textwidth]{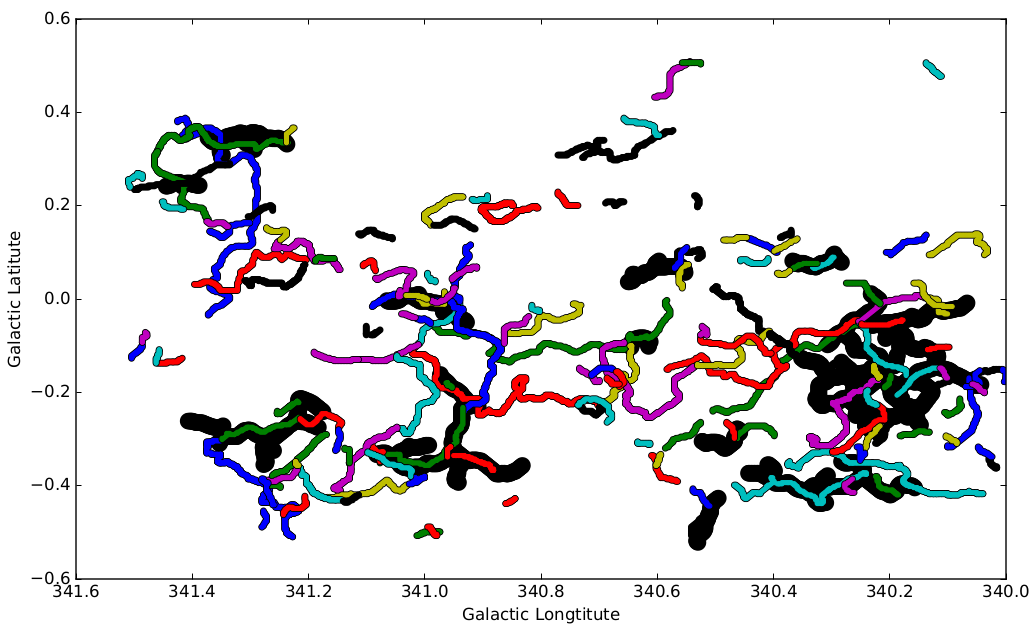}
  \caption{{\em Left}: Spines of the 145 \Disperse\ filaments overlaid on the $\rm^{13}$CO integrated intensity map. The coloured lines correspond to the sub-sample
  of 33 filaments listed in Table~\ref{tab_disperse}.
  {\em Right}: Comparison between the ATLASGAL filaments (thick black lines) and the spines of the 145 \Disperse\ structures (thin coloured lines).}\label{fig_atlasgal_disperse}\label{fig_disperse}
\end{figure*}

The gravitational stability of a filament can be estimated by its linear mass density, or mass per unit length $(M/l)$.
The virial linear mass density, above which a filament without additional support would collapse radially, is given by
$(M/l)_{vir} = 2 \sigma_v^2 /G$, where $\sigma_v$ is the 1-dimensional total (thermal plus non-thermal) velocity dispersion of the average molecular gas \citep{inutsuka1997}.
Following \citet{Hatchell2005}, we derived the velocity dispersion from the average FWHM of the line
along the spine of each filament, excluding regions that show evidence of optically thick emission.
The filamentary networks are too complex to assign a representative length. Therefore, the linear
mass density was computed only for the seven single filaments detected in the field;
the results are reported in Table\,\ref{tab:atlasgal-catalog}.

Two filaments are clearly sub-critical, three are critical and two are just at
the boundary, with $(M/l)_{\rm obs}$ $\approx$ $(M/l)_{\rm vir}$.
The ratio between the observed and virial mass per unit length, $(M/l)_{\rm obs}$/$(M/l)_{\rm vir}$,
is ranging from 0.4 to 3.5. This is somewhat high compared to results found by others
\citep[e.g.][]{Hernandez-Tan,ContrerasMNRAS}, who found $(M/l)_{\rm obs}/(M/l)_{\rm vir}$
significantly less than unity in different filamentary structures.
These studies concluded that filaments may not be gravitationally bound globally,
although star formation is occurring in local regions.
However, our measurements of masses include several uncertainties:
the $X_{\rm{^{13}CO(2-1)}}$ factor (Sect.~\ref{sec:gmc_physical_properties}), has an uncertainty of a factor of 2.
Additionally, opacity effects were not taken into account, for which \citet{Hernandez-Tan} showed
that optical depth correction factors can increase the column density by a factor 2 in the densest
clumps embedded in IRDCs.

\subsection{Identification of filamentary structures on the SEDIGISM data}\label{disperse}

The SEDIGISM data potentially includes more filaments than those identified
in the previous section from the dust emission. Indeed, the velocity
information contained in spectroscopic datasets automatically solves the
problem of blending of unrelated structures along the line of sight, which
can affect continuum surveys. Furthermore, it can reveal the presence of
sub-structures with different velocities that might be not detectable in
the continuum (e.g.\ \citealt{Hacar2013}, \citealt{Henshaw2014}).
To investigate the potential of the SEDIGISM survey to study filaments, we applied
the \Disperse\ algorithm \citep{sousbie2011} directly on the \cco\
data cube of the \testfield. In the following, we briefly describe the methodology used and the
main results; a more detailed description of the method will be given elsewhere (Suri et
al., in prep.).

We applied the \Disperse\ algorithm on a smoothed version of the $\rm^{13}$CO data cube,
convolved to an angular resolution of 50\arcsec\ to improve the signal-to-noise ratio;
the resulting \rms\ is $\sim$0.5~K. \Disperse\,
first calculates a Morse-Smale complex (see \citealt{sousbie2011})
that consists of critical points, persistence pairs, filaments, walls and voids,
with the most significant structures being associated with the highest persistence
values. In our particular case, we used a detection threshold of 3-$\sigma$
(1.5~K), and a persistence threshold of 3, in order to select only
the critical point pairs that stand out. With these criteria,
\Disperse\ creates a skeleton file that contains about 600 structures throughout
the \testfield,
%
including 145 composed of more than 15 pixels
(i.e. more than three "beams" in the data cube smoothed to 50$''$ resolution).
We will focus the rest of our analysis on these 145 filaments.

In the left panel of Fig.\,\ref{fig_disperse}, we plot the skeleton of the 145 filaments
identified by \Disperse.
Most filaments appear as substructures of the GMCs identified by \scimes\ (Sect.\,\ref{sect:gmcs}).
In the right panel of Fig.\,\ref{fig_atlasgal_disperse}, we compare the filaments identified in ATLASGAL with the structures detected by \Disperse.
Overall, there is a good agreement between the filaments identified in the continuum and in the \cco\ data cube, but \Disperse\ finds a larger number of structures compared to ATLASGAL. 
Moreover, some filaments identified in the continuum seem to be composed of several substructures in the data cube.
Thus, the search for filamentary structures directly on the data cube can reveal filaments that
are not detected in continuum images, because of projection effects or sensitivity issues.

In the following we present a basic characterization of the metrics (projected length,
width and aspect ratio) of the filaments identified by \Disperse. The length
of the filament is directly determined by adding the distance of consecutive
skeleton points within a \Disperse\ filament. The mean projected length of the filaments is
about 8.1\,arcmin, with a large dispersion ranging from 1\,arcmin to 50\,arcmin.
The width of each filament is determined from cuts perpendicular to the filament.
A Gaussian function is fitted to the intensity profile along each cut, providing
a measurement of the width at this point. The distribution of widths along a given filament
is computed, and the mean value of this distribution is considered to be
the width of the filament. The mean value and standard deviation for the widths of all
145 filaments is $2.2\pm1.1$\,arcmin. Finally, the aspect
ratio is directly determined from the length-to-width ratio. The mean aspect ratio is
4.2, and ranges from 1 to 40.

Out of the 145 \Disperse\ structures characterised in
this work, only 33 have an aspect ratio $>5$, and can, therefore, be considered
as the best filament candidates.
In Table\,\ref{tab_disperse}, we list their basic properties. Assuming distances to the
filaments in the range 2 to 6~kpc (as determined for GMCs in \,Sect.\,\ref{sec:distances}),
the mean linear length and width are 4.7--14.1~pc and 1.3--3.8~pc, respectively.
The average width is much larger than the typical 0.1~pc value measured towards nearby
star forming regions (e.g.\ \citealt{Andre2010}), which can be explained by our angular
resolution: a 50$''$ (smoothed) beam corresponds to 0.5~pc at a distance of 2~kpc.
But similar widths (in the range 0.6--3.0~pc) were measured by \cite{Wang2015} for
filaments located at 3--5~kpc.
Finally, the lengths determined for the \Disperse\ filaments are consistent with those derived
in the previous section for the ATLASGAL filaments (see Table\,\ref{tab:atlasgal-catalog}).

In summary, \Disperse\ can be applied to the SEDIGISM data to search for
filaments directly in $\ell bv$ data cubes. It is worth mentioning, however, that
some of the structures identified by \Disperse, with a relatively small aspect
ratio, may correspond to fragments of longer structures, which might not be
identified as single large-scale filaments because of, e.g. intensity variations
along the filament. A more detailed study of the spatial distribution, as well
as the velocity relation between different \Disperse\ structures, will be
performed in forthcoming papers, after combining the data of all the SEDIGISM
fields. This analysis may reveal large-scale filamentary structures, similar
to the very long and thin IRDCs `Nessie' \citep{jackson2010,goodman2014},
the Snake \citep{Wang2014} and many others \citep{Wang2015}.

\section{Conclusion and perspectives}\label{sect:conclusion}

We have completed a molecular line survey of the southern \GP, which covers 78\,deg$^2$ in the \cco\ and \coo\ lines at an angular resolution of 30$''$.
The 1-$\sigma$ sensitivity of 0.8\,K at 0.25\kms\ allows us to detect interstellar material
down to a few 10$^{21}$\,cm$^{-2}$ in H$_2$ column density.
This is well suited for mapping the structure of the Galactic ISM at an unprecedented level of detail, from the scale of giant molecular clouds and long filaments down to individual, dense molecular clumps.
The pipeline processing of this massive data set (over $10^7$ independent spectra) is in progress.

In this first overview paper, we have focussed our analysis on a 1.5\,deg$^2$ \testfield, to illustrate the potential of the survey. The main results can be summarised as follows:

\begin{enumerate}
\item Using the \scimes\ algorithm, we have extracted 182 molecular clouds from the $lbv$ data cube within 1.5\,deg$^2$, 58 of which are complexes with at least two sub-structures.
In comparison, \citet{Rice2016} extracted 1064 molecular clouds in the full Galaxy using a similar technique applied to the all-Galaxy CO survey of \citet{dame2001}; this demonstrates the power of our high-resolution data to provide a detailed view of the distribution of molecular clouds in the Milky Way.

\item We estimated the kinematical distances to all the molecular clouds and solved for the distance ambiguity thanks to the \hisa\ method for the vast majority of them.
The distance distribution of these 182 GMCs traces the spiral structure of the Galaxy, providing an accurate view on the position of the near Scutum-Centaurus arm, the near and far Norma arm, and the near and far 3\,kpc arm.

\item From an analysis of their virial parameters, we find that the molecular clouds are generally stable with a median virial parameter of $\sim$2.
However, the majority of the dense clumps within them (as traced by ATLASGAL) are unstable against gravitational collapse.
Moreover, we observe a trend of decreasing virial parameters for increasing cloud and clump masses; the most massive clouds and clumps are the most gravitationally unstable.

\item Eighteen clumps (within 12 GMCs) are associated with massive star forming tracers and these tend to be the most massive and unstable of the 140 dense clumps located in the \testfield. 

\item By combining the \cco\ data of this survey with $^{13}$CO(1\,--\,0) data from the ThrUMMS survey, we are able to solve the radiative transfer equations in order to compute excitation temperature, line opacity and column density in each voxel of the data cube.

\item From the ratio of column density to integrated line intensity, we are also able to compute a \cco-to-H$_2$ X-factor of $1.0 \times 10^{21}$\,cm$^{-2}$\,(K\,km\,s$^{-1}$)$^{-1}\,\pm\,0.2$\,dex. This is in excellent agreement with an independent estimate of this factor based on the comparison between an H$_2$ column density map derived from Hi-GAL data and the \cco\ SEDIGISM data.

\item Most of the filaments previously extracted from continuum surveys (ATLASGAL, Hi-GAL) are detected in $^{13}$CO. The velocity information allows us to confirm that they are coherent structures in $\sim$80\% cases.

\item Using the \Disperse\ algorithm directly on the $^{13}$CO data, we extracted 145 filamentary structures with lengths above 150$''$, 33 of which have an aspect ratio greater than 5.
\end{enumerate}

The \testfield\ covers only $\sim$2\% of the full survey area. Therefore, we can expect to build catalogues with several 10$^4$ molecular clouds, and several 10$^3$ filaments from the entire survey data;
we will also be able to assign distances and investigate their Galactic distribution. We may detect filamentary structures on degree-scales, which would be of prime importance to constrain the formation mechanism of filaments in the ISM. This unique data set will also allow us to put strong constraints on the star-formation efficiency as a function of environment.

Finally, the SEDIGISM survey in the 1~mm band is well complemented with the ThrUMMS survey at 3~mm wavelength,
and with other, ongoing surveys in higher J transitions. 
Combining these data sets allows us to study in 3D the excitation conditions in the Galactic ISM,
at $\la$1$'$ resolution, and to put strong constraints on the combination of excitation temperature
($T_{\rm{ex}}$), optical depth ($\tau$), and the CO-to-H$_2$ conversion factor. 
When also considering the data from the GRS in the 1$^{st}$ quadrant, this will provide us for the first
time with a global, yet detailed view of the bulk of the interstellar matter in the inner Galaxy.

The SEDIGISM data products (calibrated data cubes, catalogues of clouds and filaments) will be
made public shortly after being processed and the quality has been carefully checked.
This will give this survey a high legacy value for Milky Way studies in the southern hemisphere.

\begin{acknowledgements}
We thank the anonymous referee for his/her constructive comments, which helped
us in improving the clarity of the text.
This document was produced using the Overleaf web application, which can be found at www.overleaf.com.
This work has used the CAMELOT project \citep{Ginsburg2016}; http://camelot-project.org.
We would like to thank David Eden for useful discussions about the dense gas fraction.
We are very grateful to Dario Colombo for fruitful discussions and interactions regarding the use of the \scimes\ software.
T.Cs. and K.W. are supported by the Deutsche Forschungsgemeinschaft (DFG) priority programme 1573 ISM-SPP.
A.S.M., P.S., and S.S. are partially supported by the DFG through grant SFB~956, subprojects A4 and A6.
A.D.C. and C.L.D. acknowledge funding from the European Research Council for the FP7 ERC starting grant project LOCALSTAR. L.B. acknowledges support by CONICYT grant PFB-06.
P.J.B. acknowledges grant support from NSF AST-1312597 and NASA-ADAP NNX15AF64G.
\end{acknowledgements}

\bibliography{sedigism}

\begin{appendix}
\section{Full catalogue of clouds in the SEDIGISM \testfield}
\setlength{\tabcolsep}{5pt}
\onecolumn
\begin{longtab}
\begin{landscape}
\begin{longtable}{l l | c c c c c c c c | c c c c c c c | c c}
\caption{\label{tab:full_catalogue} Properties of all clouds from the \scimes\ extraction.
The ID number (Col.~1) shows the catalogue number associated with the cloud.
The GMC name is defined as SDG (for SEDIGISM) followed by the Galactic coordinates of the clouds' centroid.
Cols.\,3 and 4 show the intensity-weighted semi-major and semi-minor axes, $a$ and $b$, respectively;
Col.\,5 shows the position angle (P.\,A.), and Col.\,6 shows the aspect ratio (A.\,R.) defined as $a/b$.
In Cols.\,7-9 we show the centroid velocity, velocity dispersion, and average $^{13}$CO ($2-1$) integrated intensity across the area of the cloud.
Col.\,10 shows the number of dendrogram leaves, N$_{l}$, that make up each GMC.
Col.\,11 shows the adopted distance ($d$) and its uncertainty.
Cols.\,12-14 show the exact area defined by the clouds' masks, the equivalent radius ($R$, assuming circular geometry) and maximum length ($l_{\rm{max}}$).
Cols.\,15-17 show the total mass ($M$), the average surface density ($\Sigma$), and the virial parameter ($\alpha_{\rm{vir}}$).
Col.~18 shows the distance tag ($t_{d}$) that indicates the robustness of the distance determination (the lower the value, the more reliable the distance determination):
$t_{d}$=0 if all three methods agree (\hisa, Hi-GAL distance tool and ATLASGAL), $t_{d}$=1 if no
ATLASGAL counterpart was found and the other two methods agree, $t_{d}$=2 if only two out of three
methods agree, $t_{d}$=3 if distance is drawn from one out of two methods,
and $t_{d}$=4 if drawn from one out of three methods (i.e. \hisa only;
see main text for further information).
Col.~19 is a flag ($t_{edge}$) indicating whether a cloud touches the image edges (0=no, 1=yes). }\\
\hline\hline
 & & \multicolumn{8}{c |}{Measured properties} & \multicolumn{7}{c |}{Physical properties} & & \\
\hline
\multirow{2}{0.5cm}{ID} & \multirow{2}{1.0cm}{Name} & $a$ & $b$  & P.A. & A.R. & \vlsr	& $\sigma_{v}$ & $< W_{\rm{CO}} >$ & N$_{l}$ & $d$ & Area & $R$ & $l_{max}$ & $M$ & $\Sigma$	& $\alpha_{\rm vir}$  &  $t_{d}$ &  $t_{edge}$\\ 
& & (\arcsec) & (\arcsec) & ($^{\circ}$) & & (\kms) & (\kms) & (K\kms) &  & (kpc) & (pc$^{2}$) & (pc) & (pc) & (10$^{3}$M$_{\odot}$) & (M$_{\odot}$pc$^{-2}$) &   &  & \\ 
\hline
\endfirsthead
\caption{continued.}\\
\hline\hline
 & & \multicolumn{8}{c |}{Measured properties} & \multicolumn{7}{c |}{Physical properties} & & \\
\hline
\multirow{2}{0.5cm}{ID} & \multirow{2}{1.0cm}{Name} & $a$ & $b$  & P.A. & A.R. & \vlsr	& $\sigma_{v}$ & $< W_{\rm{CO}} >$ & N$_{l}$ & $d$ & Area & $R$ & $l_{max}$ & $M$ & $\Sigma$	& $\alpha_{\rm vir}$  &  $t_{d}$ &  $t_{edge}$ \\ 
& & (\arcsec) & (\arcsec) & ($^{\circ}$) & & (\kms) & (\kms) & (K\kms) &  & (kpc) & (pc$^{2}$) & (pc) & (pc) & (10$^{3}$M$_{\odot}$) & (M$_{\odot}$pc$^{-2}$) &   &  & \\ 
\hline
\endhead
\hline
\endfoot
\small
6 & SDG340.647+0.166 & 35 & 26 & -160 &  1.3 &  -132.7 &  0.8 &  1.9 & 1 & $  9.04\pm  0.35$ & 19 &    2.5 &    9.7 &      0.8 & 41 &   2.2 & 1 & 0\\
8 & SDG341.440-0.161 & 179 & 52 & 153 &  3.4 &  -130.7 &  1.1 &  3.8 & 2 & $  9.22\pm  0.29$ & 218 &    8.3 &   34.8 &     18.2 & 83 &   0.6 & 1 & 1\\
12 & SDG340.030-0.340 & 43 & 32 & 133 &  1.4 &  -124.9 &  1.4 &  4.7 & 1 & $  9.28\pm  0.30$ & 46 &    3.9 &   11.5 &      4.9 & 104 &   1.9 & 1 & 0\\
15 & SDG340.245-0.056 & 322 & 203 & -161 &  1.6 &  -122.0 &  2.3 &  9.7 & 14 & $  6.58\pm  0.28$ & 915 &   17.1 &   72.2 &    196.9 & 214 &   0.5 & 0 & 0\\
16 & SDG340.720-0.164 & 60 & 37 & 127 &  1.6 &  -126.8 &  0.8 &  3.2 & 1 & $  9.29\pm  0.29$ & 55 &    4.2 &   14.6 &      4.0 & 71 &   0.8 & 1 & 0\\
30 & SDG340.293-0.408 & 91 & 54 & -171 &  1.7 &  -124.0 &  1.1 &  3.7 & 1 & $  6.66\pm  0.29$ & 81 &    5.1 &   17.4 &      6.7 & 81 &   1.0 & 1 & 0\\
31 & SDG340.390-0.002 & 32 & 13 & -160 &  2.5 &  -126.0 &  0.5 &  2.8 & 1 & $  9.28\pm  0.30$ & 13 &    2.1 &    7.1 &      0.9 & 62 &   0.7 & 1 & 0\\
32 & SDG340.675-0.504 & 31 & 19 & -176 &  1.6 &  -125.9 &  0.5 &  2.8 & 1 & $  9.32\pm  0.28$ & 17 &    2.4 &    7.1 &      1.1 & 62 &   0.6 & 1 & 1\\
33 & SDG340.081-0.434 & 60 & 30 & 161 &  2.0 &  -125.0 &  0.5 &  2.9 & 1 & $  6.70\pm  0.30$ & 27 &    3.0 &    9.8 &      1.8 & 65 &   0.5 & 3 & 0\\
34 & SDG340.096-0.252 & 76 & 69 & 110 &  1.1 &  -123.2 &  1.8 &  4.0 & 5 & $  6.63\pm  0.29$ & 51 &    4.1 &   14.5 &      4.7 & 89 &   3.3 & 1 & 0\\
38 & SDG340.534+0.065 & 99 & 39 & 167 &  2.5 &  -123.4 &  1.0 &  2.3 & 2 & $  6.63\pm  0.28$ & 36 &    3.4 &   13.8 &      1.9 & 51 &   2.3 & 3 & 0\\
41 & SDG340.127-0.329 & 69 & 45 & 151 &  1.5 &  -123.3 &  0.9 &  2.3 & 1 & $  6.60\pm  0.29$ & 30 &    3.1 &   10.8 &      1.5 & 49 &   1.9 & 1 & 0\\
49 & SDG340.007+0.090 & 43 & 23 & 134 &  1.8 &  -124.4 &  0.6 &  4.0 & 1 & $  9.30\pm  0.30$ & 30 &    3.1 &   10.1 &      2.7 & 88 &   0.4 & 1 & 1\\
51 & SDG340.062-0.181 & 41 & 30 & -145 &  1.3 &  -122.8 &  1.4 &  5.2 & 1 & $  6.61\pm  0.29$ & 25 &    2.8 &    9.3 &      2.9 & 116 &   2.3 & 1 & 0\\
54 & SDG340.230-0.290 & 46 & 31 & -140 &  1.5 &  -122.1 &  0.8 &  3.7 & 1 & $  6.58\pm  0.28$ & 20 &    2.6 &    8.6 &      1.7 & 81 &   1.0 & 1 & 0\\
55 & SDG340.231-0.236 & 92 & 33 & 176 &  2.8 &  -121.0 &  1.3 &  5.3 & 1 & $  6.54\pm  0.27$ & 40 &    3.6 &   14.4 &      4.7 & 116 &   1.4 & 1 & 0\\
58 & SDG341.297+0.203 & 103 & 65 & 97 &  1.6 &  -122.1 &  0.7 &  3.7 & 2 & $  6.58\pm  0.25$ & 83 &    5.2 &   18.2 &      6.9 & 82 &   0.4 & 3 & 0\\
62 & SDG341.101+0.148 & 103 & 38 & 160 &  2.7 &  -121.0 &  1.2 &  3.9 & 1 & $  6.54\pm  0.25$ & 48 &    3.9 &   16.3 &      4.2 & 86 &   1.5 & 4 & 0\\
64 & SDG340.198-0.411 & 68 & 35 & 125 &  1.9 &  -120.5 &  1.1 &  4.9 & 1 & $  6.53\pm  0.27$ & 40 &    3.6 &   12.4 &      4.4 & 108 &   1.1 & 1 & 0\\
65 & SDG340.393-0.353 & 28 & 19 & 129 &  1.4 &  -122.5 &  0.4 &  2.2 & 1 & $  6.60\pm  0.28$ & 8 &    1.6 &    4.9 &      0.4 & 48 &   0.7 & 1 & 0\\
67 & SDG341.185+0.082 & 47 & 23 & 86 &  2.1 &  -122.2 &  0.6 &  2.2 & 1 & $  6.50\pm  0.25$ & 9 &    1.8 &    7.8 &      0.5 & 47 &   1.6 & 1 & 0\\
68 & SDG340.898+0.194 & 41 & 27 & -175 &  1.5 &  -121.4 &  0.6 &  3.6 & 1 & $  6.56\pm  0.26$ & 17 &    2.4 &    7.7 &      1.4 & 79 &   0.7 & 1 & 0\\
70 & SDG340.717-0.277 & 98 & 25 & 129 &  3.8 &  -120.9 &  0.7 &  4.1 & 2 & $  6.54\pm  0.26$ & 32 &    3.2 &   12.4 &      2.9 & 91 &   0.7 & 3 & 0\\
71 & SDG340.520-0.067 & 54 & 26 & 134 &  2.1 &  -119.9 &  1.2 &  3.4 & 1 & $  6.50\pm  0.26$ & 21 &    2.6 &    8.9 &      1.6 & 75 &   2.8 & 2 & 0\\
77 & SDG340.015-0.143 & 44 & 18 & -174 &  2.3 &  -120.7 &  0.7 &  4.2 & 1 & $  6.53\pm  0.28$ & 16 &    2.3 &    7.9 &      1.5 & 93 &   1.0 & 1 & 1\\
78 & SDG341.473-0.119 & 96 & 32 & 176 &  3.0 &  -120.8 &  0.5 &  3.4 & 1 & $  9.58\pm  0.25$ & 66 &    4.6 &   17.4 &      5.1 & 75 &   0.3 & 1 & 1\\
79 & SDG340.329-0.174 & 35 & 19 & 161 &  1.9 &  -120.4 &  0.6 &  3.0 & 1 & $  6.52\pm  0.27$ & 10 &    1.8 &    6.2 &      0.7 & 65 &   1.3 & 3 & 0\\
80 & SDG340.434+0.082 & 61 & 37 & 158 &  1.6 &  -119.3 &  1.0 &  2.6 & 1 & $  6.48\pm  0.26$ & 31 &    3.1 &   12.4 &      1.8 & 57 &   2.1 & 3 & 0\\
84 & SDG340.409+0.029 & 28 & 15 & 105 &  1.9 &  -119.8 &  0.8 &  4.6 & 1 & $  9.52\pm  0.27$ & 18 &    2.4 &    7.8 &      1.8 & 102 &   0.9 & 1 & 0\\
85 & SDG340.672+0.035 & 83 & 55 & 104 &  1.5 &  -119.1 &  1.5 &  2.7 & 2 & $  6.47\pm  0.26$ & 40 &    3.6 &   13.7 &      2.4 & 59 &   3.7 & 3 & 0\\
86 & SDG341.468+0.070 & 63 & 24 & 80 &  2.6 &  -119.3 &  0.7 &  3.4 & 1 & $  9.63\pm  0.24$ & 48 &    3.9 &   12.9 &      3.6 & 75 &   0.7 & 0 & 0\\
88 & SDG340.380+0.123 & 101 & 31 & -171 &  3.3 &  -118.1 &  1.6 &  3.4 & 2 & $  6.44\pm  0.26$ & 37 &    3.4 &   15.6 &      2.8 & 74 &   3.6 & 2 & 0\\
91 & SDG340.600+0.047 & 48 & 17 & 178 &  2.8 &  -119.2 &  0.7 &  1.9 & 1 & $  9.56\pm  0.26$ & 18 &    2.4 &   10.9 &      0.8 & 42 &   1.9 & 1 & 0\\
92 & SDG341.066+0.162 & 34 & 18 & 160 &  1.8 &  -117.0 &  0.8 &  3.4 & 1 & $  9.68\pm  0.24$ & 25 &    2.8 &    8.3 &      1.9 & 74 &   1.3 & 1 & 0\\
93 & SDG340.878-0.085 & 132 & 59 & -147 &  2.2 &  -116.3 &  1.1 &  3.1 & 2 & $  9.68\pm  0.25$ & 185 &    7.7 &   34.5 &     12.7 & 68 &   0.8 & 0 & 0\\
95 & SDG340.058-0.180 & 56 & 28 & 127 &  2.0 &  -116.0 &  0.6 &  4.1 & 1 & $  6.36\pm  0.26$ & 24 &    2.8 &   10.2 &      2.2 & 90 &   0.5 & 1 & 0\\
98 & SDG340.777-0.291 & 42 & 37 & -169 &  1.1 &  -115.5 &  0.9 &  2.9 & 1 & $  6.35\pm  0.25$ & 24 &    2.8 &    9.3 &      1.6 & 63 &   1.6 & 2 & 0\\
99 & SDG340.798-0.021 & 40 & 22 & -159 &  1.8 &  -115.7 &  0.7 &  3.9 & 2 & $  9.70\pm  0.25$ & 34 &    3.3 &   11.0 &      3.0 & 86 &   0.6 & 1 & 0\\
105 & SDG340.298+0.088 & 65 & 43 & -143 &  1.5 &  -110.8 &  1.9 &  5.9 & 2 & $  9.83\pm  0.25$ & 104 &    5.8 &   21.0 &     13.6 & 130 &   1.9 & 1 & 0\\
107 & SDG340.552+0.112 & 61 & 30 & 75 &  2.0 &  -113.6 &  0.8 &  4.0 & 2 & $  9.75\pm  0.25$ & 70 &    4.7 &   17.3 &      6.3 & 89 &   0.6 & 1 & 0\\
108 & SDG340.466+0.127 & 34 & 28 & -170 &  1.2 &  -113.6 &  0.8 &  3.0 & 1 & $  9.74\pm  0.25$ & 31 &    3.2 &   10.5 &      2.1 & 66 &   1.1 & 1 & 0\\
111 & SDG340.559+0.171 & 28 & 17 & 127 &  1.6 &  -113.7 &  0.5 &  2.0 & 1 & $  6.10\pm  0.25$ & 4 &    1.2 &    4.4 &      0.2 & 44 &   2.1 & 1 & 0\\
114 & SDG340.327+0.131 & 40 & 14 & -178 &  2.9 &  -104.9 &  0.6 &  1.9 & 1 & $  5.97\pm  0.24$ & 5 &    1.3 &    4.7 &      0.2 & 42 &   2.3 & 1 & 0\\
115 & SDG341.337+0.056 & 219 & 80 & -157 &  2.7 &  -102.6 &  1.4 &  2.7 & 1 & $ 10.17\pm  0.23$ & 421 &   11.6 &   56.4 &     25.3 & 60 &   1.1 & 1 & 0\\
116 & SDG340.772-0.085 & 405 & 87 & -178 &  4.6 &  -100.7 &  1.5 &  3.4 & 3 & $  5.70\pm  0.24$ & 274 &    9.3 &   50.5 &     20.6 & 75 &   1.2 & 2 & 0\\
122 & SDG340.854+0.269 & 32 & 20 & -156 &  1.6 &   -99.4 &  0.6 &  2.4 & 1 & $ 10.26\pm  0.24$ & 23 &    2.8 &    9.2 &      1.3 & 52 &   1.0 & 1 & 0\\
123 & SDG340.434+0.268 & 33 & 14 & 140 &  2.2 &   -98.4 &  0.4 &  1.8 & 1 & $ 10.27\pm  0.25$ & 14 &    2.1 &    7.8 &      0.6 & 40 &   0.8 & 1 & 0\\
124 & SDG340.164+0.122 & 91 & 35 & -159 &  2.5 &   -95.6 &  1.1 &  3.3 & 1 & $ 10.35\pm  0.25$ & 98 &    5.6 &   24.8 &      7.2 & 72 &   1.0 & 1 & 0\\
127 & SDG340.193-0.369 & 364 & 146 & 153 &  2.5 &   -90.7 &  1.6 &  5.1 & 9 & $  5.47\pm  0.26$ & 275 &    9.4 &   44.0 &     31.3 & 113 &   0.9 & 2 & 1\\
143 & SDG340.887-0.465 & 142 & 73 & 163 &  1.9 &   -89.9 &  0.9 &  2.4 & 1 & $  5.50\pm  0.25$ & 90 &    5.4 &   18.1 &      4.9 & 53 &   1.1 & 1 & 1\\
147 & SDG340.547-0.443 & 207 & 96 & 118 &  2.2 &   -90.3 &  1.1 &  4.8 & 4 & $  5.47\pm  0.25$ & 81 &    5.1 &   22.7 &      8.7 & 106 &   0.8 & 2 & 1\\
150 & SDG340.559-0.289 & 75 & 44 & -139 &  1.7 &   -90.5 &  0.9 &  2.8 & 1 & $ 10.55\pm  0.25$ & 114 &    6.0 &   20.3 &      7.1 & 61 &   0.8 & 1 & 0\\
154 & SDG340.448-0.282 & 94 & 46 & 54 &  2.0 &   -88.6 &  0.9 &  5.4 & 1 & $  5.40\pm  0.26$ & 50 &    4.0 &   15.3 &      6.1 & 120 &   0.6 & 2 & 0\\
157 & SDG340.003-0.419 & 67 & 33 & 86 &  2.0 &   -88.5 &  0.6 &  3.0 & 1 & $  5.38\pm  0.26$ & 21 &    2.6 &    9.7 &      1.4 & 65 &   0.8 & 3 & 1\\
161 & SDG340.544+0.171 & 28 & 20 & -140 &  1.4 &   -82.5 &  0.5 &  1.9 & 1 & $ 10.84\pm  0.27$ & 18 &    2.4 &    7.4 &      0.7 & 41 &   0.8 & 1 & 0\\
162 & SDG340.527+0.251 & 33 & 18 & -155 &  1.8 &   -82.4 &  0.5 &  2.7 & 1 & $ 10.84\pm  0.27$ & 26 &    2.9 &   10.2 &      1.6 & 60 &   0.5 & 1 & 0\\
163 & SDG341.495+0.068 & 56 & 21 & 119 &  2.7 &   -80.9 &  0.9 &  2.4 & 1 & $ 10.93\pm  0.26$ & 45 &    3.8 &   14.2 &      2.5 & 53 &   1.4 & 1 & 1\\
164 & SDG341.298+0.347 & 212 & 63 & 178 &  3.4 &   -78.1 &  1.5 &  6.6 & 3 & $  5.07\pm  0.27$ & 141 &    6.7 &   27.8 &     20.6 & 145 &   0.9 & 2 & 0\\
167 & SDG340.765-0.049 & 32 & 16 & 84 &  1.9 &   -79.6 &  0.6 &  4.0 & 1 & $  5.00\pm  0.27$ & 5 &    1.3 &    4.0 &      0.5 & 88 &   1.0 & 1 & 0\\
172 & SDG341.416-0.075 & 36 & 17 & 147 &  2.2 &   -78.7 &  0.7 &  2.6 & 1 & $  4.90\pm  0.27$ & 4 &    1.2 &    4.2 &      0.3 & 56 &   2.3 & 1 & 0\\
173 & SDG341.247+0.030 & 139 & 62 & -137 &  2.2 &   -76.2 &  1.1 &  5.0 & 3 & $  5.00\pm  0.27$ & 51 &    4.1 &   16.0 &      5.7 & 111 &   0.9 & 2 & 0\\
175 & SDG340.811+0.141 & 29 & 15 & 124 &  1.9 &   -77.4 &  0.5 &  2.2 & 1 & $ 11.04\pm  0.27$ & 17 &    2.4 &    9.0 &      0.8 & 47 &   0.8 & 1 & 0\\
178 & SDG341.431+0.258 & 191 & 55 & 133 &  3.5 &   -76.0 &  1.1 &  2.8 & 2 & $  5.00\pm  0.27$ & 55 &    4.2 &   20.1 &      3.5 & 62 &   1.8 & 1 & 1\\
181 & SDG340.892+0.184 & 155 & 64 & -161 &  2.4 &   -74.9 &  1.5 &  3.2 & 3 & $  4.92\pm  0.28$ & 60 &    4.4 &   18.5 &      4.4 & 71 &   2.7 & 3 & 0\\
184 & SDG340.735+0.043 & 38 & 18 & 93 &  2.0 &   -75.8 &  0.6 &  2.1 & 1 & $  4.90\pm  0.28$ & 5 &    1.3 &    4.7 &      0.3 & 46 &   2.4 & 1 & 0\\
185 & SDG340.848+0.263 & 30 & 13 & 112 &  2.3 &   -77.1 &  0.4 &  1.8 & 1 & $  5.00\pm  0.27$ & 2 &    1.0 &    3.1 &      0.1 & 40 &   1.6 & 1 & 0\\
189 & SDG340.717+0.353 & 77 & 26 & -149 &  2.9 &   -75.0 &  0.7 &  2.2 & 1 & $ 11.13\pm  0.28$ & 79 &    5.0 &   20.5 &      3.8 & 47 &   0.7 & 1 & 0\\
194 & SDG340.587+0.350 & 235 & 90 & -173 &  2.6 &   -71.9 &  1.1 &  3.4 & 4 & $  4.78\pm  0.29$ & 124 &    6.3 &   35.6 &      9.5 & 75 &   1.0 & 3 & 0\\
198 & SDG340.745+0.209 & 52 & 41 & 128 &  1.3 &   -72.8 &  0.5 &  1.8 & 1 & $  4.83\pm  0.29$ & 9 &    1.7 &    7.1 &      0.4 & 39 &   1.2 & 1 & 0\\
199 & SDG340.773+0.238 & 35 & 12 & 99 &  2.7 &   -72.6 &  0.4 &  1.7 & 1 & $ 11.23\pm  0.29$ & 17 &    2.3 &    7.9 &      0.6 & 36 &   0.7 & 1 & 0\\
204 & SDG341.295+0.065 & 105 & 26 & 171 &  4.0 &   -69.0 &  1.0 &  4.0 & 1 & $  4.90\pm  0.29$ & 16 &    2.3 &    9.8 &      1.4 & 89 &   1.7 & 2 & 0\\
205 & SDG341.275+0.381 & 86 & 50 & -161 &  1.7 &   -70.3 &  0.9 &  2.1 & 1 & $ 11.33\pm  0.29$ & 113 &    6.0 &   26.9 &      5.4 & 47 &   1.0 & 1 & 0\\
216 & SDG341.150+0.147 & 87 & 17 & -178 &  5.1 &   -62.7 &  0.7 &  2.7 & 1 & $  4.40\pm  0.31$ & 8 &    1.7 &    7.7 &      0.5 & 59 &   1.8 & 1 & 0\\
234 & SDG340.054-0.214 & 151 & 105 & 70 &  1.4 &   -51.9 &  2.6 & 15.8 & 5 & $  3.86\pm  0.36$ & 62 &    4.5 &   17.6 &     22.0 & 350 &   1.6 & 0 & 1\\
243 & SDG340.429-0.063 & 89 & 35 & 140 &  2.5 &   -56.1 &  1.5 &  3.0 & 1 & $  3.80\pm  0.34$ & 10 &    1.8 &    7.4 &      0.7 & 66 &   6.6 & 1 & 0\\
257 & SDG340.571+0.004 & 62 & 39 & 49 &  1.6 &   -53.3 &  2.5 &  3.5 & 1 & $  3.70\pm  0.35$ & 11 &    1.9 &    6.7 &      0.9 & 77 &  16.2 & 1 & 0\\
258 & SDG340.118+0.490 & 55 & 27 & 144 &  2.0 &   -54.7 &  1.3 &  9.0 & 3 & $  4.00\pm  0.35$ & 9 &    1.7 &    5.4 &      1.8 & 199 &   1.8 & 1 & 1\\
260 & SDG340.240-0.213 & 180 & 147 & 148 &  1.2 &   -48.9 &  3.5 & 17.4 & 9 & $  3.73\pm  0.37$ & 137 &    6.6 &   23.8 &     53.3 & 386 &   1.7 & 0 & 0\\
264 & SDG340.071-0.090 & 70 & 39 & -165 &  1.8 &   -53.8 &  1.0 &  4.7 & 1 & $  4.10\pm  0.35$ & 14 &    2.2 &    7.3 &      1.5 & 105 &   1.8 & 1 & 0\\
269 & SDG340.030-0.040 & 90 & 36 & 144 &  2.5 &   -52.3 &  1.9 &  6.0 & 1 & $  3.88\pm  0.36$ & 16 &    2.3 &    8.3 &      2.1 & 132 &   4.5 & 1 & 1\\
271 & SDG340.582+0.069 & 114 & 88 & -157 &  1.3 &   -47.1 &  3.6 &  3.3 & 6 & $  3.66\pm  0.38$ & 38 &    3.5 &   13.5 &      2.8 & 73 &  18.4 & 1 & 0\\
273 & SDG340.102-0.329 & 86 & 46 & 67 &  1.9 &   -53.5 &  0.8 &  9.9 & 2 & $  3.95\pm  0.35$ & 18 &    2.4 &    9.1 &      4.1 & 219 &   0.4 & 0 & 0\\
292 & SDG340.712-0.267 & 70 & 66 & 63 &  1.1 &   -49.6 &  1.6 &  7.8 & 2 & $  3.80\pm  0.37$ & 25 &    2.9 &    9.2 &      4.4 & 172 &   1.9 & 0 & 0\\
295 & SDG340.417-0.026 & 105 & 74 & 179 &  1.4 &   -49.1 &  1.6 &  7.2 & 2 & $  3.75\pm  0.37$ & 31 &    3.2 &   11.5 &      5.1 & 158 &   1.8 & 0 & 0\\
297 & SDG341.098-0.472 & 37 & 23 & -152 &  1.6 &   -50.2 &  1.7 &  4.0 & 1 & $  3.86\pm  0.37$ & 4 &    1.2 &    4.5 &      0.4 & 87 &  10.2 & 1 & 0\\
298 & SDG340.300-0.395 & 228 & 92 & 176 &  2.5 &   -49.1 &  2.9 & 14.3 & 7 & $  3.74\pm  0.37$ & 76 &    4.9 &   20.1 &     24.1 & 317 &   2.0 & 0 & 0\\
301 & SDG340.140-0.051 & 42 & 15 & 98 &  2.7 &   -51.2 &  0.9 &  4.3 & 1 & $  3.50\pm  0.36$ & 2 &    0.9 &    2.7 &      0.2 & 95 &   3.8 & 1 & 0\\
303 & SDG340.541-0.394 & 60 & 33 & -166 &  1.8 &   -49.0 &  1.7 &  6.2 & 1 & $  3.76\pm  0.37$ & 10 &    1.8 &    7.0 &      1.4 & 136 &   4.5 & 1 & 0\\
315 & SDG340.617-0.268 & 91 & 47 & 93 &  1.9 &   -47.6 &  1.9 &  5.1 & 1 & $  3.69\pm  0.38$ & 13 &    2.1 &    7.7 &      1.5 & 113 &   5.9 & 1 & 0\\
322 & SDG340.622-0.433 & 66 & 33 & 151 &  2.0 &   -48.1 &  1.6 &  3.7 & 1 & $  3.40\pm  0.38$ & 5 &    1.3 &    5.1 &      0.4 & 83 &   9.2 & 1 & 0\\
323 & SDG340.339-0.306 & 48 & 39 & 52 &  1.2 &   -48.1 &  2.6 &  9.5 & 2 & $  3.70\pm  0.38$ & 10 &    1.8 &    5.9 &      2.2 & 211 &   6.3 & 0 & 0\\
324 & SDG340.115-0.111 & 57 & 28 & 147 &  2.0 &   -49.3 &  1.2 &  6.6 & 1 & $  3.74\pm  0.37$ & 8 &    1.6 &    5.4 &      1.3 & 147 &   2.3 & 0 & 0\\
325 & SDG340.655-0.095 & 44 & 37 & -172 &  1.2 &   -46.3 &  2.6 &  6.5 & 1 & $  3.63\pm  0.38$ & 9 &    1.8 &    6.0 &      1.4 & 143 &   9.7 & 1 & 0\\
326 & SDG340.106-0.012 & 38 & 20 & 103 &  1.9 &   -49.8 &  1.1 &  4.3 & 1 & $  3.76\pm  0.37$ & 3 &    1.1 &    3.7 &      0.4 & 96 &   4.1 & 1 & 0\\
329 & SDG340.754-0.489 & 78 & 56 & -142 &  1.4 &   -46.0 &  2.3 &  4.8 & 2 & $  3.62\pm  0.39$ & 16 &    2.3 &    7.8 &      1.8 & 107 &   8.0 & 1 & 1\\
331 & SDG340.532-0.147 & 168 & 76 & 62 &  2.2 &   -46.6 &  1.4 &  9.4 & 4 & $  3.63\pm  0.38$ & 65 &    4.6 &   19.3 &     13.6 & 207 &   0.8 & 0 & 0\\
337 & SDG340.676-0.161 & 102 & 67 & 128 &  1.5 &   -46.8 &  1.4 &  4.9 & 1 & $  3.66\pm  0.38$ & 23 &    2.7 &   10.0 &      2.6 & 108 &   2.4 & 1 & 0\\
340 & SDG340.896+0.416 & 101 & 43 & 51 &  2.3 &   -48.3 &  0.6 &  2.1 & 2 & $ 12.31\pm  0.37$ & 101 &    5.7 &   27.4 &      4.7 & 45 &   0.4 & 1 & 0\\
342 & SDG340.374-0.492 & 99 & 25 & 167 &  3.9 &   -47.7 &  0.7 &  3.3 & 1 & $  3.68\pm  0.38$ & 8 &    1.7 &    7.6 &      0.7 & 73 &   1.5 & 1 & 1\\
348 & SDG341.102-0.079 & 137 & 27 & -149 &  5.0 &   -47.0 &  1.1 &  3.3 & 2 & $  3.80\pm  0.38$ & 18 &    2.4 &   11.6 &      1.3 & 72 &   2.4 & 1 & 0\\
350 & SDG340.390+0.137 & 135 & 80 & 166 &  1.7 &   -46.0 &  1.1 &  4.0 & 2 & $  3.59\pm  0.39$ & 37 &    3.5 &   13.4 &      3.4 & 89 &   1.5 & 1 & 0\\
354 & SDG340.911-0.324 & 224 & 173 & 128 &  1.3 &   -45.6 &  1.5 &  9.6 & 3 & $  3.61\pm  0.39$ & 86 &    5.2 &   23.7 &     18.3 & 212 &   0.8 & 0 & 0\\
356 & SDG340.062-0.304 & 83 & 53 & -173 &  1.6 &   -47.2 &  0.8 &  2.7 & 1 & $  3.63\pm  0.38$ & 14 &    2.1 &    8.5 &      0.9 & 60 &   1.9 & 1 & 1\\
360 & SDG340.003-0.119 & 45 & 20 & 148 &  2.2 &   -48.3 &  0.6 &  4.3 & 2 & $  3.68\pm  0.37$ & 3 &    1.0 &    3.6 &      0.3 & 96 &   1.5 & 1 & 1\\
364 & SDG341.215-0.178 & 28 & 16 & 160 &  1.8 &   -47.8 &  0.4 &  2.6 & 1 & $  3.75\pm  0.38$ & 2 &    0.8 &    2.6 &      0.1 & 57 &   1.4 & 1 & 0\\
369 & SDG340.930-0.058 & 228 & 78 & 131 &  2.9 &   -46.3 &  1.1 &  3.8 & 2 & $  3.65\pm  0.38$ & 48 &    3.9 &   18.1 &      4.1 & 84 &   1.3 & 1 & 0\\
370 & SDG340.733-0.001 & 46 & 25 & -171 &  1.8 &   -44.6 &  1.6 &  2.4 & 1 & $  3.40\pm  0.39$ & 4 &    1.1 &    4.3 &      0.2 & 52 &  16.9 & 1 & 0\\
371 & SDG340.518+0.040 & 31 & 24 & -172 &  1.3 &   -45.1 &  1.5 &  3.5 & 1 & $  3.40\pm  0.39$ & 3 &    1.0 &    3.4 &      0.2 & 78 &  11.3 & 1 & 0\\
373 & SDG340.490-0.468 & 113 & 85 & 114 &  1.3 &   -43.8 &  1.4 &  6.7 & 1 & $  3.48\pm  0.40$ & 25 &    2.8 &   11.1 &      3.7 & 148 &   1.6 & 1 & 1\\
375 & SDG341.260-0.276 & 238 & 107 & -153 &  2.2 &   -44.4 &  1.2 & 12.5 & 8 & $  3.57\pm  0.40$ & 105 &    5.8 &   23.0 &     29.2 & 277 &   0.3 & 0 & 0\\
388 & SDG340.566-0.448 & 52 & 15 & -138 &  3.4 &   -44.9 &  0.9 &  3.4 & 1 & $  3.54\pm  0.39$ & 2 &    0.9 &    3.7 &      0.2 & 75 &   4.2 & 1 & 0\\
390 & SDG341.448-0.405 & 73 & 38 & 97 &  1.9 &   -45.7 &  0.7 &  2.9 & 1 & $  3.66\pm  0.39$ & 9 &    1.8 &    6.3 &      0.6 & 64 &   1.6 & 1 & 0\\
392 & SDG340.516-0.294 & 135 & 92 & 144 &  1.5 &   -40.7 &  3.2 &  4.8 & 2 & $  3.31\pm  0.42$ & 33 &    3.3 &   12.4 &      3.6 & 106 &  11.2 & 0 & 0\\
399 & SDG341.240-0.370 & 40 & 22 & 82 &  1.7 &   -45.7 &  0.5 &  2.5 & 1 & $  3.64\pm  0.39$ & 2 &    1.0 &    3.5 &      0.2 & 55 &   2.0 & 1 & 0\\
403 & SDG340.844+0.078 & 71 & 44 & 107 &  1.6 &   -41.4 &  1.6 &  2.5 & 1 & $  3.38\pm  0.41$ & 10 &    1.8 &    6.4 &      0.6 & 54 &   9.3 & 1 & 0\\
408 & SDG341.010-0.151 & 229 & 110 & 89 &  2.1 &   -42.3 &  1.1 &  6.1 & 6 & $  3.44\pm  0.41$ & 70 &    4.7 &   20.2 &      9.5 & 134 &   0.7 & 0 & 0\\
412 & SDG341.359-0.429 & 60 & 32 & 47 &  1.9 &   -44.0 &  0.5 &  2.9 & 1 & $  3.56\pm  0.40$ & 4 &    1.2 &    4.9 &      0.3 & 65 &   1.4 & 1 & 0\\
418 & SDG341.197-0.113 & 137 & 54 & 178 &  2.5 &   -42.3 &  0.7 &  3.1 & 4 & $  3.46\pm  0.41$ & 19 &    2.5 &   11.1 &      1.4 & 68 &   1.0 & 1 & 0\\
420 & SDG341.123-0.351 & 90 & 60 & 107 &  1.5 &   -41.6 &  1.0 & 11.9 & 3 & $  3.41\pm  0.41$ & 29 &    3.0 &    9.9 &      7.7 & 263 &   0.5 & 0 & 0\\
429 & SDG340.216+0.332 & 90 & 33 & 133 &  2.7 &   -41.1 &  1.5 &  2.9 & 1 & $  3.31\pm  0.41$ & 9 &    1.7 &    6.8 &      0.6 & 64 &   7.4 & 1 & 0\\
430 & SDG340.711-0.215 & 128 & 100 & -142 &  1.3 &   -35.6 &  2.6 &  6.0 & 2 & $  3.03\pm  0.45$ & 41 &    3.7 &   14.6 &      5.6 & 133 &   5.3 & 0 & 0\\
433 & SDG340.938+0.050 & 111 & 54 & 64 &  2.1 &   -41.9 &  0.7 &  4.2 & 3 & $  3.41\pm  0.41$ & 20 &    2.6 &   10.0 &      1.9 & 92 &   0.8 & 1 & 0\\
434 & SDG340.441+0.012 & 32 & 21 & 105 &  1.5 &   -41.8 &  0.7 &  2.2 & 1 & $  3.40\pm  0.41$ & 2 &    0.9 &    3.0 &      0.1 & 49 &   4.0 & 1 & 0\\
437 & SDG340.616+0.096 & 29 & 22 & 99 &  1.4 &   -42.5 &  0.6 &  1.6 & 1 & $  3.40\pm  0.41$ & 1 &    0.7 &    2.3 &      0.1 & 35 &   5.7 & 1 & 0\\
446 & SDG340.854-0.174 & 207 & 111 & 82 &  1.9 &   -40.1 &  1.2 &  3.2 & 1 & $  3.30\pm  0.42$ & 40 &    3.6 &   15.0 &      2.9 & 71 &   2.0 & 1 & 0\\
450 & SDG340.985+0.182 & 130 & 65 & 154 &  2.0 &   -40.1 &  0.6 &  2.5 & 2 & $  3.32\pm  0.42$ & 27 &    3.0 &   11.8 &      1.6 & 56 &   0.9 & 1 & 0\\
451 & SDG340.101-0.500 & 97 & 21 & 176 &  4.5 &   -40.1 &  1.7 &  2.8 & 2 & $  3.30\pm  0.42$ & 6 &    1.4 &    5.9 &      0.4 & 63 &  12.1 & 1 & 1\\
458 & SDG340.790+0.019 & 35 & 13 & 127 &  2.5 &   -41.3 &  0.7 &  1.4 & 1 & $  3.37\pm  0.41$ & 0 &    0.5 &    2.3 &      0.0 & 31 &  10.3 & 1 & 0\\
462 & SDG340.578-0.106 & 100 & 76 & 132 &  1.3 &   -39.6 &  0.8 &  3.0 & 1 & $  3.25\pm  0.42$ & 22 &    2.7 &    9.4 &      1.5 & 65 &   1.2 & 1 & 0\\
463 & SDG341.025+0.009 & 65 & 29 & -163 &  2.3 &   -40.1 &  0.7 &  4.5 & 1 & $  3.30\pm  0.42$ & 6 &    1.5 &    5.6 &      0.7 & 100 &   1.1 & 0 & 0\\
465 & SDG340.816-0.421 & 86 & 51 & 119 &  1.7 &   -39.4 &  0.8 &  2.5 & 1 & $  3.20\pm  0.42$ & 14 &    2.1 &    7.5 &      0.8 & 56 &   1.9 & 1 & 0\\
466 & SDG340.950-0.282 & 74 & 44 & -170 &  1.7 &   -37.9 &  1.6 &  3.7 & 1 & $  3.19\pm  0.43$ & 8 &    1.7 &    6.9 &      0.7 & 82 &   6.6 & 0 & 0\\
470 & SDG340.364-0.133 & 334 & 64 & -137 &  5.2 &   -37.1 &  1.2 &  5.6 & 2 & $  3.09\pm  0.44$ & 51 &    4.1 &   22.0 &      6.4 & 123 &   1.1 & 1 & 0\\
472 & SDG340.630+0.383 & 55 & 37 & 49 &  1.5 &   -39.1 &  0.5 &  1.5 & 1 & $  3.23\pm  0.43$ & 4 &    1.2 &    4.7 &      0.1 & 34 &   2.1 & 3 & 0\\
473 & SDG340.685-0.326 & 29 & 17 & 60 &  1.7 &   -39.1 &  0.7 &  2.4 & 1 & $  3.24\pm  0.43$ & 2 &    0.8 &    2.5 &      0.1 & 52 &   3.8 & 1 & 0\\
475 & SDG340.088-0.020 & 120 & 69 & 138 &  1.7 &   -38.3 &  0.7 &  3.1 & 1 & $  3.14\pm  0.43$ & 16 &    2.3 &   10.0 &      1.1 & 68 &   1.3 & 3 & 0\\
477 & SDG340.712+0.336 & 54 & 43 & 90 &  1.3 &   -39.6 &  0.5 &  1.8 & 1 & $ 12.78\pm  0.42$ & 98 &    5.6 &   19.7 &      4.0 & 40 &   0.5 & 1 & 0\\
478 & SDG341.386-0.428 & 211 & 94 & 45 &  2.2 &   -36.4 &  1.0 &  2.8 & 1 & $  3.13\pm  0.45$ & 39 &    3.6 &   16.2 &      2.5 & 61 &   1.9 & 1 & 1\\
479 & SDG340.674+0.059 & 66 & 61 & 179 &  1.1 &   -39.2 &  0.7 &  2.0 & 2 & $  3.10\pm  0.43$ & 8 &    1.6 &    5.5 &      0.4 & 45 &   2.6 & 1 & 0\\
481 & SDG340.541+0.439 & 75 & 50 & 55 &  1.5 &   -38.0 &  0.5 &  2.1 & 3 & $ 12.87\pm  0.43$ & 128 &    6.4 &   25.6 &      5.9 & 45 &   0.4 & 1 & 0\\
482 & SDG341.398+0.262 & 180 & 103 & -166 &  1.7 &   -36.9 &  1.0 &  3.3 & 2 & $  3.16\pm  0.44$ & 61 &    4.4 &   16.7 &      4.5 & 73 &   1.0 & 1 & 1\\
484 & SDG341.109-0.430 & 137 & 87 & -174 &  1.6 &   -34.7 &  0.8 &  4.4 & 2 & $  3.00\pm  0.46$ & 27 &    3.0 &   11.2 &      2.7 & 98 &   0.9 & 1 & 0\\
485 & SDG340.062-0.238 & 36 & 15 & 152 &  2.3 &   -37.4 &  0.8 &  3.4 & 1 & $  3.09\pm  0.44$ & 2 &    0.8 &    2.7 &      0.2 & 74 &   3.5 & 1 & 0\\
487 & SDG341.342+0.156 & 76 & 58 & -166 &  1.3 &   -36.5 &  1.3 &  3.6 & 1 & $  3.14\pm  0.45$ & 14 &    2.2 &    8.4 &      1.2 & 79 &   3.7 & 1 & 0\\
492 & SDG341.140+0.078 & 315 & 119 & 157 &  2.6 &   -32.0 &  3.7 &  3.1 & 3 & $  2.83\pm  0.48$ & 60 &    4.4 &   20.3 &      4.2 & 69 &  17.2 & 1 & 0\\
497 & SDG341.105+0.064 & 56 & 32 & 128 &  1.7 &   -36.5 &  0.5 &  3.6 & 1 & $  2.70\pm  0.45$ & 4 &    1.2 &    3.9 &      0.4 & 79 &   1.1 & 1 & 0\\
501 & SDG340.969-0.122 & 66 & 41 & 161 &  1.6 &   -34.8 &  1.1 &  2.4 & 1 & $  3.00\pm  0.46$ & 6 &    1.5 &    4.8 &      0.4 & 53 &   5.4 & 1 & 0\\
502 & SDG340.009-0.371 & 56 & 47 & 113 &  1.2 &   -35.0 &  1.4 &  4.0 & 1 & $  2.94\pm  0.45$ & 7 &    1.6 &    5.0 &      0.7 & 89 &   5.4 & 1 & 1\\
504 & SDG340.505-0.076 & 63 & 34 & 159 &  1.8 &   -34.2 &  1.0 &  2.2 & 1 & $  2.70\pm  0.46$ & 4 &    1.2 &    4.8 &      0.2 & 49 &   6.1 & 1 & 0\\
505 & SDG341.287-0.260 & 51 & 19 & 91 &  2.7 &   -35.3 &  0.5 &  2.3 & 1 & $  3.05\pm  0.46$ & 2 &    0.9 &    3.3 &      0.1 & 52 &   2.3 & 1 & 0\\
508 & SDG341.212-0.345 & 146 & 42 & -136 &  3.4 &   -30.4 &  1.6 &  5.5 & 5 & $  2.00\pm  0.49$ & 7 &    1.5 &    7.7 &      0.9 & 121 &   5.2 & 1 & 0\\
509 & SDG341.041-0.360 & 148 & 33 & 148 &  4.4 &   -32.3 &  1.0 &  4.9 & 4 & $ 13.23\pm  0.48$ & 199 &    8.0 &   36.8 &     21.8 & 109 &   0.4 & 1 & 0\\
511 & SDG340.643+0.046 & 36 & 18 & 90 &  2.0 &   -34.4 &  0.7 &  1.9 & 1 & $  2.70\pm  0.46$ & 1 &    0.6 &    2.2 &      0.1 & 42 &   6.2 & 1 & 0\\
512 & SDG341.122+0.133 & 61 & 28 & 174 &  2.2 &   -34.1 &  0.5 &  2.0 & 1 & $  2.97\pm  0.46$ & 4 &    1.2 &    4.1 &      0.2 & 44 &   1.9 & 1 & 0\\
517 & SDG340.237-0.231 & 108 & 36 & 119 &  3.0 &   -33.5 &  0.7 &  2.5 & 1 & $  2.86\pm  0.46$ & 9 &    1.7 &    7.1 &      0.5 & 54 &   2.0 & 1 & 0\\
518 & SDG340.284+0.079 & 42 & 15 & 79 &  2.8 &   -33.7 &  0.7 &  2.1 & 1 & $  2.88\pm  0.46$ & 1 &    0.7 &    2.3 &      0.1 & 47 &   5.5 & 1 & 0\\
520 & SDG341.290-0.471 & 121 & 57 & -135 &  2.1 &   -31.8 &  1.0 &  3.3 & 2 & $  2.84\pm  0.48$ & 9 &    1.8 &    6.8 &      0.7 & 72 &   3.0 & 1 & 1\\
525 & SDG341.327+0.219 & 422 & 219 & 104 &  1.9 &   -23.6 &  2.2 &  6.2 & 13 & $  2.27\pm  0.56$ & 147 &    6.9 &   26.7 &     20.4 & 138 &   1.9 & 0 & 1\\
528 & SDG341.090-0.389 & 69 & 40 & 140 &  1.7 &   -30.9 &  0.9 &  2.3 & 1 & $  2.00\pm  0.49$ & 3 &    1.0 &    4.1 &      0.2 & 50 &   5.9 & 1 & 0\\
530 & SDG340.334-0.096 & 36 & 34 & 157 &  1.1 &   -30.1 &  1.3 &  2.8 & 1 & $  2.70\pm  0.49$ & 3 &    1.0 &    3.3 &      0.2 & 62 &  10.0 & 1 & 0\\
538 & SDG340.016+0.214 & 60 & 43 & 159 &  1.4 &   -30.6 &  0.9 &  3.3 & 1 & $  2.66\pm  0.48$ & 6 &    1.4 &    4.3 &      0.5 & 73 &   2.7 & 1 & 1\\
541 & SDG340.063+0.116 & 125 & 93 & 176 &  1.3 &   -27.8 &  1.8 &  4.4 & 2 & $  2.48\pm  0.51$ & 24 &    2.8 &    9.2 &      2.3 & 97 &   4.3 & 3 & 1\\
545 & SDG340.991+0.007 & 92 & 55 & 136 &  1.7 &   -27.3 &  2.3 &  3.2 & 1 & $  2.51\pm  0.52$ & 10 &    1.8 &    6.2 &      0.7 & 71 &  14.8 & 1 & 0\\
547 & SDG341.210-0.444 & 65 & 44 & -158 &  1.5 &   -27.8 &  1.5 &  2.5 & 1 & $  2.56\pm  0.52$ & 5 &    1.3 &    4.3 &      0.3 & 54 &  11.5 & 1 & 0\\
550 & SDG340.995-0.499 & 53 & 28 & -176 &  1.9 &   -29.2 &  1.0 &  4.0 & 1 & $  2.64\pm  0.50$ & 3 &    1.1 &    3.5 &      0.3 & 89 &   3.9 & 3 & 1\\
551 & SDG341.286+0.064 & 48 & 15 & 157 &  3.1 &   -29.7 &  0.6 &  1.9 & 1 & $  2.69\pm  0.50$ & 1 &    0.7 &    2.8 &      0.1 & 41 &   4.7 & 1 & 0\\
553 & SDG341.032-0.367 & 76 & 59 & 122 &  1.3 &   -25.6 &  2.1 &  3.4 & 1 & $  2.39\pm  0.54$ & 7 &    1.6 &    5.3 &      0.6 & 75 &  13.4 & 1 & 0\\
560 & SDG340.011+0.024 & 50 & 40 & -154 &  1.3 &   -26.7 &  1.2 &  2.9 & 1 & $  2.40\pm  0.52$ & 2 &    0.9 &    3.2 &      0.2 & 63 &   9.9 & 3 & 1\\
561 & SDG340.346-0.194 & 31 & 24 & 100 &  1.3 &   -27.7 &  0.3 &  2.1 & 1 & $  2.49\pm  0.51$ & 1 &    0.6 &    2.2 &      0.1 & 47 &   1.1 & 3 & 0\\
576 & SDG341.273-0.466 & 42 & 28 & 130 &  1.5 &   -23.7 &  0.6 &  3.3 & 1 & $  2.00\pm  0.56$ & 1 &    0.8 &    2.5 &      0.1 & 72 &   2.1 & 1 & 0\\
593 & SDG340.961-0.180 & 39 & 23 & 120 &  1.7 &   -14.0 &  1.1 &  5.5 & 1 & $  2.00\pm  0.67$ & 1 &    0.7 &    2.4 &      0.2 & 122 &   5.3 & 0 & 0\\
596 & SDG340.912-0.035 & 86 & 33 & 174 &  2.5 &   -13.8 &  1.1 &  2.9 & 1 & $  2.00\pm  0.67$ & 3 &    1.0 &    4.1 &      0.2 & 64 &   6.7 & 1 & 0\\
603 & SDG340.163-0.189 & 83 & 26 & 46 &  3.1 &    -8.5 &  1.0 &  3.1 & 1 & $  0.91\pm  0.74$ & 0 &    0.4 &    1.5 &      0.0 & 67 &  16.0 & 0 & 0\\
607 & SDG341.234-0.321 & 31 & 21 & 47 &  1.5 &    -5.5 &  0.9 &  4.4 & 1 & $ 15.46\pm  8.18$ & 61 &    4.4 &   16.1 &      6.0 & 97 &   0.7 & 1 & 0\\
609 & SDG340.017-0.102 & 65 & 37 & -143 &  1.8 &    -4.6 &  0.9 &  2.8 & 1 & $ 15.47\pm  8.23$ & 168 &    7.3 &   25.7 &     10.3 & 61 &   0.6 & 1 & 1\\
611 & SDG340.133-0.191 & 38 & 32 & 85 &  1.2 &    -3.3 &  1.0 &  4.5 & 1 & $  0.38\pm  8.39$ & 0 &    0.2 &    0.5 &      0.0 & 100 &  24.6 & 0 & 0\\
612 & SDG340.841-0.440 & 62 & 35 & 159 &  1.8 &     1.4 &  1.1 &  3.2 & 1 & $ 16.24\pm  8.28$ & 166 &    7.3 &   26.1 &     11.8 & 70 &   0.9 & 1 & 0\\
613 & SDG340.316-0.104 & 43 & 18 & -159 &  2.3 &     1.7 &  0.7 &  2.9 & 1 & $ 16.21\pm  8.28$ & 83 &    5.1 &   17.6 &      5.3 & 64 &   0.5 & 1 & 0\\
614 & SDG340.481+0.365 & 71 & 34 & 146 &  2.1 &     0.6 &  0.2 &  1.6 & 1 & $ 16.09\pm  8.15$ & 179 &    7.5 &   26.4 &      6.2 & 34 &   0.0 & 1 & 0\\
616 & SDG340.064-0.024 & 43 & 18 & 148 &  2.3 &     4.2 &  0.9 &  2.5 & 1 & $ 16.50\pm  8.59$ & 68 &    4.7 &   15.8 &      3.8 & 55 &   1.1 & 1 & 0\\
618 & SDG340.491-0.036 & 51 & 33 & -174 &  1.5 &     4.1 &  0.1 &  1.2 & 1 & $ 16.54\pm  8.61$ & 127 &    6.4 &   22.4 &      3.3 & 26 &   0.0 & 1 & 0\\
619 & SDG340.544+0.473 & 145 & 78 & -153 &  1.9 &     5.1 &  0.3 &  2.1 & 2 & $ 16.68\pm  8.75$ & 815 &   16.1 &   61.9 &     37.4 & 45 &   0.0 & 1 & 1\\
\end{longtable}
\end{landscape}
\end{longtab}

\section{Properties of the \Disperse\ filaments}

\begin{table*}[!h]
\centering
\caption{Properties of the 33 \Disperse\ filaments with aspect ratios $>$5.
Cols. 2 to 5 give the positions of the extrema (minimum and maximum in Galactic coordinates).
Cols. 6 and 7 indicate the projected length and width; the aspect ratio is given in Col. 8.}\label{tab_disperse}
\begin{tabular}{l l l l l l l l l}
\hline\hline
\multicolumn{1}{c}{Name}
& \multicolumn{1}{c}{gl1}
& \multicolumn{1}{c}{gb1}
& \multicolumn{1}{c}{gl2} 
& \multicolumn{1}{c}{gb2}
& \multicolumn{1}{c}{Length}
& \multicolumn{1}{c}{Width}
& \multicolumn{1}{c}{R}
\\

& \multicolumn{1}{c}{$^\circ$}
& \multicolumn{1}{c}{$^\circ$}
& \multicolumn{1}{c}{$^\circ$}
& \multicolumn{1}{c}{$^\circ$}
& \multicolumn{1}{c}{(arcmin)}
& \multicolumn{1}{c}{(arcmin)}
& \multicolumn{1}{c}{}
\\
\hline
Dis001 & 341.4252 & $+$0.3797 & 341.3725 & $-$0.0346 & 48.85    & 1.27 & 38.6 \\
Dis002 & 340.9872 & $-$0.0557 & 340.5834 & $-$0.0029 & 43.08    & 1.18 & 36.4 \\
Dis003 & 340.5043 & $-$0.2404 & 340.2457 & $-$0.0478 & 25.98    & 0.81 & 32.2 \\
Dis004 & 340.3749 & $-$0.3565 & 340.0398 & $-$0.4172 & 28.93    & 1.70 & 17.0 \\
Dis005 & 341.0901 & $-$0.0161 & 340.9318 & $-$0.2404 & 35.07    & 2.19 & 16.0 \\
Dis006 & 341.0162 & $-$0.1032 & 340.6995 & $-$0.2299 & 40.39    & 2.88 & 14.0 \\
Dis007 & 340.8051 & $+$0.1950 & 340.8473 & $+$0.2003 & 13.50    & 0.96 & 14.0 \\
Dis008 & 341.2273 & $-$0.5096 & 341.3514 & $-$0.3038 & 27.86    & 2.15 & 13.0 \\
Dis009 & 341.2062 & $+$0.0736 & 341.3012 & $+$0.0314 & 10.96    & 0.88 & 12.4 \\
Dis010 & 340.5834 & $-$0.1507 & 340.2536 & $-$0.0478 & 36.53    & 2.83 & 12.9 \\
Dis011 & 340.6573 & $-$0.0927 & 340.5333 & $-$0.1428 & 27.58    & 2.62 & 10.5 \\
Dis012 & 341.3962 & $+$0.0314 & 341.2036 & $+$0.0841 & 20.11    & 1.91 & 10.5 \\
Dis013 & 340.6995 & $-$0.2325 & 340.6468 & $-$0.1006 & 15.65    & 1.62 & 9.7 \\
Dis014 & 340.7708 & $+$0.3085 & 340.5729 & $+$0.3612 & 17.30    & 1.78 & 9.7 \\
Dis015 & 340.1005 & $-$0.3170 & 340.0636 & $-$0.1692 & 12.20    & 1.35 & 9.1 \\
Dis016 & 340.2245 & $-$0.3090 & 340.0081 & $-$0.1507 & 22.71    & 2.53 & 9.0 \\
Dis017 & 340.2245 & $-$0.1560 & 340.2984 & $-$0.2220 & 25.69    & 2.92 & 8.8 \\
Dis018 & 340.5254 & $-$0.0900 & 340.4040 & $-$0.1164 & 11.41    & 1.56 & 7.3 \\
Dis019 & 341.4595 & $+$0.3190 & 341.2352 & $+$0.3348 & 21.29    & 2.94 & 7.2 \\
Dis020 & 340.5861 & $-$0.3196 & 340.5254 & $-$0.2642 & 6.18 & 0.86 & 7.2 \\
Dis021 & 341.5097 & $+$0.2478 & 341.5070 & $+$0.2398 & 5.22 & 0.74 & 7.1 \\
Dis022 & 341.0769 & $-$0.3513 & 340.9344 & $-$0.2325 & 17.41    & 2.56 & 6.8 \\
Dis023 & 341.4648 & $+$0.2847 & 341.3725 & $+$0.1660 & 16.94    & 2.59 & 6.5 \\
Dis024 & 341.1904 & $-$0.1164 & 341.1138 & $-$0.1296 & 5.71 & 0.91 & 6.3 \\
Dis025 & 341.0109 & $-$0.0003 & 340.9661 & $+$0.0155 & 5.33 & 0.89 & 6.0 \\
Dis026 & 340.4673 & $-$0.3988 & 340.2456 & $-$0.3724 & 24.93    & 4.40 & 5.7 \\
Dis027 & 341.0980 & $-$0.3302 & 341.0241 & $-$0.3803 & 7.72 & 1.37 & 5.6 \\
Dis028 & 340.9397 & $-$0.0056 & 340.9212 & $+$0.1158 & 9.66 & 1.80 & 5.4 \\
Dis029 & 340.5306 & $-$0.1428 & 340.4066 & $-$0.0742 & 13.91    & 2.65 & 5.2 \\
Dis030 & 340.1322 & $+$0.0947 & 340.0372 & $+$0.0947 & 11.27    & 2.21 & 5.1 \\
Dis031 & 341.2484 & $-$0.4040 & 341.1693 & $-$0.2906 & 12.49    & 2.51 & 5.0 \\
Dis032 & 340.2984 & $-$0.3275 & 340.2193 & $-$0.2272 & 10.87    & 2.17 & 5.0 \\
Dis033 & 340.2061 & $+$0.1000 & 340.1375 & $+$0.1369 & 6.28 & 1.26 & 5.0 \\
\hline
\end{tabular}
\end{table*}

\end{appendix}

\end{document}